\documentclass[letterpaper,11pt]{article}
\pdfoutput=1
\usepackage{jheppub}
\usepackage{slashed}
\usepackage{graphicx}
\usepackage{subfigure}
\usepackage{soul}
\usepackage{amsmath}
\usepackage{multirow}
\usepackage{tablefootnote}
\usepackage{epstopdf}

\usepackage[dvipsnames]{xcolor}

\newcommand{\beq}{\begin{equation}}
\newcommand{\eeq}{\end{equation}}
\newcommand{\bea}{\begin{eqnarray}}
\newcommand{\eea}{\end{eqnarray}}

\def\figureautorefname~#1\null{Fig.\,#1\null}

\def\equationautorefname~#1\null{Eq.\,(#1)\null}

\def\m1{M_1}
\def\m2{M_2}
\def\m3{M_3}

\def\ch10{\tilde \chi^0_1}

\def\tev{\,{\rm TeV}}
\def\gev{\,{\rm GeV}}

\def\to{\rightarrow}

\newcommand{\lsim}{\mathrel{\mathop{\kern 0pt \rlap
  {\raise.2ex\hbox{$<$}}}
  \lower.9ex\hbox{\kern-.190em $\sim$}}}
\newcommand{\gsim}{\mathrel{\mathop{\kern 0pt \rlap
  {\raise.2ex\hbox{$>$}}}
  \lower.9ex\hbox{\kern-.190em $\sim$}}}

\definecolor{pink}{RGB}{255,105,180}

\newcommand{\ee}{{e^{-} e^{+}}}

\newcommand{\bpm}{\begin{pmatrix}}
\newcommand{\epm}{\end{pmatrix}}

\newcommand{\eehz}{e^+e^- \to hZ}
\newcommand{\eevvh}{e^+e^- \to \nu \bar{\nu} h}
\newcommand{\eeww}{e^+e^- \to WW}

\newcommand{\inab}{\,{\rm ab}^{-1}}
\newcommand{\infb}{\,{\rm fb}^{-1}}

\title{
Learning from Higgs Physics at Future Higgs Factories}

\author[\dagger,\ddagger]{Jiayin Gu,}
\author[*,\star]{Honglei Li,}
\author[\diamond]{Zhen Liu,}
\author[\star]{Shufang Su,}
\author[\star,\circ,\#]{Wei Su}

\affiliation[*]{School of Physics and Technology, University of Jinan, Jinan, Shandong 250022, China }
\affiliation[\dagger]{DESY, Notkestra{\ss}e 85, D-22607 Hamburg, Germany}
\affiliation[\ddagger]{Center for Future High Energy Physics, Institute of High Energy Physics, 19B YuquanLu, Chinese Academy of Sciences, Beijing 100049, China}
\affiliation[\star]{Department of Physics, University of Arizona, Tucson, AZ 85721}
\affiliation[\diamond]{Theoretical Physics Department, Fermi National Accelerator Laboratory, Batavia, IL 60510}
\affiliation[\#]{ CAS Key Laboratory of Theoretical Physics, Institute of Theoretical Physics,\\
Chinese Academy of Sciences, Beijing 100190, China}

\affiliation[\circ]{School of Physics, University of Chinese Academy of Sciences, Beijing 100049, China}

\emailAdd{jiayin.gu@desy.de, sps$\_$lihl@ujn.edu.cn, shufang@email.arizona.edu, weisv@itp.ac.cn, zliu2@fnal.gov}

\abstract{
Future Higgs factories can reach impressive precision on Higgs property measurements.  In this paper, instead of conventional focus of Higgs precision in certain interaction bases, we explored its sensitivity to new physics models at the   electron-positron colliders.  In particular, we studied two categories of new physics models, Standard Model (SM) with a real scalar singlet extension, and Two Higgs Double Model (2HDM) as   examples of weakly-interacting models, Minimal Composite Higgs Model (MCHM) and three typical patterns of the more general operator counting for  strong interacting models as examples of strong dynamics.  We performed a global fit to various Higgs search channels to obtain the 95\%\,C.L. constraints on the model parameter space.  In the SM with a singlet extension, we obtained the limits on the singlet-doublet mixing angle $\sin\theta$, as well as the more general Wilson coefficients of the induced higher dimensional operators.
 In the 2HDM, we analyzed tree level effects in  $\tan\beta$ vs. $\cos(\beta-\alpha)$  plane, as well as the one-loop contributions from the heavy Higgs bosons in the alignment limit to obtain the constraints on heavy Higgs masses for different types of 2HDM. In strong dynamics models, we obtained lower limits on the strong dynamics scale.  In addition, once deviations of Higgs couplings are observed, they can be used to distinguish different models.  We also compared the sensitivity of various future Higgs factories, namely Circular Electron Positron Collider (CEPC), Future Circular Collider (FCC)-ee and International Linear Collider (ILC).  }


\preprint{
\begin{flushright}
DESY 17-130  \\
FERMILAB-PUB-17-348-T
\end{flushright}
}

\begin{document}
\maketitle
\flushbottom


\section{Introduction}
\label{sec:intro}
The discovery of a Standard Model (SM)-like Higgs boson at the Large Hadron Collider (LHC) strongly motivates the studies of its properties.  Lepton colliders running at center of mass energies of around 240\,GeV or above are ideal Higgs factory machines for the precision studies of the Higgs boson properties.
Several plans for future lepton colliders, in particular, Higgs factories,  have been proposed, including the Circular Electron Positron Collider (CEPC) in China \cite{CEPC-SPPCStudyGroup:2015csa}, the electron-positron stage of the Future Circular Collider (FCC-ee) at CERN (previously known as TLEP \cite{Gomez-Ceballos:2013zzn}), and the International Linear Collider (ILC) in Japan \cite{Baer:2013cma}.   Through the Higgsstrahlung process, $\eehz$, Higgs cross sections can be measured to an astonishing precision: about 0.2\%$-$0.5\% for ``inclusive'' $\eehz$ cross section and $\eehz, h\to b\bar b$ exclusive cross section, and about a few percent for many other exclusive cross sections, under typical running scenarios.
The Compact Linear Collider (CLIC), while only planning to run at energies 350\,GeV and above, could also probe the Higgs couplings very well by measuring the $WW$ fusion process at high energies \cite{CLIC:2016zwp, Abramowicz:2016zbo}, as well as measuring  the top Yukawa and the triple Higgs couplings.  These colliders could provide exciting opportunities in Higgs studies and greatly improve our understanding of the physics at the TeV scale.  Any deviation of the Higgs couplings from their SM predictions is a strong indication of new physics beyond the SM.  Even if no observation is observed, it will provide tight constraints on new physics models.

In the study of Higgs boson properties and the sensitivity of Higgs precision measurements to new physics models, two ``model-independent" approaches are usually taken.   One is the so-called ``\emph{kappa}'' framework, in which $\kappa_i$ is defined as the Higgs coupling to SM particles, normalized to its SM value. The kappa framework features the simple and direct connection to measurable cross sections and a straight forward inclusion of  new light degree of freedom beyond the SM, e.g., exotic Higgs decays~\cite{Curtin:2013fra,Liu:2016zki,Liu:2017lpo}.  Deviation of $\kappa_i$ from their SM expectation indicates contributions from new physics \cite{Dawson:2013bba,deFlorian:2016spz,Han:2013kya,Peskin:2013xra,Giardino:2013bma,Belanger:2013xza,Bechtle:2014ewa,Cheung:2014noa,Barger:2014qva,Fichet:2015xla,Reece:2016sdb,Lafaye:2017kgf}.
 Another approach using the language of the Standard Model Effective Field Theory (SMEFT) for Higgs couplings has also been extensively studied, both at the LHC and at future colliders \cite{Elias-Miro:2013mua, Pomarol:2013zra, Corbett:2012dm, Corbett:2012ja, Corbett:2013pja, Falkowski:2015fla, Butter:2016cvz, Craig:2014una, Beneke:2014sba, Craig:2015wwr, Ellis:2015sca, Ge:2016zro, Ellis:2017kfi, Durieux:2017rsg, Barklow:2017suo, Barklow:2017awn}.   The fitting parameters are the coefficients of various higher dimensional operators.    For both approaches, global fits to the SM observed values of $\sigma\times{\rm BR}$ for various Higgs search channels are usually performed to obtain robust constraints and correlations on $\kappa_i$ couplings in the kappa framework,  or Wilson coefficients in EFT framework. These fitting results are then further used to impose constraints on parameter spaces of a specific model when model parameters are mapped to either those $\kappa_i$ couplings or Wilson coefficients.

Such model independent approaches are very useful in evaluating the performance of measurements and providing general constraints on new physics models.  However,   they have certain disadvantages.  The precision reaches of the global fits in such approaches usually suffer from the large level of degeneracy among fitting parameters.  For specific models, the parameter space is usually much smaller, with various Higgs couplings determined by a small set of parameters.    Furthermore, while the coupling fit results  are provided in the official documents for each of the colliders,  the corresponding correlation matrix is often not provided.   For poorly determined quantities, linear approximation might not be sufficient.    Directly applying these coupling fitting results alone on specific models  without including correlation matrix  may result in overly conservative estimations.  A specific example of this is shown later in \autoref{sec:2hdm}.   Moreover, in most model independent approaches,  certain model assumptions and simplifications are usually made, which may not be valid for a particular new physics model.  Typical examples are the omission of Higgs couplings with Lorentz structures different from the SM ones ({\it e.g.} $hZ_{\mu\nu}Z^{\mu\nu}$ vs.  the SM $hZ^\mu Z_\mu$) in the  kappa framework, or the absence of possible light beyond the SM (BSM) states in SMEFT.  The latter may also lead to Higgs exotic decays, which is hard to study in a model independent way, while many appealing BSM scenarios with light states can be sufficiently probed at $\ee$ colliders~\cite{Liu:2016zki}.
 These disadvantages severely limit the usefulness of  model independent methods when applied to specific BSM models.

To demonstrate the sensitivity of Higgs factories to BSM new physics models and access the capability of various machine options, it is thus important to directly study the implication of Higgs precision measurements on model parameter space. We focused on a few generic types of new physics models in our study.   For weakly coupled models, we used the SM plus a real scalar singlet, and   Two Higgs doublet Model (2HDM) as two prototypes,  which are incorporated as the Higgs sector for many BSM scenarios such as supersymmetric models, and left-right symmetric models.  In particular, we studied the tree level effects in general case, as well as the one-loop contributions from the heavy Higgs bosons.   For models with strong dynamics, we studied the Minimal Composite Higgs Model (MCHM), including all ten different Fermion embeddings,  as well as three typical patterns  of the more generic operator counting for strong interacting models with a light Higgs.    With these two classes of Higgs extension models covering both weakly interacting theories and strong dynamics, we could develop a better understanding of the physics potential of future Higgs factories.

We performed a global fit to various Higgs search channels in those two classes of specific models.  In the case of no deviations   observed in the future Higgs factories,  a 95\% C.L. constrained parameter space was obtained.   For the SM with a singlet extension, we obtained limits on the singlet-doublet mixing angle $\sin\theta$, as well as the more general Wilson coefficients of the induced higher dimensional operators.   For the 2HDM tree level effects, it tells how much deviation from the alignment limit we can tolerate.  For loop effects under the alignment limit, a constrained range of heavy Higgs masses can be obtained. In the MCHM, we translated the Higgs precision measurements into constraints on the vacuum misalignment parameter $\xi$, which further constrain the composite scale $f$.  In the three cases of strong interacting models with a light Higgs, we  obtained the lower limits on the strong dynamics scale $m_*$ as a function of the strong interaction coupling $g_*$.    In cases of an observed deviation of the SM Higgs couplings, our studies also demonstrated how to distinguish different models.

While our analyses mostly used the CEPC results, comparison with the reach of FCC-ee and ILC was also performed.  Run-I, Run-II and high luminosity LHC Higgs precision measurements were included as well.   We also compared the sensitivity of fitting using $\Delta(\sigma\times {\rm BR})/(\sigma\times {\rm BR})_{\rm SM}$ of variously Higgs search channels directly, with the fitting results using the constraints on effective couplings $\kappa_i$ from existing studies.

The rest of this paper is organized as follows.  In \autoref{sec:input} we summarized the run scenarios and the estimated precisions of Higgs measurements for various future lepton collider Higgs factories, gathered from the corresponding official documents.  These  inputs were used to obtain constraints in the model parameter spaces.  In  \autoref{sec:fit}, we presented the global fitting method used in our analyses.  In \autoref{sec:singlet}, we started with the simple case of the SM plus a singlet as a warm up.  The results for Two Higgs doublet Models were presented in \autoref{sec:2hdm} and the results for Composite Higgs Models were presented in \autoref{sec:comp}.  We concluded in \autoref{sec:conclusion}.  The LHC Run-I Higgs measurement results and the projected precisions for future LHC runs were collected in \autoref{app:lhcinput}.  We also listed the formulae of loop corrections to various Higgs couplings in the 2HDM along the alignment limit in \autoref{app:2HDMloop}.


\section{The Higgs precision measurements at future lepton colliders}
\label{sec:input}

At future lepton colliders, the dominant channel to measure the Higgs boson properties is the Higgsstrahlung process, $\eehz$, at center of mass energies of around 240$-$250\,GeV.  Due to the nature of lepton colliders, both the inclusive cross section, $\sigma(hZ)$, and the exclusive ones of different Higgs decays in terms of $\sigma(hZ)\times {\rm BR}$, can be measured to remarkable precisions.  The invisible decay width of the Higgs can also be very well constrained.  In addition, the  cross section of $WW$ fusion process for Higgs production  grows with energy.  While it can not be measured very well at 240$-$250\,GeV, at higher center of mass energies (in particular at linear colliders), such fusion process becomes significantly more important and can provide crucial complementary information. For $\sqrt{s}>500$ GeV, $tth$ production can also be used as well.

\begin{table}[h]
 \begin{center}
 \begin{tabular}{|l|r|r|r|r|r|r|r|r|r|r|}
   \hline
   collider& \multicolumn{1}{c|}{CEPC}& \multicolumn{1}{c|}{FCC-ee}&\multicolumn{6}{c|}{ILC} \\
   \hline
   $\sqrt{s}$     &  $\text{240\,GeV} $ &  $\text{240\,GeV}$  &  \text{250\,GeV}  &
   \multicolumn{2}{c|}{\text{350\,GeV}}  & \multicolumn{3}{c|}{\text{500\,GeV}} \\
   $\int{\mathcal{L}}dt $     &  $\text{5 ab}^{-1} $ &  $\text{10 ab}^{-1}$   &  $\text{2 ab}^{-1} $  &
   \multicolumn{2}{c|}{$\text{200 fb}^{-1}$}  & \multicolumn{3}{c|}{$\text{4 ab}^{-1}$} \\
   \hline
    \hline
production& $Zh$  & $Zh$   & $Zh$      & $Zh$     & $\nu\bar{\nu}h$     & $Zh$     & $\nu\bar{\nu}h$ & $t\bar{t}h$ \\
   \hline
  $\Delta \sigma / \sigma$ & 0.51\%  & 0.4\% & 0.71\% & 2.1\% & - & 1.06 & - & - \\ \hline \hline
   decay & \multicolumn{8}{c|}{$\Delta (\sigma \cdot BR) / (\sigma \cdot BR)$}  \\
  \hline
   $h \to b\bar{b}$              &  0.28\%               & 0.2\%                      &  0.42\%    & 1.67\%         & 1.67\%                  & 0.64\%    & 0.25\%           & 9.9\%        \\

   $h \to c\bar{c}$              & 2.2\%                   & 1.2\%                   & 2.9\%          & 12.7\%    & 16.7\%                   & 4.5\%     & 2.2\%           & -           \\

   $h \to gg$                    & 1.6\%                   & 1.4\%                  & 2.5\%           & 9.4\%    & 11.0\%                  & 3.9\%     & 1.5\%           & -           \\

   $h \to WW^*$                  & 1.5\%                   & 0.9\%                   & 1.1\%          & 8.7\%    & 6.4\%                  & 3.3\%    & 0.85\%           & -           \\

   $h \to \tau^+\tau^-$         & 1.2\%                   & 0.7\%                   &2.3\%           &4.5\%          & 24.4\%                  & 1.9\%    & 3.2\%           & -           \\

   $h \to ZZ^*$                  & 4.3\%                   & 3.1\%                   & 6.7\%        & 28.3\%     & 21.8\%                   & 8.8\%     & 2.9\%           & -           \\

   $h \to \gamma\gamma$          & 9.0\%                   & 3.0\%                   & 12.0\%     & 43.7\%     &50.1\%                   & 12.0\%   &6.7\% & - \\

   $h \to \mu^+\mu^-$           & 17\%                   & 13\%                  & 25.5\%        & 97.6\%     & 179.8\%                  & 31.1\%     & 25.5\%            & -           \\
   \hline
    $(\nu\bar\nu)h \to b\bar{b}$  & 2.8\%       &   2.2\%       &       3.7\% & -  & -  & -  & -  & -  \\
    \hline
  \end{tabular}
  \caption{Estimated statistical precisions for Higgs measurements obtained at  the proposed CEPC program with 5 ab$^{-1}$ integrated luminosity~\cite{CEPC-SPPCStudyGroup:2015csa},     FCC-ee program with 10 ab$^{-1}$ integrated luminosity~\cite{Gomez-Ceballos:2013zzn},  and  ILC with various center of mass energies~\cite{Barklow:2015tja}.   }
\label{tab:mu_precision}
  \end{center}
\end{table}

To set up the baseline of our study, we hereby listed the run scenarios of various machines in terms of center of mass energy and the corresponding integrated luminosity, as well as the estimated precisions of relevant Higgs measurements that we used in our global analyses:
\begin{itemize}

    \item  {\bf CEPC~~}  According to the preCDR \cite{CEPC-SPPCStudyGroup:2015csa}, CEPC plans to collect $5\inab$ data at 240\,GeV.  The estimations on the measurements of the Higgsstrahlung process $\eehz$ with various final states, as well as the $WW$ fusion process with Higgs decaying to bottom pairs ($\eevvh, h\to b\bar{b}$), are summarized in \autoref{tab:mu_precision}.
     \item  {\bf FCC-ee~~}  The FCC-ee CDR is expected to be finished by  2018 \cite{fccplan}. At the current moment, the TLEP whitepaper \cite{Gomez-Ceballos:2013zzn} still provides the most updated estimations, which assumes  total luminosities of $10\inab$ at 240\,GeV and $2.6\inab$ at 350\,GeV.  The estimations for the $\eehz$ measurements at 240\,GeV, as well as $h\rightarrow b\bar{b}$ channel in $WW$ fusion are listed   in \autoref{tab:mu_precision}.    Assuming statistical uncertainties only, Ref.~\cite{Gomez-Ceballos:2013zzn} also estimated that the  $WW$ fusion process with $h\rightarrow bb$ can be measured to a precision of   0.6\% at  350 \,GeV.   We did not include 350 GeV $WW$ fusion process in our global fit   since it has  little impact on the constrained model parameter spaces considered in our study.

    \item  {\bf ILC~~}  The proposed run scenarios in the ILC TDR  \cite{Baer:2013cma} have been updated in recent documents \cite{Fujii:2015jha, Barklow:2015tja}, which suggested that the ILC could collect $2\inab$ data at 250\,GeV, $200\infb$ at 350\,GeV, and $4\inab$ at 500\,GeV.  However, the estimations of signal strengths, summarized in Ref.~\cite{Barklow:2015tja}, are only available for smaller benchmark luminosities for which the full detector studies were performed.  We took these estimations and scaled them up to the current run scenarios, assuming statistical uncertainties dominate~\cite{Asner:2013psa}.     The scaled estimations are summarized in \autoref{tab:mu_precision}.   Such scaling provides a reasonable approximation as long as the luminosities are not excessively large and the systematic uncertainties are under control.

\end{itemize}
With large center of mass energies up to 3\,TeV, CLIC is also able to measure the Higgs properties very well through the $WW$ fusion process \cite{CLIC:2016zwp, Abramowicz:2016zbo}.  On the other hand, with its extensive coverage of energy scales, the primary goal of CLIC is to directly search for new particles, in particular the ones coupled to SM particles only through electroweak interactions.
CLIC is advantageous in the possibility of directly producing   new particles that modifies Higgs properties at low energy, and a large span of observables at different scale. Consequently, a study on the CLIC physics potential for various models would require additional considerations beyond Higgs precision physics.
 A comprehensive study of the CLIC physics potential including both the direct and indirect searches of new physics is beyond the scope of this paper.

In our global fit to the Higgs measurements, we only included the rate information for the Higgsstrahlung as well as the $WW$ fusion process.   The measurements of the angular distributions in the Higgsstrahlung process can provide important information in addition to the rate measurements alone~\cite{Beneke:2014sba,Craig:2015wwr}.  The diboson process, $\eeww$, can be measured to a great precision, which imposes very strong constraints on the anomalous triple gauge couplings (TGCs)~\cite{Falkowski:2015jaa, Marchesini:2011aka, Wells:2015eba, Bian:2015zha}.  These measurements are very helpful in probing new physics.  In particular, it is shown in Ref.~\cite{Durieux:2017rsg} that the inclusion of these measurements are crucial for constraining new physics in a global EFT framework, which exhibits large flat directions with the Higgs rate measurements alone.  However,  their impacts are significantly smaller for specific models like the SM plus a singlet,  2HDM and MCHM,  due to a much smaller model parameter space (compared with SMEFT).  Therefore, they were   included in our global fit of operator approach of strong dynamics models only.    The electroweak (EW) precision measurements at the $Z$-pole also impose strong constraints on the new physics~\cite{Gori:2015nqa,Su:2016ghg}.  The current constraints from the Large Electron Positron Collider (LEP) could be significantly improved by a $Z$-pole run at any of the future lepton colliders.  While these constraints were not explicitly considered in our study, we did restrict ourselves to models with suppressed EW precision corrections ({\it e.g.} by imposing custodial symmetries) such that these constraints are automatically satisfied.

It is also important to compare the reaches of the future Higgs factories to that of the LHC.  For the LHC Run-I Higgs measurements with 5 fb$^{-1}$ integrated luminosity at $\sqrt{s} = 7$ TeV and 20 fb$^{-1}$ at $\sqrt{s} = 8$ TeV, we used the results in Ref.~\cite{Khachatryan:2016vau}.  For the  LHC with $300\infb$ and $3000\infb$ luminosities,  we used the ATLAS projections in Ref.~\cite{ATL-PHYS-PUB-2014-016}, which collects the information from several other studies.
The detailed inputs are listed in \autoref{app:lhcinput}, with the LHC Run-I results in \autoref{tab:HIGGS_datarun12} and the ATLAS projections for LHC $300\infb$ and $3000\infb$ summarized in \autoref{tab:HIGGS_datarun3000}.



\section{Global fit framework}
\label{sec:fit}

To transfer the estimated error on the experimental measurements to the constraints on the model parameters, we made a global fit by constructing the $\chi^2$ with the profile likelihood method
\begin{equation}\label{eq:chi2}
  \chi^2 = \sum_{i} \frac{(\mu_i^{\rm BSM}-\mu_i^{\rm obs})^2}{\sigma_{\mu_i}^2} .
\end{equation}
 Here $\mu_i^{\rm BSM} = \frac{(\sigma\times \text{Br})_{\rm BSM}}{(\sigma\times \text{Br})_{\rm SM}}$ for various Higgs search channels and $\sigma_{\mu_i}$ is the experimental precision on a particular channel.   We note that the correlations among different $\sigma\times {\rm BR}$ are usually not provided, and are thus assumed to be zero in the fits.   $\mu_i^{\rm BSM}$ is predicted in each specific model, depending on model parameters.   For the LHC Run-I, the measured $\mu_i^{\rm obs}$ and corresponding $\sigma_{\mu_i}$ are given in  Table~\ref{tab:HIGGS_datarun12}. In our analyses, for the future colliders, $\mu_i^{\rm obs}$ are set to be the SM value: $\mu_i^{\rm obs}=1$, assuming no deviation to the SM observables are observed.   The corresponding  $\sigma_{\mu_i}$ are the estimated error for each process, as shown in Table~\ref{tab:mu_precision} for the CEPC, FCC-ee, ILC and Table~\ref{tab:HIGGS_datarun3000} for the LHC.   For the ILC with three different center of mass energies, we summed the contribution from each individual channel. For one or two parameter fit, the corresponding $\Delta \chi^2=\chi^2 -\chi^2_{\rm min}$ for 95\% C.L. is 3.84 or 5.99,  respectively.

We fitted  directly to the signal strength $\mu_i$, instead of the effective couplings $\kappa_i$.  The latter are usually presented in most experimental papers.  While using $\kappa$-framework is easy to map to specific models, unlike $\mu_i$, various $\kappa_i$ are not independent experimental observables.  Ultimately, fitting to either  $\mu_i$ or $\kappa_i$  should give the same results, if the correlations between $\kappa_i$ are properly included.  Those correlation matrices, however, are typically not provided.  Therefore, fit to $\kappa_i$ only, assuming that they are uncorrelated, usually leads to a more relaxed constraints.  Comparison of $\mu$-fit versus $\kappa$-fit results is given later in the example of the 2HDM.

In the last example of our study for future collider constraints on generic strong dynamics,   we adopted the fitting results  in EFT coefficients given in Ref.~\cite{Durieux:2017rsg} since various scenarios can only be meaningfully discussed in power counting of the structure of the induced EFT operators.   EFT provides more complex and rich structure for the Higgs couplings, which requires the inclusion of additional
measurements to ensure the parameters are reasonably well-constrained.  In particular,  diboson process ($\eeww$) and  the angular observables in $\eehz$ were included in the global fit  in addition to the Higgs rate measurements.
 In the EFT framework, the $\chi^2$ from the Higgs rate measurements and other measurements can be constructed in a similar way as \autoref{eq:chi2}, with $\mu_i^{\rm BSM}$ being a function of the EFT parameters ({\it i.e.}, Wilson coefficients).

\section{SM with a real singlet extension}
\label{sec:singlet}
We first applied this global fit to the simplest extension to the SM with a real scalar singlet. The general Lagrangian for this model is,
\beq
\mathcal{L} = \mathcal{L}_{\rm SM}+ \frac 1 2 (\partial_\mu S)^2 - \frac 1 2 m_{S}^2 S^2 - \Lambda_{SH} S (H^\dagger H) - \frac 1 2 \lambda_{SH} S^2 (H^\dagger H) - \frac 1 {3!} \Lambda_{S} S^3 - \frac 1 {4!} \lambda_S S^4,
\label{eq:singletL}
\eeq
where $H$ is the SM Higgs doublet and $S$ is the new real singlet field.

This simplest extension to the SM could already induce many test scenarios. Such model is of particular interest as it can in general help enhance the electroweak phase transition to be a strong first order one~\cite{Grojean:2004xa},  which is needed for electroweak baryogengesis.   With certain constraints on the model parameters,  it is also a good description of the scalar section for the next to minimal supersymmetric Standard Model~\cite{Ellis:1988er,Drees:1988fc} in the decoupling regime.

The  Lagrangian given in Eq.~(\ref{eq:singletL}) can be categorized into two scenarios:  $Z_2$ preserving and $Z_2$ breaking.   In the   $Z_2$ preserving scenario,   the Lagrangian is further simplified:  $\Lambda_{SH} = \Lambda_{S}=0$, reducing the number of free parameters in this model to be three.   Note that in the $Z_2$ limit,  it is still possible to have spontaneous $Z_2$ breaking once $S$ acquires a vacuum expectation value (VEV), leaving us same number of terms for the interaction Lagrangian comparing to the $Z_2$ breaking case,  but still with only three free parameters.   Since the purpose of this study is to focus on the Higgs physics implications, instead of the  extraction of singlet model parameters, we did not single out spontaneous $Z_2$ breaking as a separate scenario.

The biggest difference between the $Z_2$ preserving and the $Z_2$ breaking is that the latter enables the singlet-Higgs mixing.     The $125~\gev$ Higgs is one of the mass eigenstates,   which is a mixture of real component of the SM doublet $h_{\rm SM}$ and the singlet $S$:
\beq
h_{125}=\cos\theta~h_{\rm SM} +\sin\theta~S,
\eeq
where $\theta$ is the mixing angle. All of the SM-like Higgs couplings to other SM particles are scaled down by $\cos\theta$ at tree level: $\kappa_i =  {g_{i}^{\rm SM+singlet}}/{g_{i}^{SM}}=\cos\theta$.   Since $\cos\theta$ does not exceed unity,   the modifications of the Higgs couplings through mixing with the singlet  always lead to a reduction of the Higgs couplings to SM particles.  This mixing angle description remains the same for both a light and a heavy singlet-like scalar and is used widely due to its simplicity.

\begin{figure}[t]
  \centering
  \includegraphics[width=6.07cm]{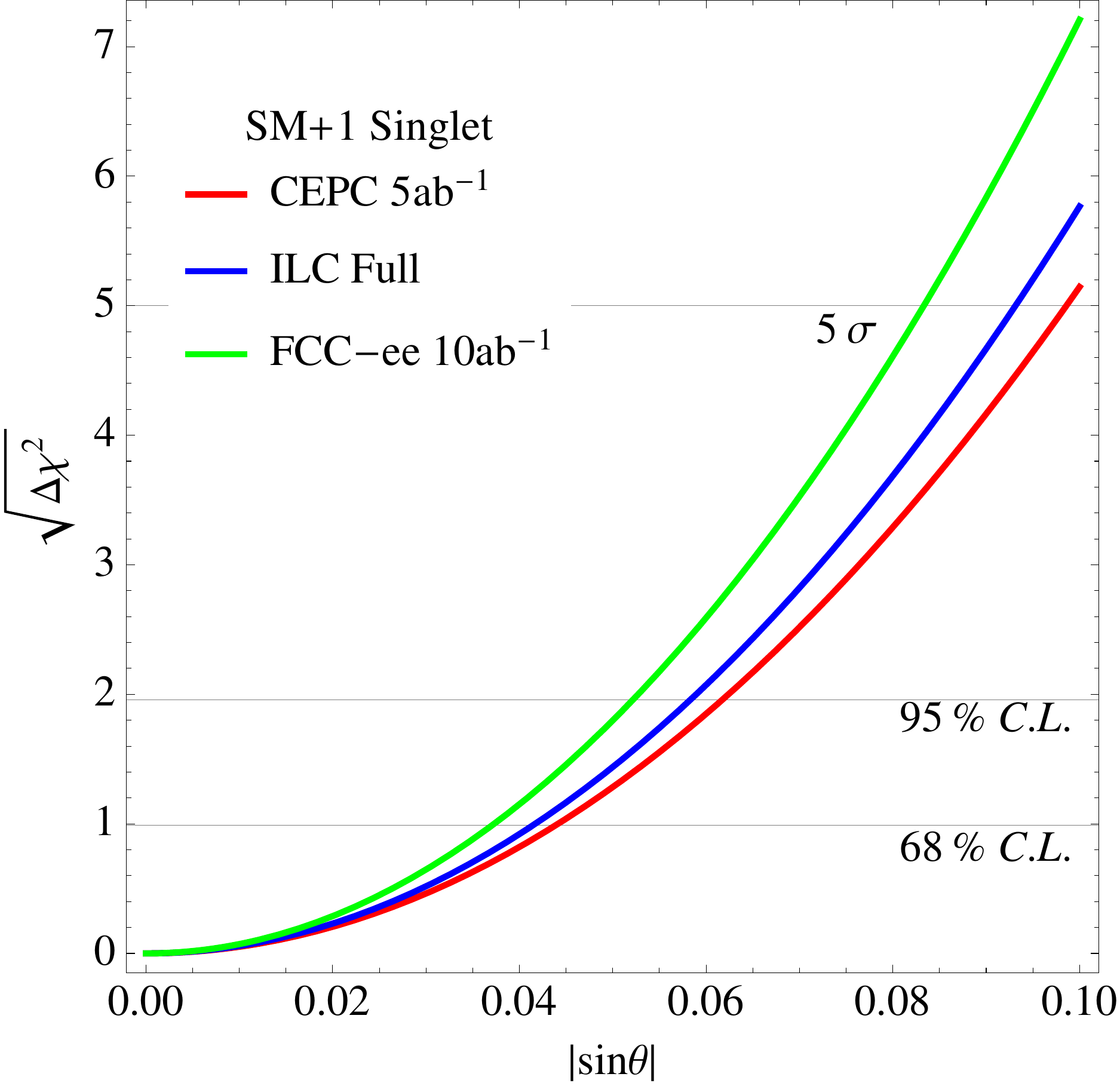}
  \includegraphics[width=7cm]{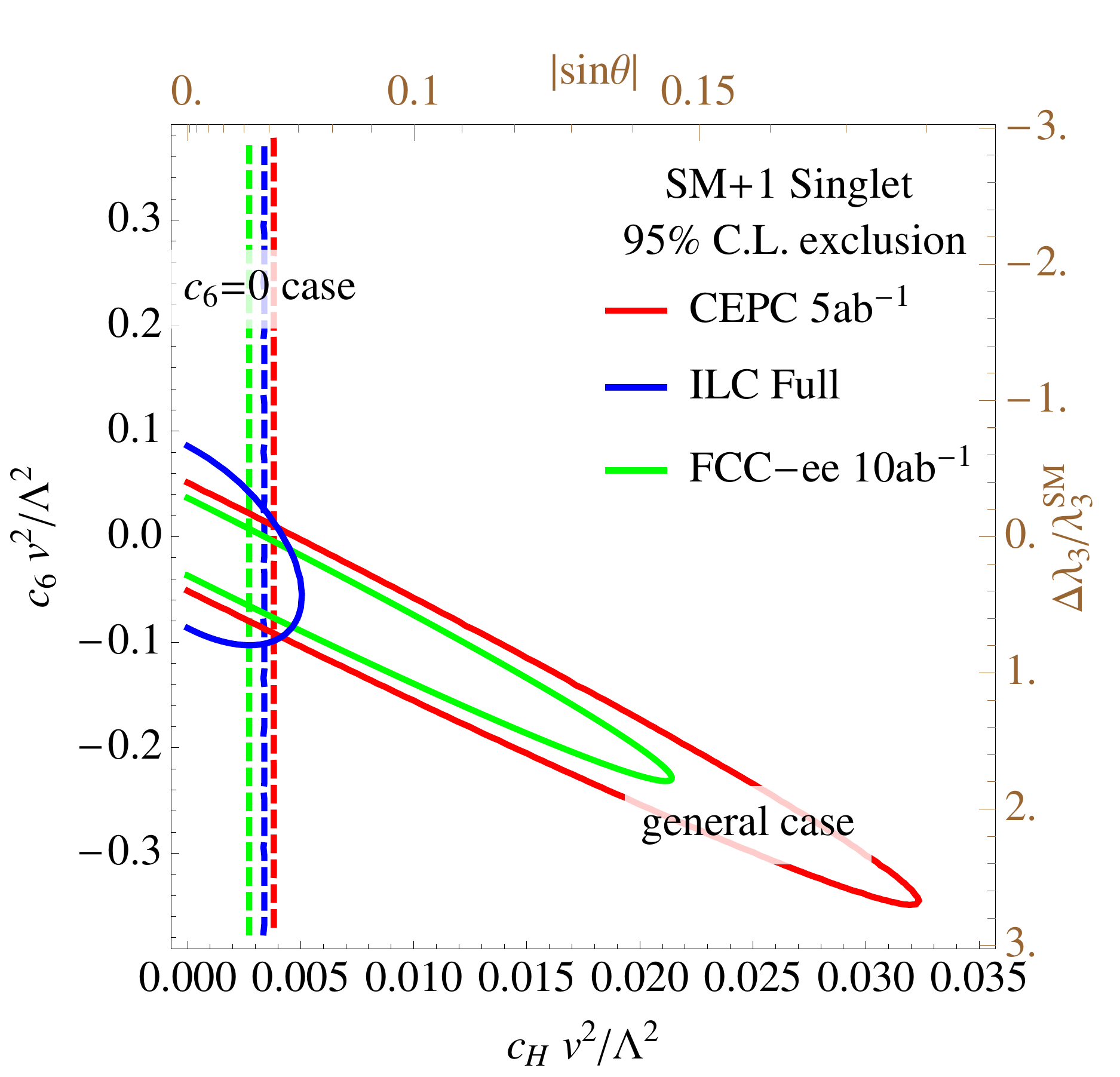}
    \caption{For SM plus one real singlet model, left panel shows the $\Delta\chi^2$   as a function of the singlet-SM Higgs doublet mixing angle $|\sin\theta|$ in the $Z_2$-breaking case for CEPC (red), FCC-ee(green) and ILC (blue) from Higgs precision measurements.     Right panel shows the  95\% C.L. limit on the Wilson coefficient $c_6v^2/\Lambda^2$ vs. $c_Hv^2/\Lambda^2$.  The vertical dashed lines indicate the one parameter fit limit on $c_Hv^2/\Lambda^2$  with $c_6$ set to be 0. The corresponding values of the mixing angle $|\sin\theta|$ and changes to trilinear Higgs coupling $\Delta \lambda_3/\lambda_3^{\rm SM}$ are shown in the upper and right axes of this figure.
    }
  \label{fig:SM1S}
\end{figure}

As the first and simplest  application of the global fit, we applied the global $\Delta_\chi^2$ fit to the SM plus a singlet model with the only fitting parameter being $\sin\theta$.  The $\Delta\chi^2$ distribution is shown in the left panel of Fig.~\ref{fig:SM1S} for the CEPC, ILC and FCC-ee precisions.   The mixing angle $|\sin\theta|$ is constrained to be 0.62, 0.058 and 0.052 for the CEPC, ILC and FCC-ee, respectively, at 95\% C.L.

While the mixing angle $\sin\theta$ captures the most important tree-level effect for the Higgs properties, it does not characterize the changes to the Higgs trilinear coupling, and neither the loop corrections to Higgs physics from the singlet field, e.g. Ref.~\cite{McCullough:2013rea}.   To fully explore the changes of the SM Higgs property in models with an extra real scalar singlet, we adopted  the EFT language to examine all the possible effects.

After integrating out the singlet field, the general EFT with dimension-six operators  can be written as,
\beq
\Delta \mathcal{L} = \frac {c_H} {\Lambda^2} \mathcal{O}_H+\frac {c_6} {\Lambda^2} \mathcal{O}_6,\ \ \ \
{\rm with}\ \mathcal{O}_H\equiv \frac 1 2 (\partial_\mu |H^\dagger H|)^2, \ \ \  \mathcal{O}_6\equiv |H^\dagger H|^3.
\eeq
The operator $\mathcal{O}_H$ induces a universal shift in the Higgs couplings through the Higgs wave-function renormalization.   The mixing angle $\theta$ can be mapped with $c_H$ as $1-\cos\theta \simeq \theta^2/2\simeq 1/2\times c_H v^2 /\Lambda^2$, for $v=246$ GeV.

The Wilson coefficients for $\mathcal{O}_H$ and $\mathcal{O}_6$ can be mapped through the tree-level processes in the $Z_2$ breaking scenario as~\cite{Henning:2014gca,Henning:2014wua}
\beq
\frac{c_H} {\Lambda^2} = \frac {\Lambda_{SH}^2} {m_S^4},~~~~\frac{c_6} {\Lambda^2}=\left (-\frac {\lambda_{SH} \Lambda^2_{SH}} {2 m_S^4} + \frac {\Lambda_S \Lambda^3_{SH}} {6 m_S^6} \right),
\label{eq:SM1Stree}
\eeq
and   through loop-level processes in both the $Z_2$ preserving and the $Z_2$ breaking scenario as~\cite{Henning:2014wua},
\beq
\frac{c_H} {\Lambda^2} = \frac {\lambda_{SH}^2} {48 m_S^2 \pi^2} ,~~~~\frac{c_6} {\Lambda^2}=-\frac {\lambda_{SH}^3} {48 m_S^2 \pi^2}.
\label{eq:SM1Sloop}
\eeq
The sign of the $\mathcal{O}_H$ operator is positive semi-definite, for both $Z_2$ breaking and $Z_2$ preserving scenario, which is consistent with the mixing angle description that $\cos\theta\leq 1$ and the Higgs couplings are always reduced comparing to that of the SM.     The $\mathcal{O}_6$ operator modifies mostly the Higgs couplings to electroweak gauge bosons and top quarks at loop-level.  In our analysis we took into account the loop corrections to Higgs to weak gauge boson couplings via $\mathcal{O}_6$ operators using calculations detailed in Ref.~\cite{cepchhh}.

We performed our global fit to this singlet scenario with CEPC, ILC, and FCC-ee  Higgs precision measurements, with  95\% C.L. limits  in the $c_6 v^2/\Lambda^2$-$c_H v^2/\Lambda^2$ plane   shown in the right panel of Fig.~\ref{fig:SM1S}.\footnote{It should be noted that, the ILC 500\,GeV could directly probe the $\mathcal{O}_6$ operator by measuring the double Higgsstrahlung process, $e^+e^- \to Z hh$, which is not included in our analysis.  See Refs.~\cite{Duerig:2016dvi, Barklow:2017awn} for more details.}  The corresponding values of the mixing angle $\sin\theta$  are shown in the upper axis of the figure, and the corresponding relative change to the SM trilinear Higgs coupling $\Delta \lambda_3 /\lambda_3^{\rm SM}$ is shown in the right axis of the figure.  The one-parameter fit results with $c_H v^2 / \Lambda^2$ (or equivalently, $\sin\theta$)   are  also shown in the vertical dashed lines.

From Fig.~\ref{fig:SM1S},  the allowed range in $c_6 v^2/\Lambda^2$ is almost one  order of magnitude larger than  that of the $c_H v^2/\Lambda^2$.   In the general case,  at 95\% C.L., the maximum allowed Wilson coefficient $c_H v^2/\Lambda^2$ is 0.032, 0.021 and 0.005, corresponding to the max value of the singlet-double mixing parameter $\sin\theta$ of 0.18, 0.15 and 0.07, for the CEPC, FCC-ee and   ILC, respectively. The loop-modifications to Higgs to gauge boson couplings through the modifications of the trilinear Higgs coupling have sizable energy dependence. Hence, the ILC reaches a much better precision in this two-parameter fit via measuring the Higgs processes through different production modes well at separated center of mass energies of 250~GeV and 500~GeV. On the other hand, in the restrict one parameter fit case with $c_6$ set to zero, the limit on $c_H v^2/\Lambda^2$ improves to 0.0038, 0.0034 and 0.0028, for the CEPC, ILC and FCC-ee, respectively. These limits on $c_H v^2/\Lambda^2$ correspond to the maximum value of the singlet-double mixing parameter $\sin\theta$ derived from the left panel of this figure.\footnote{The electroweak precision observables also provide relevant constraints on the parameter space of this model, mainly from the running and mixing of  the $\mathcal{O}_H$  operator. For detailed analysis in various cases, see e.g., Refs.~\cite{Henning:2014gca,Chen:2014ask}.}


\section{Two Higgs doublet model}
\label{sec:2hdm}

2HDMs are very generic BSM Higgs extension of the SM,  including MSSM, gauge extensions (such as Left-Right symmetric model), and flavor models~\cite{Martin:1997ns,Mohapatra:1974hk}.  Understanding the Higgs physics potential of 2HDM at future lepton colliders provides unique information covering a broad class of BSM.

Two ${\rm SU}(2)_L$ scalar doublets $\Phi_i$, $i=1,2$ are introduced in 2HDM,
\begin{equation}
\Phi_{i}=\begin{pmatrix}
  \phi_i^{+}    \\
  (v_i+\phi^{0}_i+iG_i)/\sqrt{2}
\end{pmatrix}.
\end{equation}
Each obtains a VEV  $v_1$ or $v_2$ after electroweak symmetry breaking (EWSB) with $v_1^2+v_2^2 = v^2 = (246\ {\rm GeV})^2$, and $v_1/v_2=\tan\beta$.

The 2HDM lagrangian for 
Higgs sector can be written as
\begin{equation}\label{}
\mathcal{L}=\sum_i |D_{\mu} \Phi_i|^2 - V(\Phi_1, \Phi_2) + \mathcal{L}_{Yuk},
\end{equation}
 with the Higgs potential  
 \begin{eqnarray}
 V(\Phi_1, \Phi_2)=&& m_{11}^2\Phi_1^\dag \Phi_1 + m_{22}^2\Phi_2^\dag \Phi_2 -m_{12}^2(\Phi_1^\dag \Phi_2+ h.c.) + \frac{\lambda_1}{2}(\Phi_1^\dag \Phi_1)^2 + \frac{\lambda_2}{2}(\Phi_2^\dag \Phi_2)^2  \notag \\
 & &+ \lambda_3(\Phi_1^\dag \Phi_1)(\Phi_2^\dag \Phi_2)+\lambda_4(\Phi_1^\dag \Phi_2)(\Phi_2^\dag \Phi_1)+\frac{1}{2}\Big[ \lambda_5(\Phi_1^\dag \Phi_2)^2 + h.c.\Big],
\end{eqnarray}
assuming CP-conserving and a soft ${Z}_2$ symmetry breaking term $m_{12}^2$.

After EWSB, one of the four neutral components and two of the four charged components are eaten by the SM $Z$, $W^\pm$, providing their masses.  The remaining physical mass eigenstates are the two CP-even Higgses $h$ and $H$, with $m_h<m_H$, one CP-odd Higgs $A$, as well as a pair of charged ones $H^\pm$.  Instead of the eight parameters   appearing in the Higgs potential: $(m_{11}^2, m_{22}^2, m_{12}^2, \lambda_{1,2,3,4,5})$, a more convenient choice of the parameters is: $(v, \tan\beta, \alpha, m_h, m_H, m_A, m_{H^\pm}, m_{12}^2)$, in which  $\alpha$ is the rotation angle  diagonalizing the CP-even Higgs mass matrix.

The  CP-even Higgs couplings to the SM gauge bosons are:  $g_{hVV} \propto \sin(\beta-\alpha)$, and $g_{HVV} \propto \cos(\beta-\alpha)$.  For $\cos(\beta-\alpha)=0$, the light CP-even Higgs $h$ couples to the gauge boson with full SM coupling strength while $H$ decouples.  This is usually referred to the ``alignment limit"~\cite{Carena:2013ooa} with $h$ identified as the SM Higgs \footnote{Yukawa couplings of $h$ to the SM fermions are also identical to the SM predictions under alignment limit.}.   For $\sin(\beta-\alpha)=0$, the opposite occurs with the heavy $H$ being identified as the SM Higgs. While it is still a viable option for the heavy Higgs being the observed 125 GeV SM-like Higgs \cite{Coleppa:2014cca}, the parameter spaces are squeezed with the tight direct and indirect experimental constraints.  Therefore, in our analyses below, we   identified the light CP-even Higgs $h$ as the SM-like Higgs with $m_h$ fixed to be 125 GeV.

The most general Yukawa interactions of $\Phi_{1,2}$ with the SM fermions under the $Z_2$ symmetry is
\begin{equation}\label{}
  -\mathcal{L}_{Yuk}=Y_{u}{\overline Q}_Li\sigma_2\Phi^*_uu_R^{} +Y_{d}{\overline Q}_L\Phi_dd_R^{}+Y_{e}{\overline L}_L\Phi_e e_R^{}+\text{h.c.}
\end{equation}
where $\Phi_{u,d,e}$ are either $\Phi_1$ or $\Phi_2$.
 Depending on the interactions of $\Phi_i$ coupling to the fermion sector, there are typically four types of 2HDM:
\begin{itemize}
\item{Type-I:} $\Phi_1$ couples to all the fermions while $\Phi_2$ does not couple to fermions at all.
\item{Type-II:} $\Phi_1$ couples to the up-type quark, and $\Phi_2$ couples to the down-type quarks and leptons.
\item{Type-L:} $\Phi_1$ couples to the quarks and $\Phi_2$ couples to the leptons.
\item{Type-F:} $\Phi_1$ couples to the up-type quarks and leptons while $\Phi_2$ couples to the down-type quarks.
\end{itemize}
For a  review on different types of 2HDM as well as the  phenomena, see Ref.~\cite{Branco:2011iw}.

\begin{table}[h]
\begin{center}
{\renewcommand\arraystretch{1.2}
\begin{tabular}{c|ccccccccc}\hline\hline
&\multicolumn{9}{c}{Normalized Higgs couplings}\\\cline{2-10}
&$\kappa_h^u$&$\kappa_h^d$&$\kappa_h^e$&$\kappa_H^u$&$\kappa_H^d$&$\kappa_H^e$&$\kappa_A^u$&$\kappa_A^d$&$\kappa_A^e$\\\hline
Type-I&$\frac{\cos\alpha}{\sin\beta}$&$\frac{\cos\alpha}{\sin\beta}$&$\frac{\cos\alpha}{\sin\beta}$&$\frac{\sin\alpha}{\sin\beta}$&$\frac{\sin\alpha}{\sin\beta}$&$\frac{\sin\alpha}{\sin\beta}$&$\cot\beta$&$-\cot\beta$&$-\cot\beta$\\\hline
Type-II&$\frac{\cos\alpha}{\sin\beta}$&$-\frac{\sin\alpha}{\cos\beta}$&$-\frac{\sin\alpha}{\cos\beta}$&$\frac{\sin\alpha}{\sin\beta}$&$\frac{\cos\alpha}{\cos\beta}$&$\frac{\cos\alpha}{\cos\beta}$&$\cot\beta$&$\tan\beta$&$\tan\beta$\\\hline
Type-L&$\frac{\cos\alpha}{\sin\beta}$&$\frac{\cos\alpha}{\sin\beta}$&$-\frac{\sin\alpha}{\cos\beta}$&$\frac{\sin\alpha}{\sin\beta}$&$\frac{\sin\alpha}{\sin\beta}$&$\frac{\cos\alpha}{\cos\beta}$&$\cot\beta$&$-\cot\beta$&$\tan\beta$\\\hline
Type-F&$\frac{\cos\alpha}{\sin\beta}$&$-\frac{\sin\alpha}{\cos\beta}$&$\frac{\cos\alpha}{\sin\beta}$&$\frac{\sin\alpha}{\sin\beta}$&$\frac{\cos\alpha}{\cos\beta}$&$\frac{\sin\alpha}{\sin\beta}$&$\cot\beta$&$\tan\beta$&$-\cot\beta$\\\hline\hline
\end{tabular}}
\caption{Higgs couplings to the SM fermions in the four different types of 2HDM, normalized to the corresponding SM value. }
\label{tab:yukawa_tab}
\end{center}
\end{table}

After EWSB, the effective lagrangian for the light CP-even Higgs couplings to the SM particles can be parameterized as
 \begin{eqnarray}\label{}
&&\mathcal{L}= \kappa_Z \frac{m_Z^2}{v}Z_{\mu}Z^{\mu}h+\kappa_W \frac{2m_W^2}{v}W_{\mu}^+ W^{\mu-}h + \kappa_g \frac{\alpha_s}{12 \pi v} G^a_{\mu\nu}G^{a\mu\nu}h + \kappa_{\gamma} \frac{\alpha}{2\pi v} A_{\mu\nu}A^{\mu\nu}  \nonumber \\
&& + \kappa_{Z\gamma}\frac{\alpha}{\pi v}A_{\mu\nu}Z^{\mu\nu}h -\Big( \kappa_u \sum_{f=u,c,t} \frac{m_f}{v}f \bar f + \kappa_d \sum_{f=d,s,b} \frac{m_f}{v}f \bar f + \kappa_{e} \sum_{f=e,\mu,\tau} \frac{m_f}{v}f \bar f \Big)h,
\end{eqnarray}
 in which $\kappa_i$ is the SM-like Higgs coupling normalized to the corresponding SM value: $\kappa_i =  {g_{hii}^{BSM}}/{g_{hii}^{SM}}$ (same as $\kappa_h^i$ in Table~\ref{tab:yukawa_tab}).   Table~\ref{tab:yukawa_tab} summarizes all the tree-level non-zero $\kappa_i$ of the light CP-even Higgs $h$  for the four different types of 2HDM, as well as the normalized couplings of the non-SM Higgses $H$ and $A$.  Note that while $\kappa_g$ and $\kappa_\gamma$ are zero at tree level for both the SM and 2HDM, they are  generated at the loop level.  In the SM, both $\kappa_g$ and $\kappa_\gamma$ receive contribution from fermions (mostly top quark) running in the loop, while $\kappa_\gamma$ receives contribution from $W$-loop in addition\cite{Henning:2014wua}.  In the 2HDM, the corresponding $hff$ and $hWW$ couplings that enter  the loop corrections need to be modified to the corresponding 2HDM values.   Expressions for the dependence of $\kappa_g$, $\kappa_\gamma$ and $\kappa_{Z\gamma}$ on $\kappa_V$ and $\kappa_f$ can be found in Ref.~\cite{Heinemeyer:2013tqa}.  There are, in addition, loop corrections to $\kappa_g$, $\kappa_\gamma$ from extra Higgses.

In the alignment limit of $\cos(\beta-\alpha)=0$, the light CP-even Higgs couplings are exactly identical to the SM ones.  However, deviations of the Higgs couplings from the alignment limit are still allowed given the current LHC Higgs measurements \cite{Coleppa:2013dya,Barger:2013ofa,Belanger:2013xza}.  Note that all the tree-level deviations to the SM couplings in Table~\ref{tab:yukawa_tab} are parametrized by only two parameters: $\beta$ and $\alpha$, or more conveniently, $\tan\beta$ and $\cos(\beta-\alpha)$.   In sec.~\ref{sec:2HDMtree}, we studied the constraints on the 2HDM tree level effects, namely, in the parameters space of $\tan\beta$ vs. $\cos(\beta-\alpha)$ with the Higgs precision measurements.

There are, of course, loop corrections to all the $\kappa_i$s above with non-SM heavy Higgses running in the loop.   While their contributions are typically small comparing to the tree deviations, in the alignment limit, their contributions could be manifest given the high precision Higgs coupling measurements achieveable at future Higgs factories.  The masses of the heavy Higgses will enter the loop corrections, as well as the soft $Z_2$ symmetry breaking parameter $m_{12}^2$.  We studied the constraints on the 2HDM loop effects in Sec.~\ref{sec:2HDMloop} under the alignment limit.

 \subsection{2HDM tree-level results}
 \label{sec:2HDMtree}

\begin{figure}[h]
\begin{center}
\includegraphics[width=7.5cm]{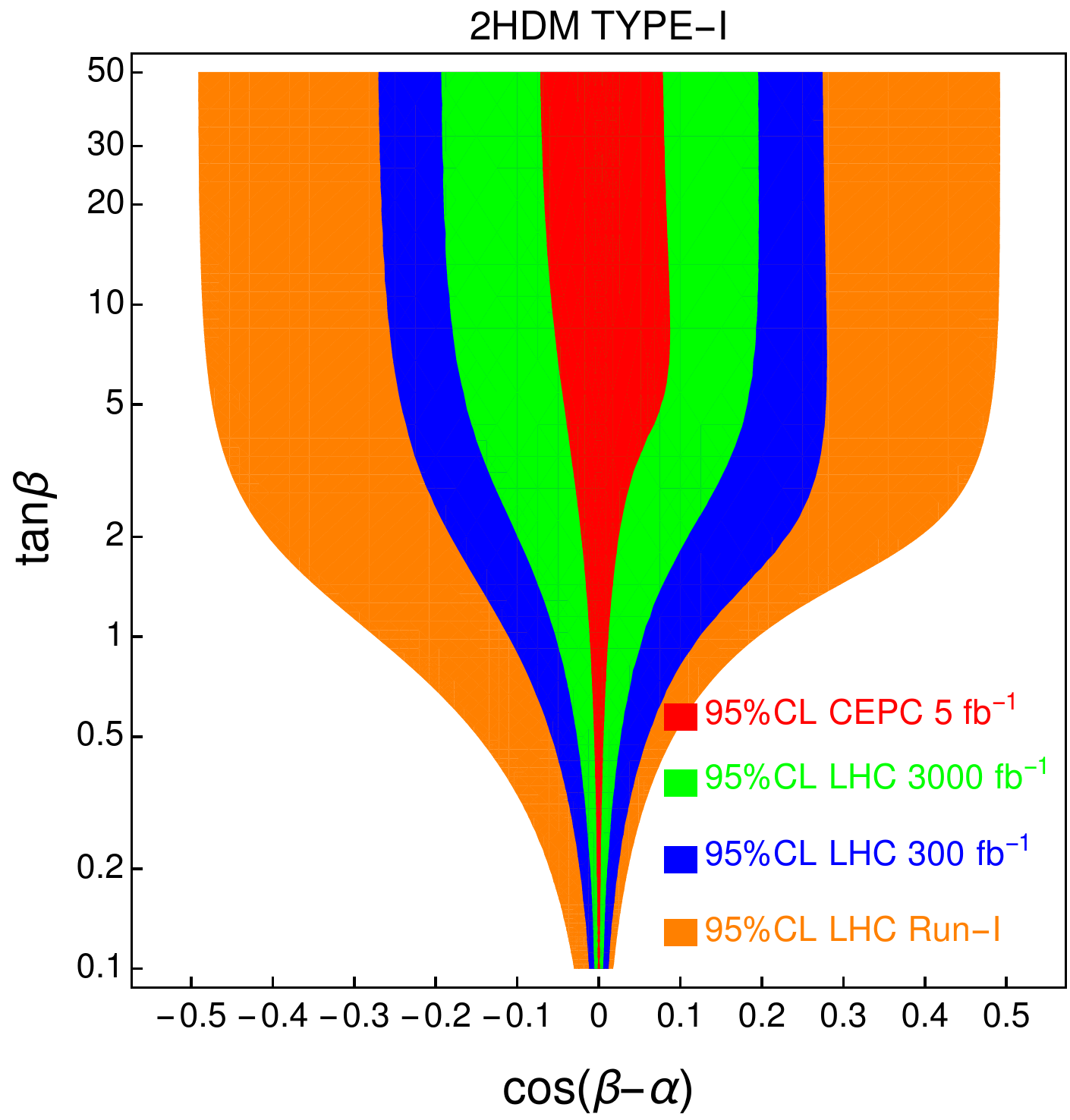}
\includegraphics[width=7.5cm]{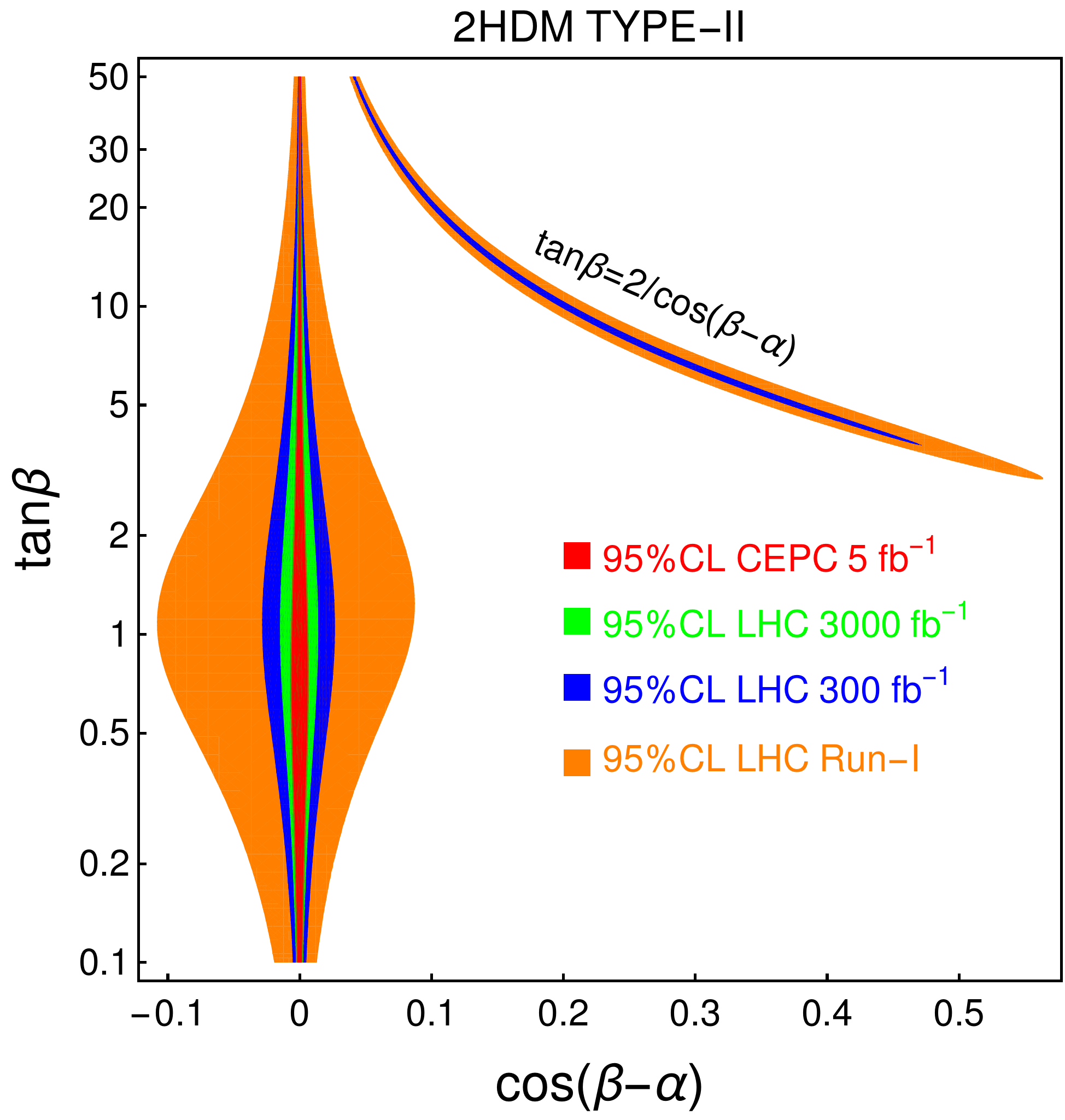}\\\vspace{3mm}
\includegraphics[width=7.5cm]{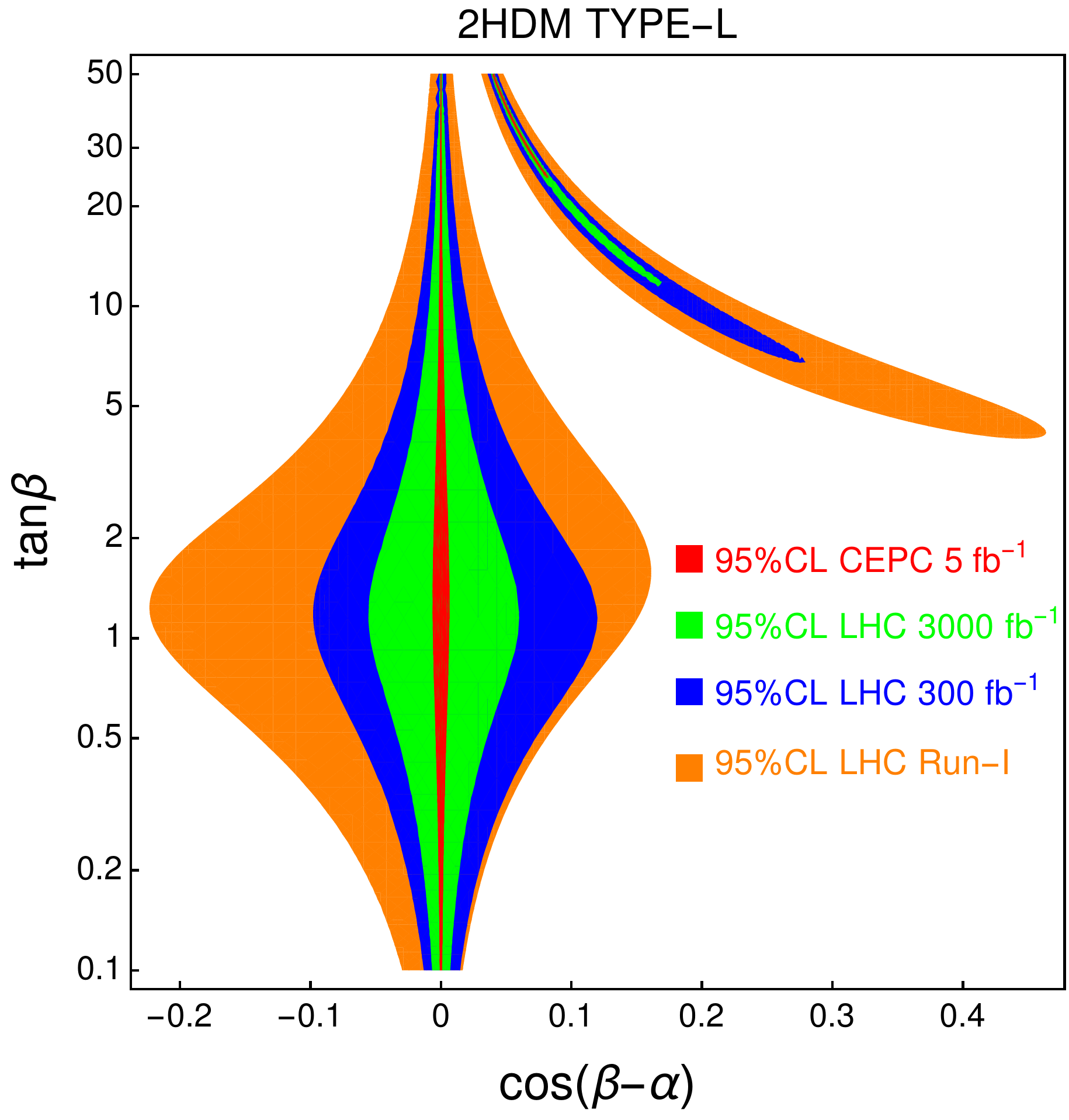}
\includegraphics[width=7.5cm]{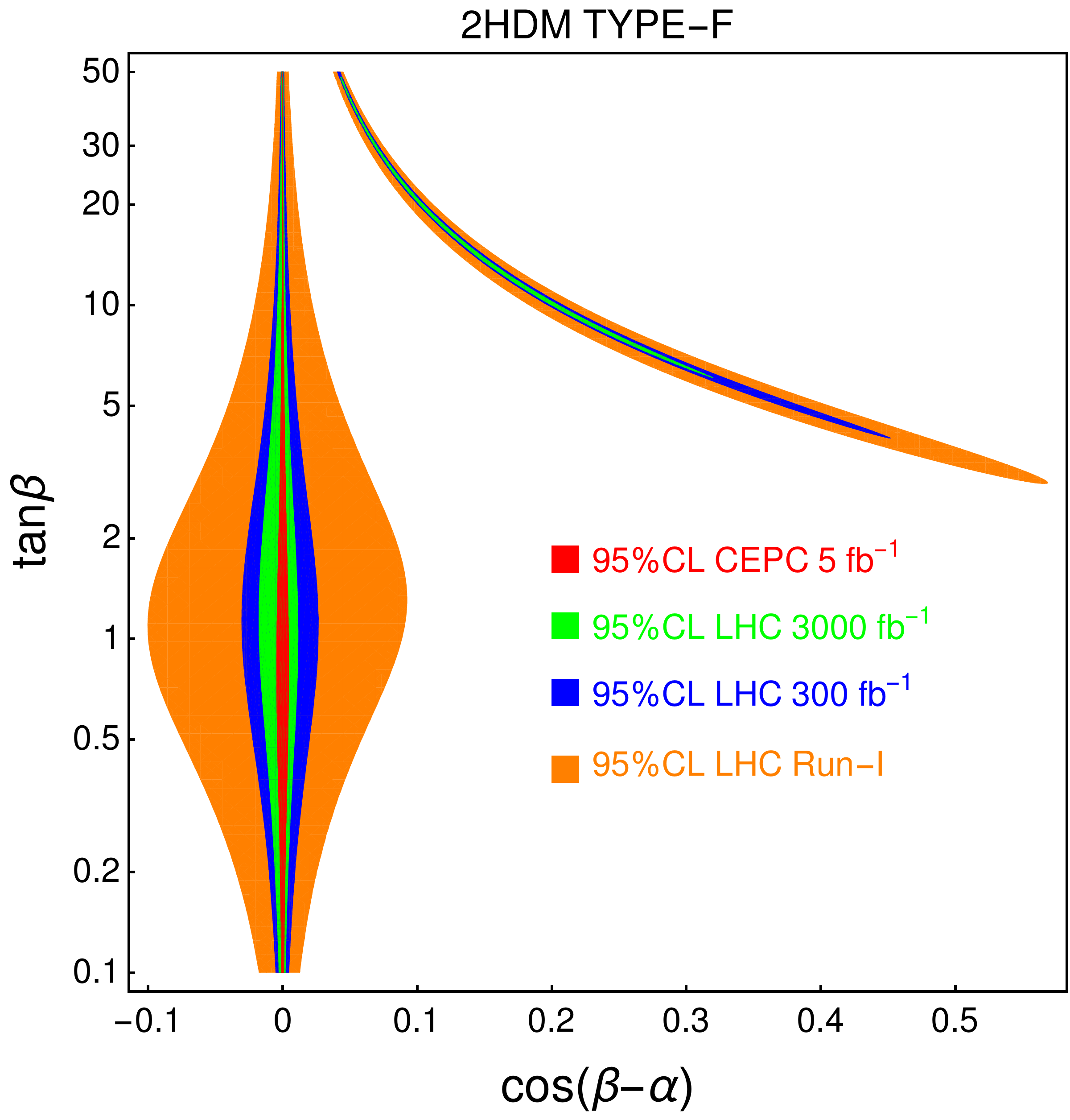}
\caption{The allowed region in the  plane of $\tan \beta$ vs. $\cos(\beta-\alpha)$ at 95\% C.L. for the four types of 2HDM, given LHC and CEPC Higgs precision measurements.  For future measurements, we have assumed that the measurements agree with SM predictions.  The special  ``arm" regions for the Type-II, L and F are the wrong-sign Yukawa regions.  See text for more details. }
\label{fig:cepc-tree}
\end{center}
\end{figure}

Performing a global fit to the Higgs rate measurements at the LHC as well as the CEPC\footnote{For future measurements, we have assumed that no deviation to the SM values is observed.}, we   obtained the 95\% C.L. region in the $\tan\beta$ vs. $\cos(\beta-\alpha)$ plane for various types of 2HDM, as shown in Fig.~\ref{fig:cepc-tree}.   For Type-I 2HDM with $\tan\beta\gtrsim 2$, $|\cos(\beta-\alpha)|$ is constrained to be less than about 0.5 with the current LHC Run-I data.  With full LHC luminosity of 300 ${\rm fb}^{-1}$, the range in $|\cos(\beta-\alpha)|$ can be shrunk by about a factor of 2.  At the HL-LHC with 10 times   luminosity, the range can be further constrained  to be less than  0.2.  At the CEPC with 5 ${\rm ab}^{-1}$, $|\cos(\beta-\alpha)|$ is constrained to be  less than about $0.08$.  The preferred range in $\cos(\beta-\alpha)$ is quickly shrunk at small $\tan\beta$, given that
\begin{equation}
\Delta  \kappa_{u,d,e} = \frac{\cos \alpha}{\sin \beta}-1 =-\frac{1}{2} \cos^2 (\beta-\alpha) + \frac{\cos (\beta-\alpha)}{\tan \beta}.
\label{eq:dkappa_u}
\end{equation}
The deviation of the Yukawa   couplings from the experimental centre value   is proportional to  $1/{\tan \beta}$,  resulting a reduced survival parameter space at small $\tan \beta$.    The up-type Yukawa couplings are the same for all the four types of 2HDM.  Therefore, the small $\tan\beta$ behavior is similar.

For the Type-II, Type-L and Type-F 2HDM, the correction to the down quark and/or lepton Yukawa couplings receive $\tan \beta$ enhancement at large value,
\begin{equation}
\Delta \kappa_x = -\frac{\sin \beta}{\cos \alpha} -1
=-\frac{1}{2} \cos^2 (\beta-\alpha) - \cos (\beta-\alpha)\times \tan \beta.
\label{eq:dkappa_x}
\end{equation}
Here $x$ is  $d, e$ in the Type-II, $e$ in the Type-L and $d$ in the Type-F.   Therefore, the survival parameter space at large $\tan\beta$ is reduced significantly in all these three types.

For the Type-II at the upper right panel of Fig.~\ref{fig:cepc-tree}, as a result of larger $\tan \beta$ enhancement from $\Delta \kappa_{d,e}$ and small $\tan \beta$ enhancement from $\Delta \kappa_{u}$, the region around $\tan \beta =1$ accommodates the largest deviation from the alignment. The current LHC Run-I constrain $|\cos(\beta-\alpha)|$ less than 0.1 around $\tan \beta =1$ (except for the wrong-sign Yukawa couplings region~\cite{Ferreira:2014naa, Han:2017etg}), and the LHC Run-II 300 (3000) $\text{fb}^{-1}$ can reduce it to be less than 0.03(0.015).  CEPC precision measurements further reduce it to be less than 0.006.   Similar behavior appears in Type-L and Type-F, with the small difference mostly coming from $\kappa_b$ and $\kappa_\tau$ parameter dependence.  A summary of the 95\% C.L. allowed maximum $|\cos(\beta-\alpha)|$ range is given in Table~\ref{tab:tree_results}.

\begin{table}[h]
\begin{center}
\begin{tabular}{|c|c|c|c|c|}
\hline
Type                    & LHC Run-I
& $\begin{array}{l} \text{LHC Run-II} \\ ~~~~300\  \text{fb}^{-1} \end{array}$
&$\begin{array}{l}  \text{LHC Run-II} \\ ~~~3000\  \text{fb}^{-1}  \end{array}$     &CEPC
\\ \hline
Type-I $\tan\beta \gtrsim 5$    & 0.5       & 0.27                             & 0.2                         &0.08
\\ \hline
Type-II $\tan\beta\sim1$   & 0.1       & 0.03                              & 0.015                         &0.006
\\ \hline
Type-L $\tan\beta\sim1$  & 0.2       & 0.1                               &   0.06                        & 0.007
\\ \hline
Type-F $\tan\beta\sim1$   & 0.1       & 0.03                              & 0.02                         &0.005
\\ \hline
\end{tabular}
\end{center}
\caption{Maximally allowed $|\cos(\beta-\alpha)|$ range at 95\% C.L. given LHC and CEPC Higgs precision measurements.    }
\label{tab:tree_results}
\end{table}

Because of those large or small $\tan \beta$ enhanced Higgs couplings deviation, we can examine which Higgs decay channel provides the best constraints in certain region.    Conservative 7-parameters fit with CEPC Higgs measurements shows that  $\sigma_{\kappa_V}\leq 0.16\% $, $\sigma_{\kappa_b} \leq 1.2\% $, $\sigma_{\kappa_c} \leq 1.6\% $, $\sigma_{\kappa_{\tau}} \leq 1.3\% $~\cite{CEPC-SPPCStudyGroup:2015csa}.      $\kappa_V$ and $\kappa_b$ typically provide the strongest constraints.    For the Type-I,  $hbb$ coupling provides the strongest constrains at $\tan \beta\leq1$, while $hZZ$ coupling constraints dominate for  $\tan \beta \gtrsim  5$.  For Type-II,  $hbb$ coupling dominates at large $\tan\beta$ and $hgg$ coupling dominates at small $\tan\beta$. For Type-L, while the small $\tan\beta$ case is similar to that of Type-I, $h\tau\tau$ dominates at large $\tan\beta$.   Type-F is very similar to Type II, only that $h\tau\tau$ enters at small $\tan\beta$ instead.

 The  special ``arm" region survived for the Type-II, L and Type-F in Fig.~\ref{fig:cepc-tree}  is the so-called the wrong-sign Yukawa couplings~\cite{Ferreira:2014naa}, coming from fermion couplings with $\Phi_1$: $ {\cos \alpha}/{\sin \beta}$,   as shown in Eq.~(\ref{eq:dkappa_u}).    For $\tan\beta =  {2}/{\cos (\beta-\alpha)}$, there is no deviation of the corresponding couplings,   which corresponds to exactly the  ``arm" central lines. For regions near this line, $\Delta\kappa$ flips its sign while the absolute value is small.    The $\kappa_i$ for the other fermions that couple to $\Phi_2$ is near $-1$, therefore ``wrong-sign" while still  surviving the  Higgs measurements at the LHC.   Actually, there would be another ``arm" to make zero deviation for Eq.~(\ref{eq:dkappa_x}): $\tan \beta = -\frac{1}{2}\cos (\beta-\alpha)$.  For Type-I, there is no such kind of coupling, and for other types, such region does not show up in our plot since we consider region $\tan\beta>0.1$.   Smaller region of $\tan\beta$ is tightly constrained by the perturbativity of top Yukawa couplings.

\begin{figure}[h]
\begin{center}
\includegraphics[width=7.5cm]{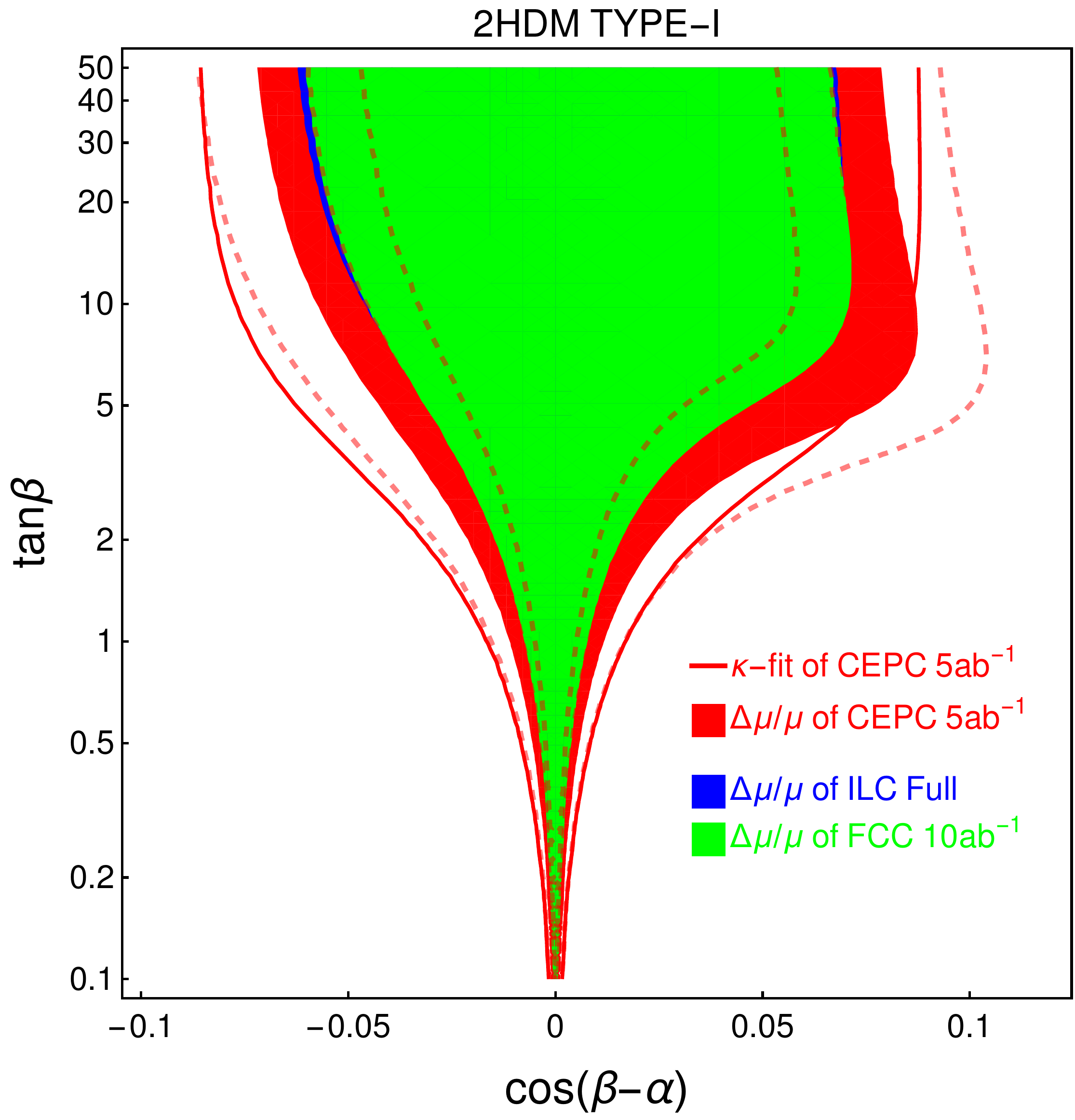}
\includegraphics[width=7.5cm]{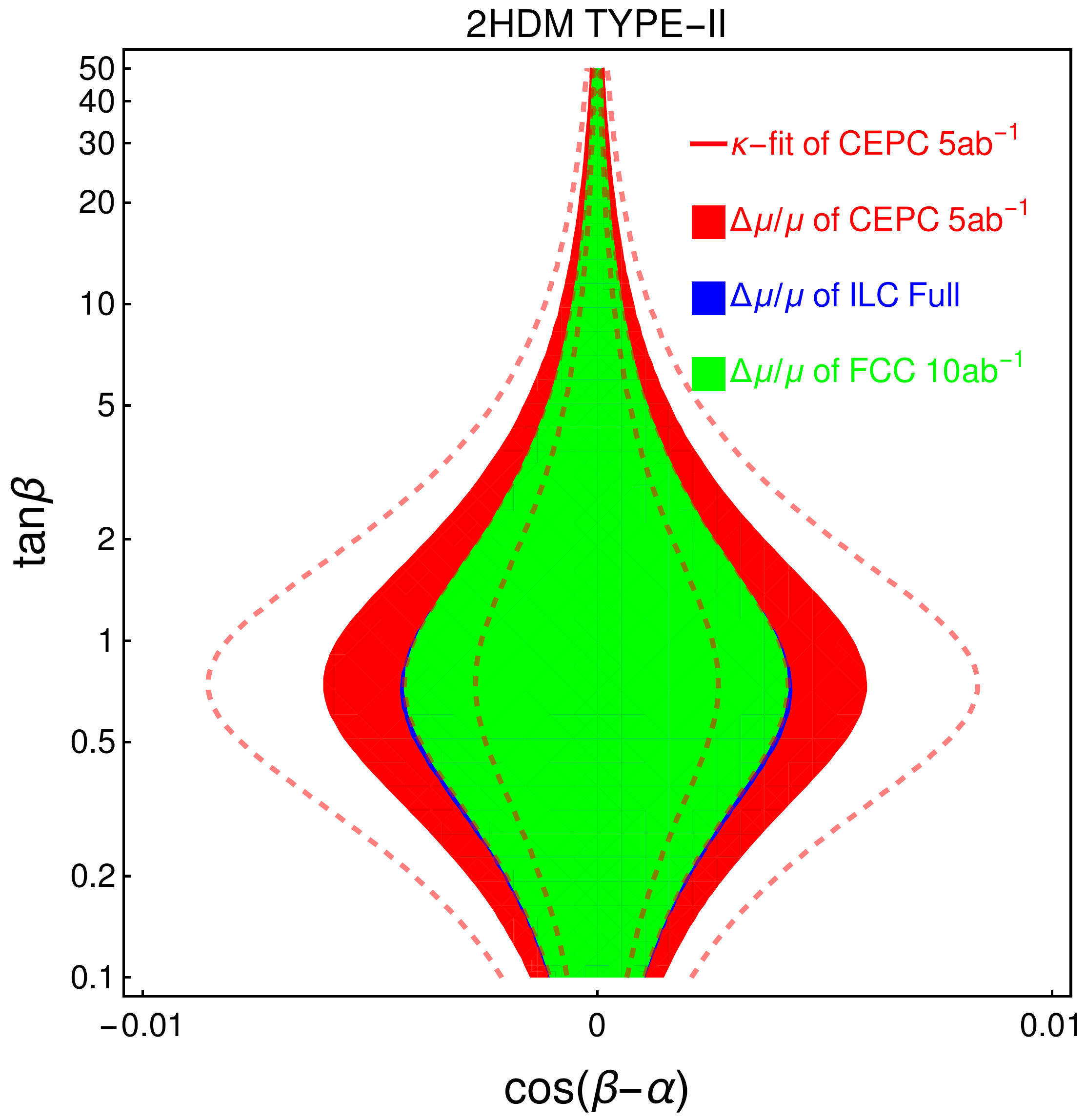}\\\vspace{3mm}
\includegraphics[width=7.5cm]{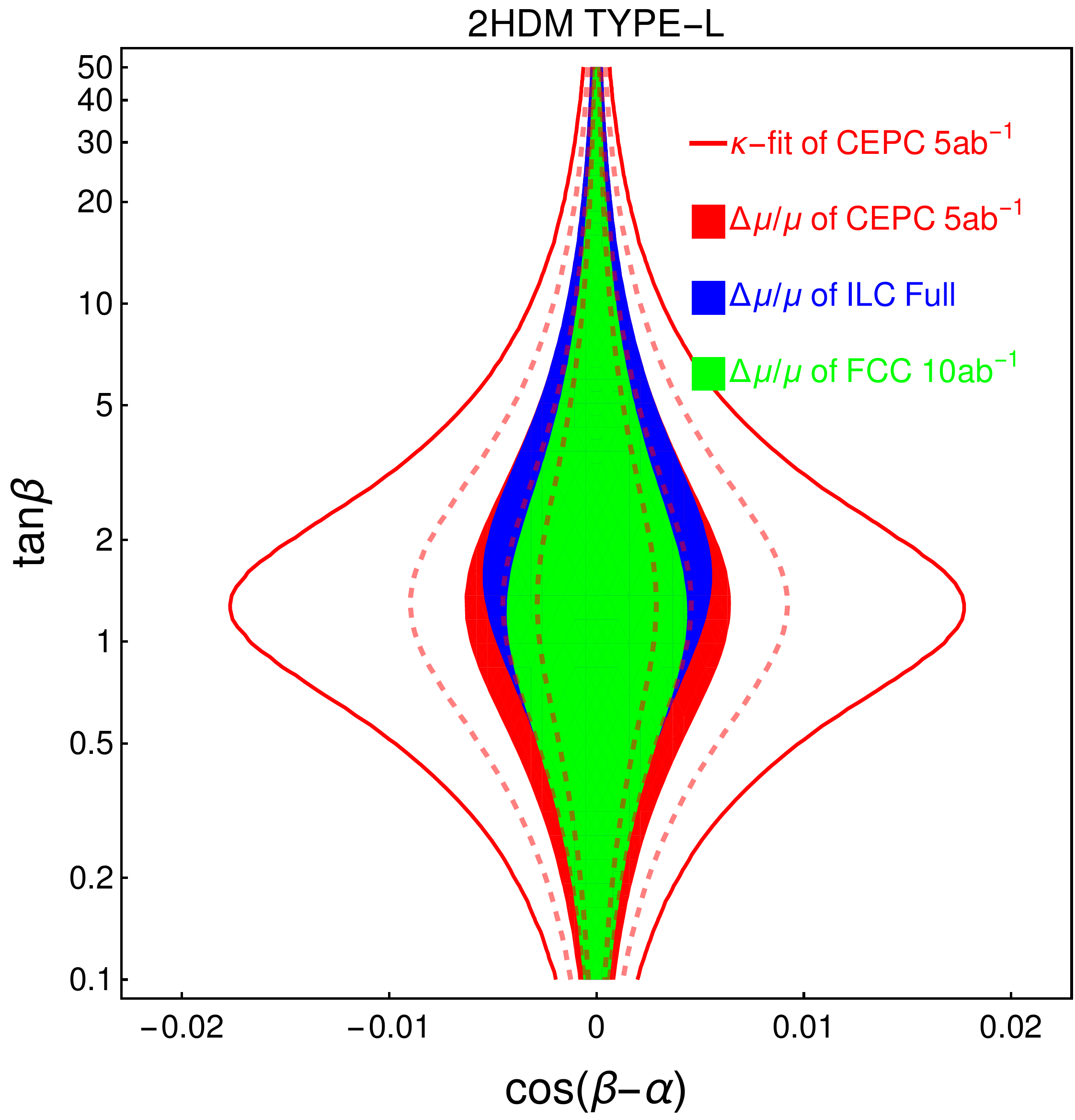}
\includegraphics[width=7.5cm]{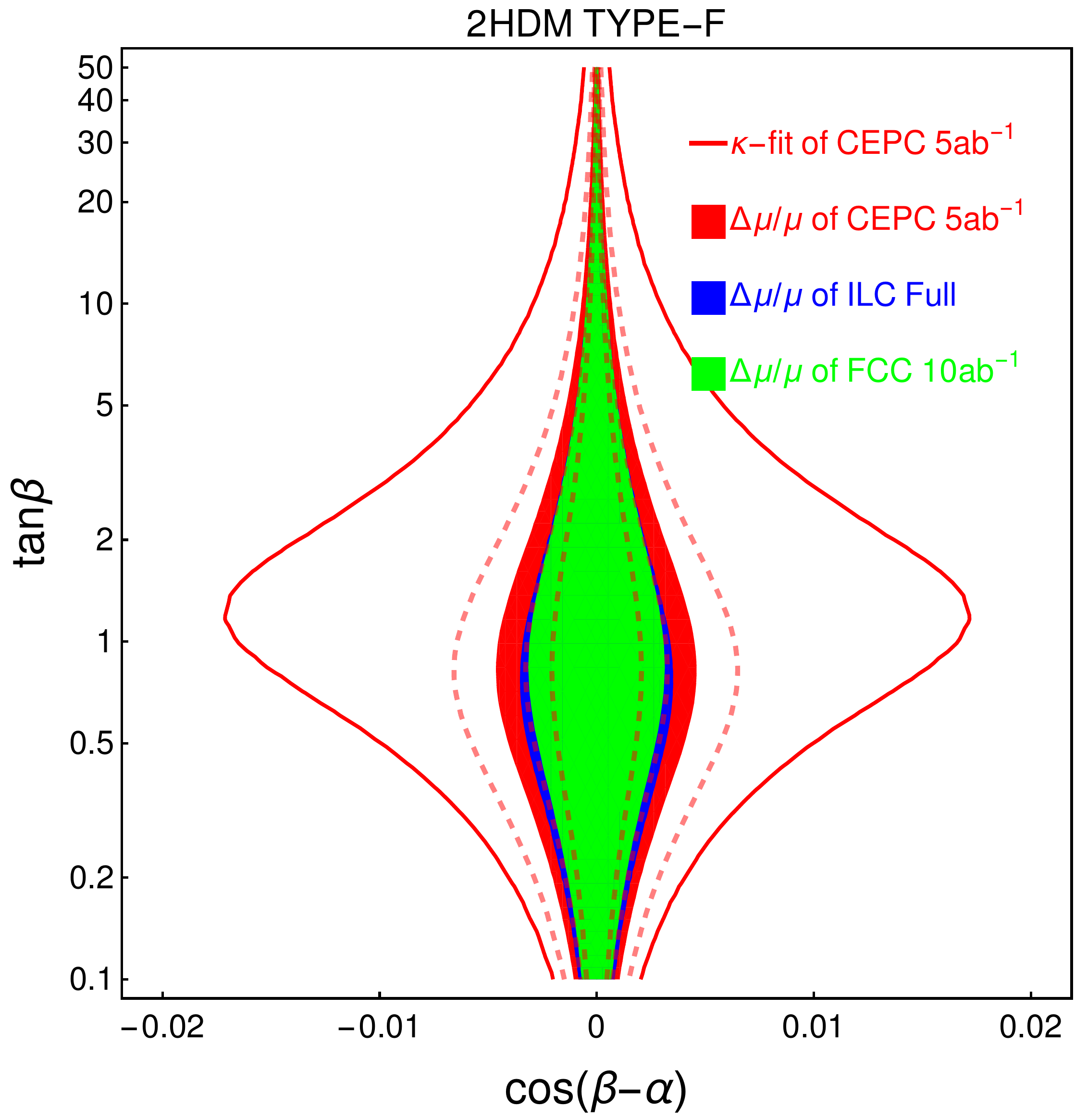}
\caption{The comparison between the CEPC (red region), ILC (blue region) and FCC-ee (green region) reach in the plane $\cos(\beta-\alpha)$ vs. $\tan \beta$.   A tiny arm region for Type-L is omitted for clarity.  We also show the global fitting results to effective couplings from the 7 parameter fit of CEPC,  instead of fitting to $\Delta\mu/\mu$, in red solid  line.  Scaled CEPC results with  2.5 $ab^{-1}$, 10 $ab^{-1}$, 25 $ab^{-1}$  are shown in dashed lines, from outer to inner region.}
\label{fig:ee_comparison}
\end{center}
\end{figure}

To compare the potential reach for different Higgs factories, we show the 95\% C.L. reach in $\tan \beta$ vs. $\cos(\beta-\alpha)$ plane based on the estimated Higgs measurement precision at CEPC (red), ILC (blue), and FCC-ee (green) in Fig.~\ref{fig:ee_comparison}, for four different types of 2HDM.     The red CEPC region is  the zoom-in of the red region in Fig.~\ref{fig:cepc-tree}.  The 95\% C.L.  green and blue regions are almost the same, showing that ILC and FCC-ee has about the same constraining power, and both are  slightly better than the CEPC results, with about $70\% - 90\%$ of the maximum $|\cos(\beta-\alpha)|$ range.

In Fig.~\ref{fig:ee_comparison}, we also studied the CEPC results with different luminosity to get a knowledge of various running scenarios.   With the dashed red lines from the outer to inner, we presented the CEPC 2.5, 10, and 25 ab$^{-1}$ luminosity reach. In particular, results with CEPC 10 ab$^{-1}$ are almost the same as the FCC-ee 10 ab$^{-1}$.

We also show the comparison between the results using the signal strength $\mu$-fit (red region) or the effective coupling $\kappa$-fit (red line) for the CEPC precision. We adopted the precision for $\kappa$ using CEPC 7-parameter fit~\cite{CEPC-SPPCStudyGroup:2015csa}.    The results with $\kappa$-fit are less restrained than that of the $\mu$-fit since no correlations between $\kappa_i$ have been taken into account.   Numerically, except for the Type-I, the   range of $|\cos(\beta-\alpha)|$ at $\tan \beta \approx 1$ with $\kappa$-fit  is about 3 times that of  the $\mu$-fit for all three types of 2HDM.  For Type II and Type-F, the CEPC $\kappa$-fit results are similar to those of the HL-LHC, while for Type-I and -L, the CEPC $\kappa$-fit results are still better than those of the HL-LHC.  This demonstrates the under estimation of the Higgs physics potential  if using $\kappa$ results without full correlation information.

 \begin{figure}[h]
\begin{center}
\includegraphics[width=7.5cm]{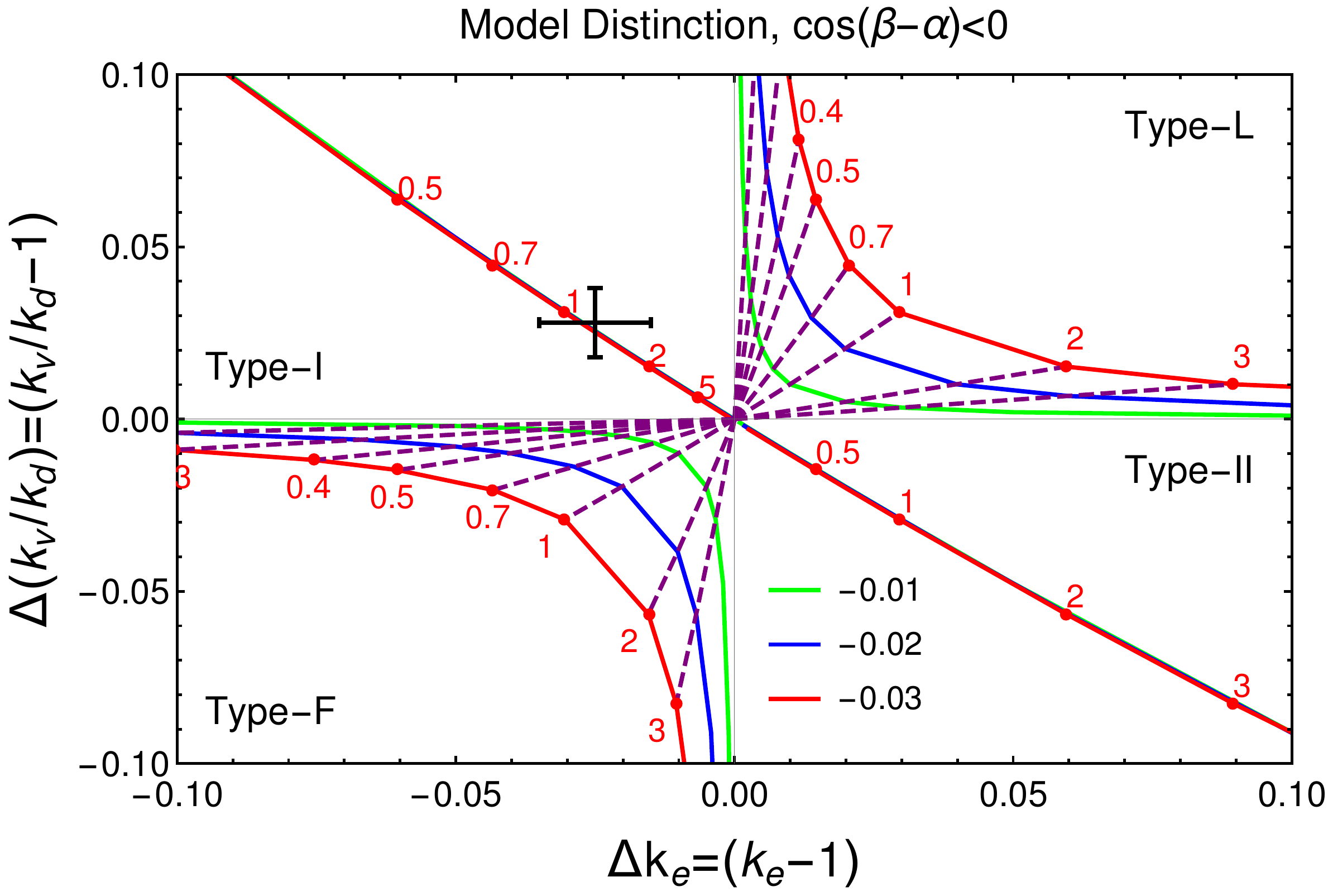}
\includegraphics[width=7.5cm]{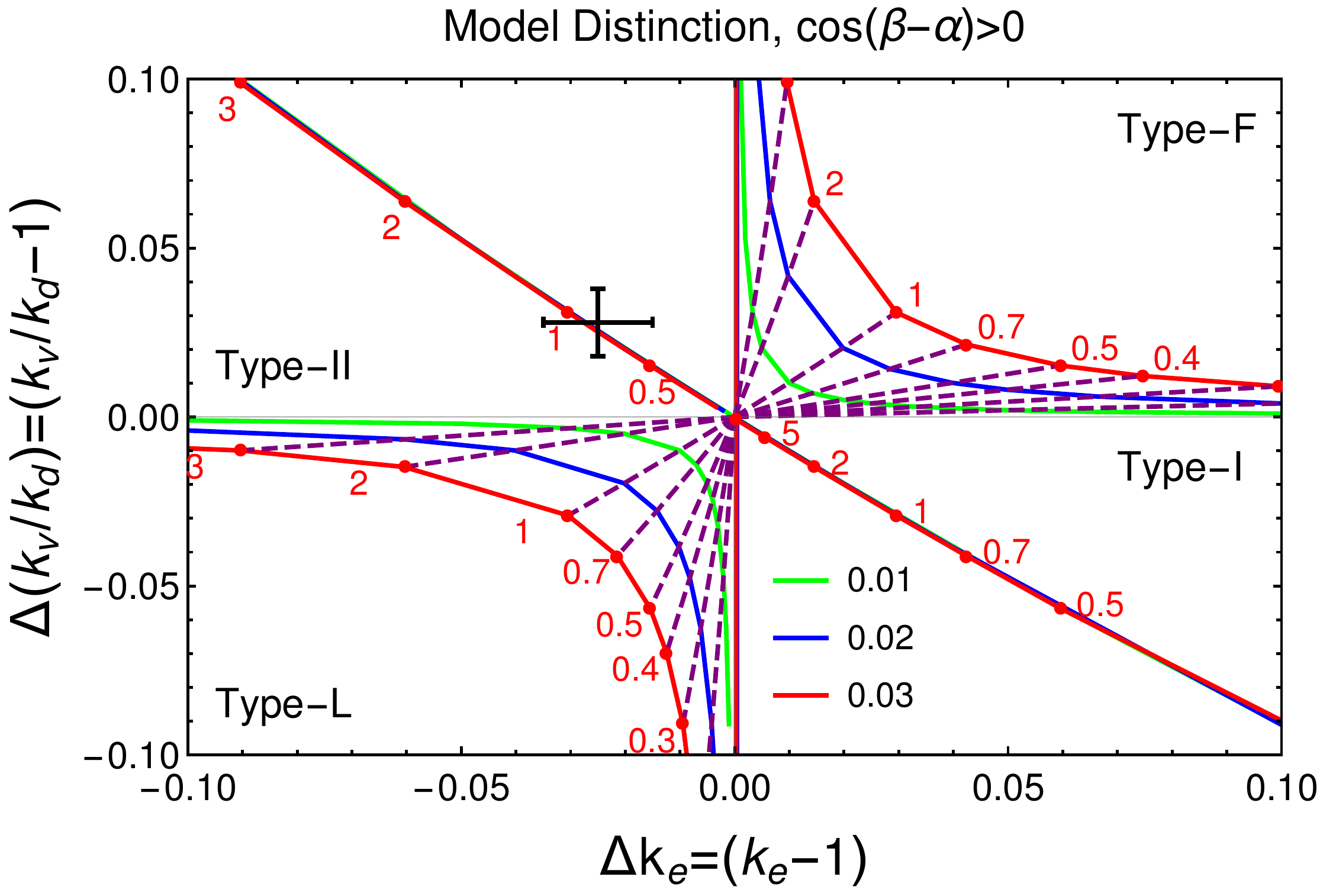}
\caption{$\Delta  ({\kappa_V}/{\kappa_d})$ vs $\Delta \kappa_e$ for four different types of 2HDM, varying $\tan \beta$ and $\cos (\beta-\alpha)$. The left panel  is for $\cos (\beta-\alpha) < 0$ and the right panel is for $\cos (\beta-\alpha) > 0$. The green, blue and red lines are for $|\cos(\beta-\alpha)| = 0.01, 0.02, 0.03$ respectively.  The dashed   lines are for different $\tan \beta$ values, as labeled in the plot.  The black cross indicates  the estimated experimental errors with a random central point.}
\label{fig:four_quadrant}
\end{center}
\end{figure}

Once a deviation of the Higgs couplings to the SM is observed, simultaneous measurements of various couplings can be used to distinguish different types of 2HDM.  In Fig.~\ref{fig:four_quadrant},  we plotted $\Delta \kappa_e$ vs. $\Delta  ({\kappa_V}/{\kappa_d})$ for four different types of 2HDM, for $|\cos(\beta-\alpha)|=0.01$ (green), 0.02 (blue) and 0.03 (red).  Left and right panels are for negative and positive  $\cos(\beta-\alpha)$, respectively.   Dashed lines are for different values of $\tan\beta$, as labeled next to the lines.   We also indicated the experimental precision of those couplings by the black cross,  with a random central point, for the purpose of comparison.

Type-I, II, L and Type-F are well separated, occupying the second, fourth, first and third quadrant,  respectively, for $\cos(\beta-\alpha)<0$.     For the $\cos(\beta-\alpha) > 0$, the behavior is similar, except for exchange of the quadrants of Type-I $\leftrightarrow$ Type-II and Type-L $\leftrightarrow$ Type-F.
   For Type-I and II,  $\Delta {\kappa_e} = \Delta {\kappa_d}$.  Therefore, the variations of the coupling deviation with $\cos(\beta-\alpha)$ are small, given the small allowed $\cos(\beta-\alpha)$ range that we picked.   Smaller  $\tan \beta$ value   leads to larger deviation from the SM value for the Type-I, while the opposite occurs for Type-II.    For the Type-L and Type-F,  $\Delta \kappa_e$ and $\Delta  ({\kappa_V}/{\kappa_d})$ spreads over the whole region, depending on the values of $\tan\beta$ and $\cos(\beta-\alpha)$.      

\subsection{2HDM loop-level results in the alignment limit}
\label{sec:2HDMloop}

Other than the tree level deviation of the light Higgs couplings (as well as the loop generated $hgg$ and $h\gamma\gamma$ couplings with SM particles) in the 2HDM away from the alignment limit, heavy Higgses in the 2HDM could also provide loop corrections to those couplings~\cite{Kanemura:2015mxa,Kanemura:2014dja,Kanemura:2004mg}. While these contributions are typically small given the loop-suppression and heavy Higgs masses suppression,  they become the dominant correction to Higgs physics  in or close to the alignment limit,  $\cos(\beta-\alpha)\sim 0$.  Heavy Higgs contributions to $hgg$ and  $h\gamma\gamma$ could also be important, given the loop suppressed SM values at the leading order.

In this section, we analyzed the implication of Higgs precision measurements on the heavy Higgs loops, and explored the sensitivity to the heavy Higgs masses, as well as the Higgs self-couplings that enter the loop corrections.    For simplicity, we  set $m_{H}=m_{H^{\pm}}=m_{A}$ in the following discussion, which satisfies the EW precision $\rho$-parameter constraint automatically.  We also worked under the most challenging scenario of 2HDM with  tree-level alignment limit $\cos(\beta-\alpha)= 0$ to show the relevance and importance of these loop corrections.

\subsubsection{Theoretical and experimental constraints}
Heavy Higgs loop corrections would involve the Higgs self-couplings, which are constrained by various theories considerations and experimental measurements, such as vacuum stability,  perturbativity and unitarity, as well as heavy flavor,  electroweak precision measurements, and LHC direct searches.  We briefly summarized below the constraints we adopted in our analyses.
\begin{itemize}
\item \textbf{Vacuum stability}

In order to have a stable vacuum, the following conditions on the quartic couplings need to be satisfied~\cite{Gunion:2002zf}:
\begin{align}
\lambda_1>0, \quad \lambda_2>0,\quad \lambda_3 > -\sqrt{\lambda_1\lambda_2},\quad \lambda_3+\lambda_4-|\lambda_5|> -\sqrt{\lambda_1\lambda_2}.
\end{align}
 \item \textbf{Perturbativity and unitarity}

We adopted a general perturbativity condition of $|\lambda_i| \leq 4\pi$ and the tree-level unitarity of the scattering matrix in the 2HDM scalar sector~\cite{Ginzburg:2005dt}.
\end{itemize}

EW precision measurements from the LEP constrain the $\rho$-parameter, or equivalently, the amount of custodial symmetry breaking in the 2HDM.   Studies \cite{Kling:2016opi,Haber:2015pua} showed that the charged Higgs mass is constrained to be close to the mass of either of the neutral Higgses ($H$ or $A$) in order  to satisfy the EW precision measurements.  In our analyses below, we adopted the simplification of $m_{H^{\pm}}=m_{H}=m_{A}\equiv m_{\phi}$ so that $\rho$-parameter constraint is automatically satisfied.   The results we obtained below show characteristic features of the Higgs factory sensitivities to heavy Higgs mass, even though the numerical values  might differ if   masses for heavy Higgses deviate from the simplified relation.

\begin{figure}[h]
\begin{center}
 \includegraphics[width=9cm]{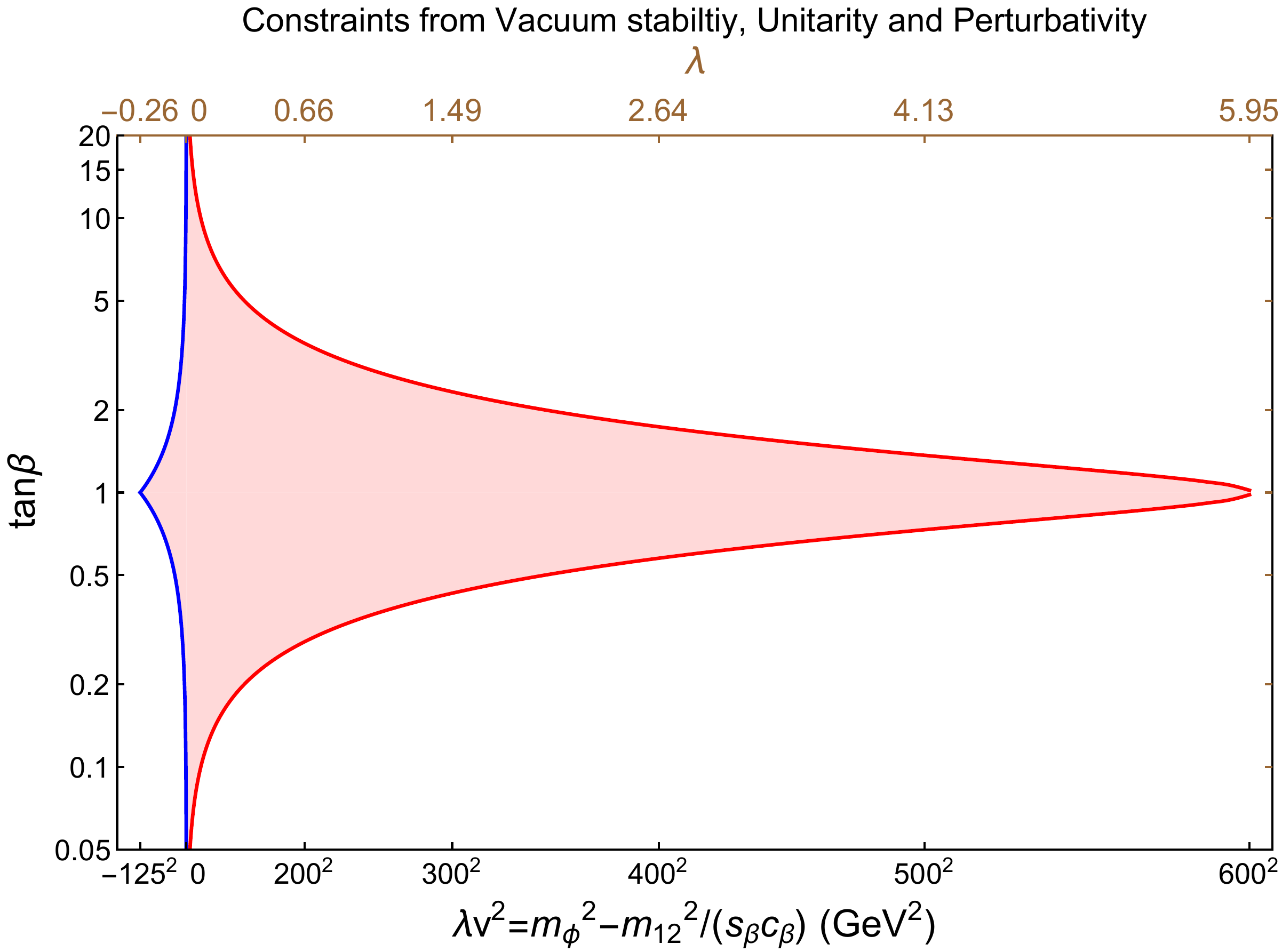}
\caption{The shaded region indicates the surviving region of 2HDM parameter space of  $\tan \beta$ vs. $\lambda v^2$, after  vacuum stability (blue lines), unitarity and perturbativity (red lines) are taken into account.  The corresponding values of $\lambda$ are shown in the upper axis.  Here we took the simplification of  $m_{H^{\pm}}=m_{H}=m_{A}\equiv m_{\phi}$ and alignment limit of $\cos(\beta-\alpha) = 0 $.}
\label{fig:constraint}
\end{center}
\end{figure}

\begin{table}[h]
\begin{center}
\begin{tabular}{|c|c|c|c|c|c|}
    \hline
   $\lambda v^2 (\text{GeV}^2) $   & $-100^2$ & 0 & $100^2$ &$300^2$ & $500^2$ \\ \hline
  $\tan \beta_{min}$   & 0.80 &0& 0.14 & 0.43 & 0.73 \\
  $\tan \beta_{max}$   & 1.25 &+$\infty$& 7.01 & 2.33 & 1.37 \\
  \hline
\end{tabular}
\caption{A few benchmark values for $\lambda v^2$, and the corresponding range of $\tan\beta$ given the theoretical considerations.   }
\label{tab:lam_cons}
\end{center}
\end{table}

Under the assumption of alignment limit and equal mass for all the heavy Higgses, all the Higgs  quartic couplings are related to a particular linear combination of $m_\phi^2$ and $m_{12}^2$: $\lambda v^2 \equiv m_{\phi}^2- {m_{12}^2}/({\sin \beta \cos \beta})$ and we have $\lambda v^2 = \lambda_3 v^2 - m_h^2 = - \lambda_4 v^2 = - \lambda_5 v^2$. The above theoretical considerations can be translated to
\begin{equation}
m_h^2 + \lambda v^2 \tan^2 \beta > 0, \ \ \ m_h^2 + \frac{\lambda v^2 }{\tan^2 \beta } > 0,
\label{eq:constn}
\end{equation}
for $\lambda v^2 < 0$, and
\begin{equation}\label{eq:constp}
 \tan^2 \beta + \frac{1}{\tan^2 \beta} < \frac{64 \pi^2 v^4 + 5 m_h^4 - 48 \pi v^2 m_h^2 + 8 \lambda^2 v^4 - 4 m_h^2 \lambda v^2}{3\lambda v^2(8 \pi v^2 - 3 m_h^2)},
\end{equation}
for $\lambda v^2>0$.
In Fig.~\ref{fig:constraint}, we show the allowed shaded  region in $\tan\beta$ vs.  $\lambda v^2$ plane given the theoretical considerations.    Region in $\tan\beta$ and $1/\tan\beta$ is symmetric, which is obvious from Eqs.~(\ref{eq:constn}) and (\ref{eq:constp}). A few representative values of $\lambda v^2$ that we used in our later analyses and the corresponding acceptable region of $\tan\beta$ are shown in Tab.~\ref{tab:lam_cons} as well.   Note that for $\lambda=0$, i.e. $m_{\phi}^2= {m_{12}^2}/({\sin \beta \cos \beta})$, $\tan\beta$ is unconstrained,  which is consistent with the results of Ref.~\cite{Kling:2016opi}.
 Given the symmetry between $\tan \beta$ and $ {1}/{\tan \beta}$ in the above conditions, the largest region on $\lambda v^2$ occurs at $\tan \beta =1$:
\begin{equation}
  -m_h^2 < \lambda v^2 < (600\  \text{GeV})^2,
     \label{eq:lam_cons}
\end{equation}
which gives $ -0.258 < \lambda =- \lambda_4=- \lambda_5  < 5.949$ and $0 <  \lambda_3  < 6.207$.

There are direct searches of non-SM heavy Higgses at the LHC \cite{ATLAS:2017mpg}, with the dominant search channel being $A/H\rightarrow \tau\tau$.  In the framework of MSSM, $m_{A/H}$ is excluded to about 250 GeV for $\tan\beta \ge 1.0$, and about 1.5 TeV for $\tan\beta \ge 45$.  Since the branching ratio of the  dominant search channel $A/H\rightarrow \tau\tau$ could be highly suppressed once other exotic decay channels of the non-SM Higgs opens up \cite{Coleppa:2013xfa,Coleppa:2014hxa,Kling:2015uba}, the current exclusion limits could   depend highly  on the non-SM Higgs spectrum.   The direct search limits on the heavy charged Higgs $H^\pm$ above $m_t$ are relatively weak given the large SM backgrounds for the dominant $H^\pm\rightarrow tb$ channel, and the relatively small branching fraction of $H^\pm \rightarrow \tau \nu$~\cite{Arbey:2017gmh}.

Flavor physics consideration usually constrains the charged Higgs mass to be larger than about 600 GeV for the Type-II  2HDM~\cite{Arbey:2017gmh}.  However, the charge Higgs contributions to various flavor observables can be cancelled by other new particles in a specific model \cite{Han:2013mga} and  be relaxed consequently. In our analyses, we focused on the indirect search potential of the Higgs factories on the masses of heavy Higgses.   Therefore, we did not impose  flavor constrains on the 2HDM parameter space, as well as the LHC direct search limits.


\subsubsection{2HDM loop effects}

We defined the normalized Higgs coupling including loop effects as:
\begin{equation}
\kappa^{\rm 2HDM}_{\rm loop}  \equiv  \frac{g_{\rm  tree}^{\rm 2HDM}+g_{\rm loop}^{\rm 2HDM}}{g_{\rm tree}^{\rm SM}+g_{\rm loop}^{\rm SM}},
\label{eq:loop}
\end{equation}
in which $g_{\rm  loop}^{\rm 2HDM}$ is 2HDM loop correction involving both SM loop corrections and non-SM parts.

To the leading order of 1-loop correction,  Eq.~(\ref{eq:loop}) simplifies to
\begin{equation}
\kappa^{\rm 2HDM}_{\rm 1-loop}  = \kappa^{\rm tree} + \Delta \kappa^{\rm 2HDM}_{\rm 1-loop},
 \end{equation}
with $\kappa^{\rm tree}\equiv g_{\rm tree}^{\rm 2HDM}/g_{\rm tree}^{\rm SM}$, $\Delta \kappa^{\rm 2HDM}_{\rm 1-loop}\equiv g_{\rm 1-loop}^{\rm 2HDM}/g_{\rm tree}^{\rm SM}$, where  $g_{\rm 1-loop}^{\rm 2HDM}$ is the one loop corrections with non-SM particles running in the loop.   In the alignment limit of $\cos(\beta - \alpha) =0$, $\kappa^{\rm tree}=1$, the expression is simplified to be
\begin{equation}
\kappa^{\rm 2HDM}_{\rm 1-loop}|_{\rm alignment} =1+ \Delta \kappa^{\rm 2HDM}_{\rm 1-loop} .
 \end{equation}

The expressions of the non-SM Higgs loop correction to the Higgs couplings are summarized in Appendix~\ref{app:2HDMloop} \cite{Kanemura:2015mxa,Kanemura:2014dja,Kanemura:2004mg}  under  the alignment limit $\cos (\beta-\alpha)=0$ and with the mass simplification relation $m_{H^{\pm}} = m_{A^0}=m_{H^0}\equiv m_{\phi} $.   Note that the tree level relations of $\kappa_W = \kappa_Z $ and $\kappa_{\mu} = \kappa_{\tau}$ are still approximately valid at 1-loop level.

\begin{figure}[h]
\begin{center}
\includegraphics[width=7.5cm]{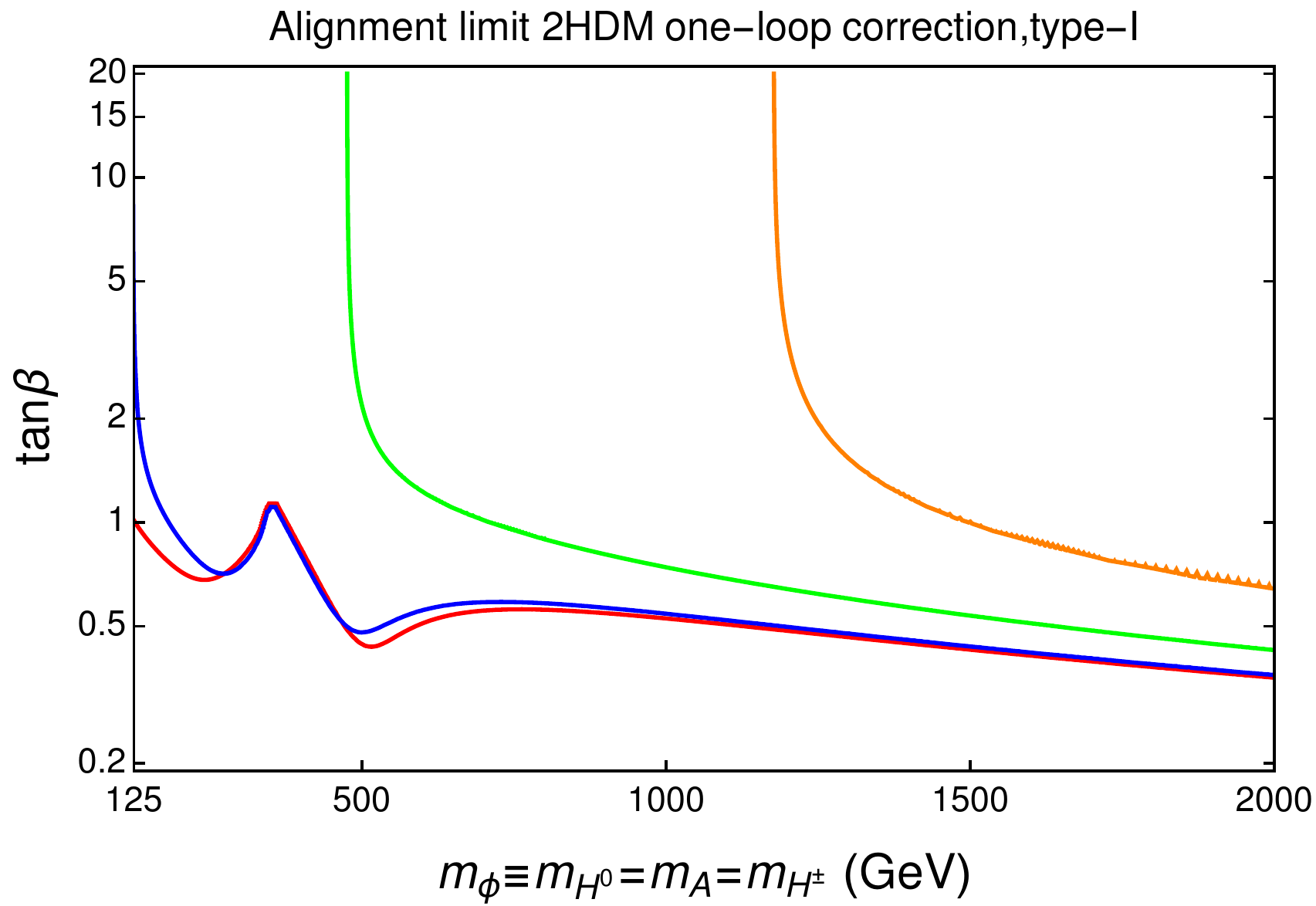}
\includegraphics[width=7.5cm]{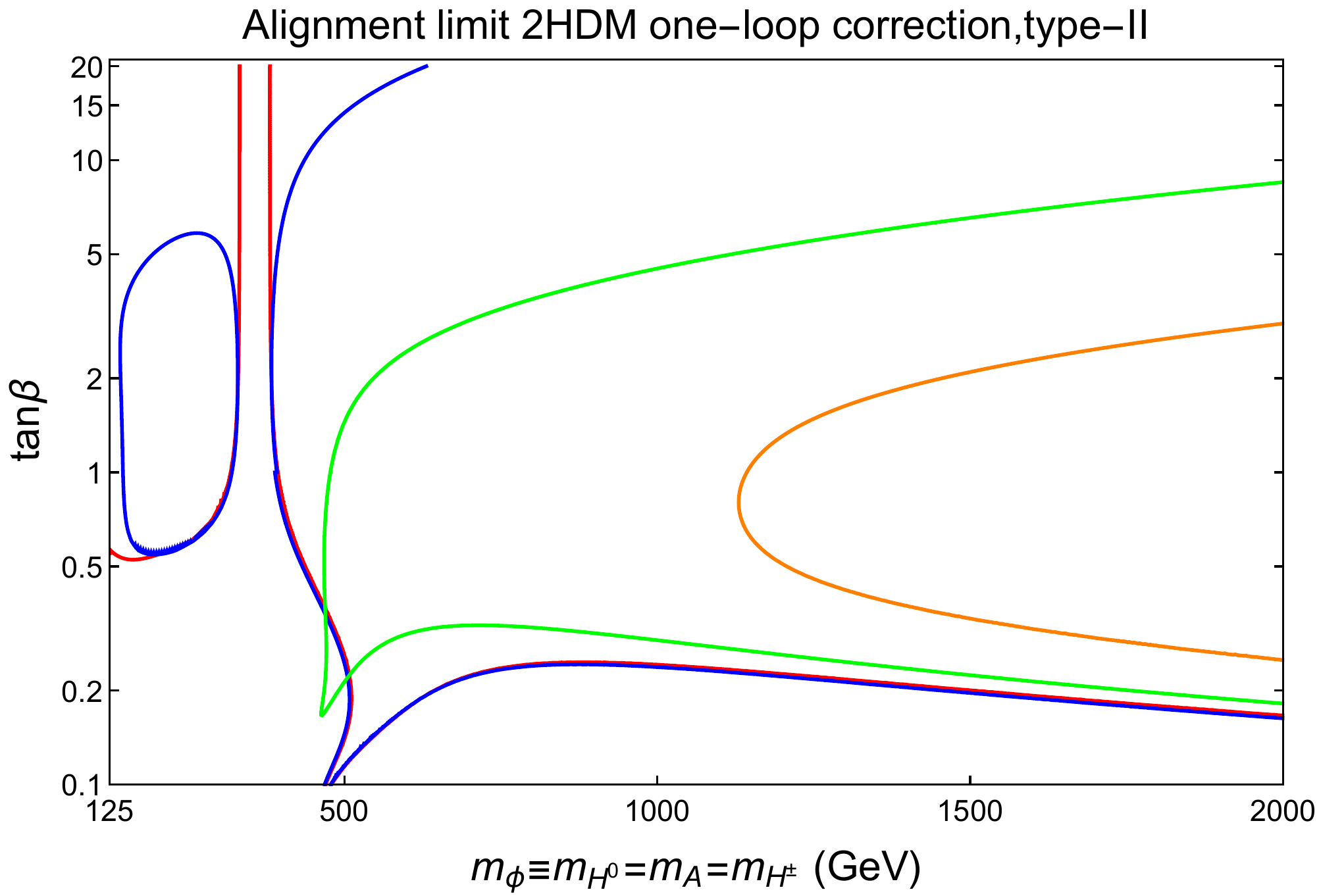}\\\vspace{3mm}
\includegraphics[width=7.5cm]{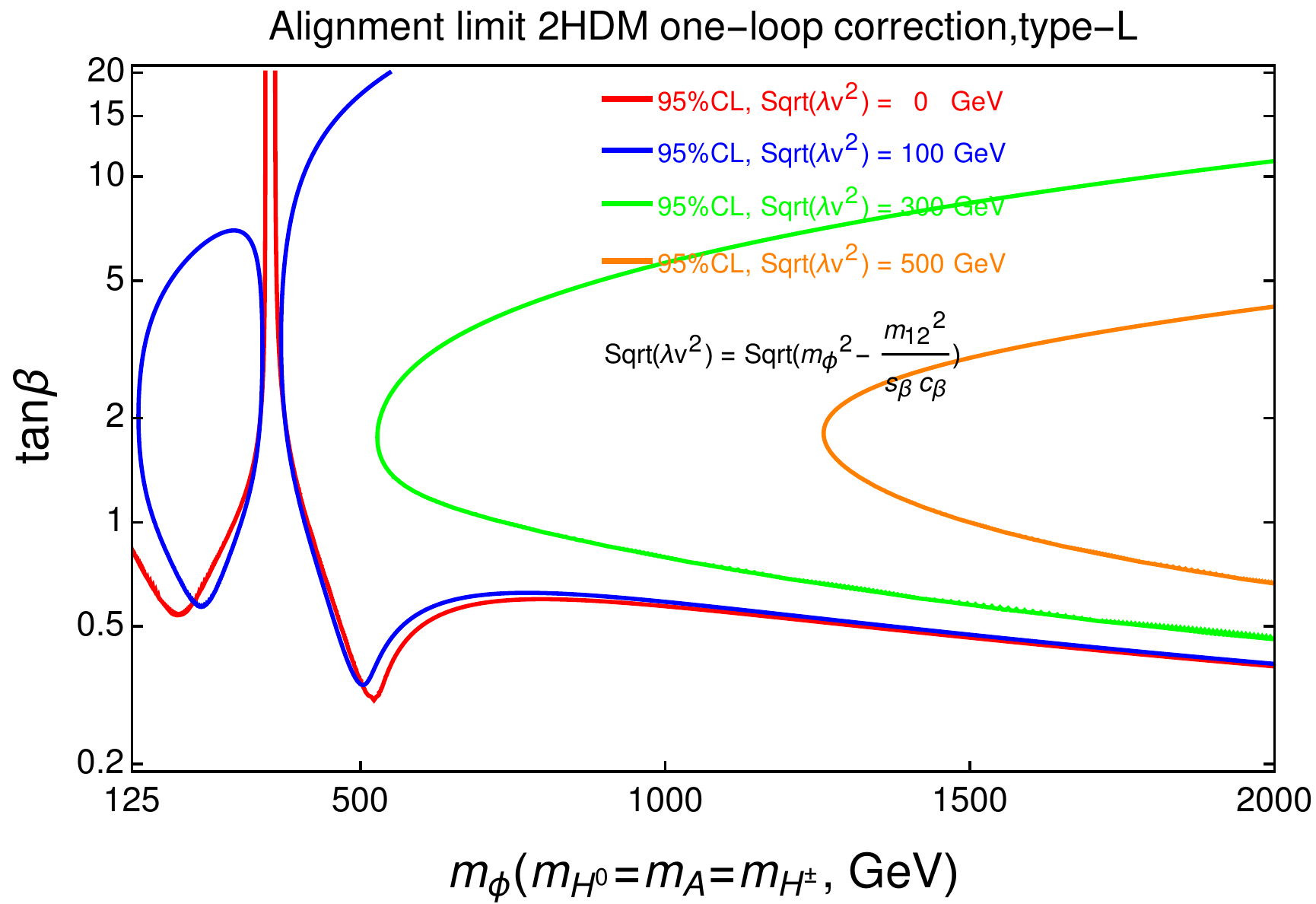}
\includegraphics[width=7.5cm]{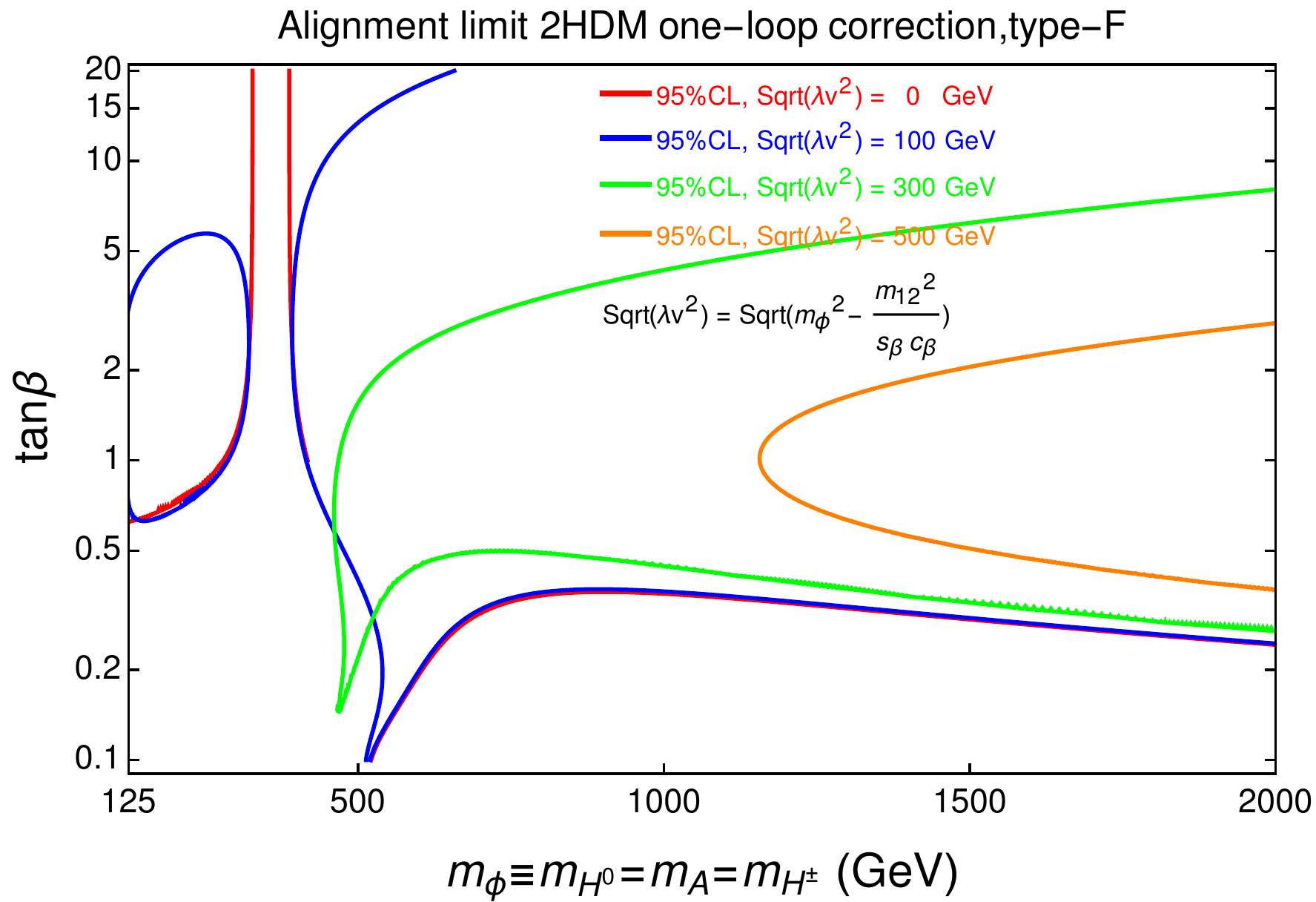}
\caption{95\% C.L. constraints on $\tan\beta$ vs.  $m_{\phi}$  plane based on CEPC Higgs precision measurements.  The orange, green, blue and red (from right to the left at the large $m_\phi$ region) are for $\sqrt{\lambda v^2} \equiv \sqrt{m_{\phi}^2 - \frac{m_{12}^2}{s_{\beta} c_{\beta}}}=$500, 300, 100, 0 GeV, respectively.   Regions to the right of the curves for large $m_\phi$ or to the left of the curves for small $m_\phi$ (enclosed region for the blue curves) are allowed.  Four panels are for Type-I, II, L, and F 2HDM, as labeled at the top of each figure.}
\label{fig:cepc_loop}
\end{center}
\end{figure}

Following the same global fitting techniques as in the tree level case,  we obtained the 95\% C.L. constraints on  the $\tan\beta$ vs. $m_{\phi}$ plane for four different types of 2HDM given CEPC precisions, as shown in Fig.~\ref{fig:cepc_loop}.   Four benchmark values of $\sqrt{\lambda v^2}$ are chosen: 0, 100, 300, and 500 GeV, which correspond to red, blue, green, and orange curves.   Regions to the right of the curves for large $m_\phi$ or to the left of the curves for small $m_\phi$ (enclosed region for the blue curves) are allowed by the CEPC precision measurements assuming no deviations from the SM predictions are observed.

For Type-I, small values of $\tan\beta \lesssim 0.5$ are excluded since all the non-SM Higgs Yukawa couplings are proportional to $1/\tan\beta$.  The dependence on $\tan\beta$ is   weak once $\tan\beta \gtrsim 2$.   While for smaller values of $\lambda v^2$, $m_\phi$ as low as 125 GeV are allowed,   $m_\phi$ is constrained to be larger than 500 GeV for $\sqrt{\lambda v^2}=300$ GeV, and 1200 GeV for $\sqrt{\lambda v^2}=500$ GeV, given that heavy Higgs loop corrections enhance as  $\lambda v^2$ increases.  The small kink around $m_\phi\sim 350$ GeV is due to the top quark threshold effects in Yukawa couplings.     The relaxed constraints on $\tan\beta$ around $m_\phi\sim 500$ GeV are caused by the smallness of $\Delta\kappa_{b, c, \tau}$, which flips sign near that region.

For Type-II, while the generic features of the excluded region are the same as those in Type-I, there are three major differences.  Firstly, the constraints at large $\tan\beta$ get tighter since both the down-type Yukawa and lepton Yukawa (in particular, those of bottom and tau) are proportional to $\tan\beta$.  Therefore, loop contributions are more tightly constrained at large $\tan\beta$.  Secondly, the top quark threshold effects are stronger since the relevant terms do not have the    $ {1}/{\tan^2 \beta}$ suppression as  in Type-I.   Thirdly, the constraints at small $\tan\beta$ get weaker.   This is because only up-type Yukawa couplings are proportional to $1/\tan\beta$, while the precision on that  is   worse than the $bb$ and $\tau\tau$ channels.

The results for Type-L are similar to those  of Type-II, with small $\tan\beta$ constraints getting  stronger since the down-type Yukawa couplings are now proportional to $1/\tan\beta$ as well.  The sensitivity at large $\tan\beta$ becomes a bit weaker  since the lepton Yukawa coupling structure is the same as that of Type-II.    The results for Type-F are almost the same as that of Type-II, since for the most precisely measured couplings, $hZZ$ and $hbb$, the dominant loop contributions (from bottom quark and heavy Higgses) in these two types are identical.

\begin{figure}[h]
\begin{center}
\includegraphics[width=7.5cm]{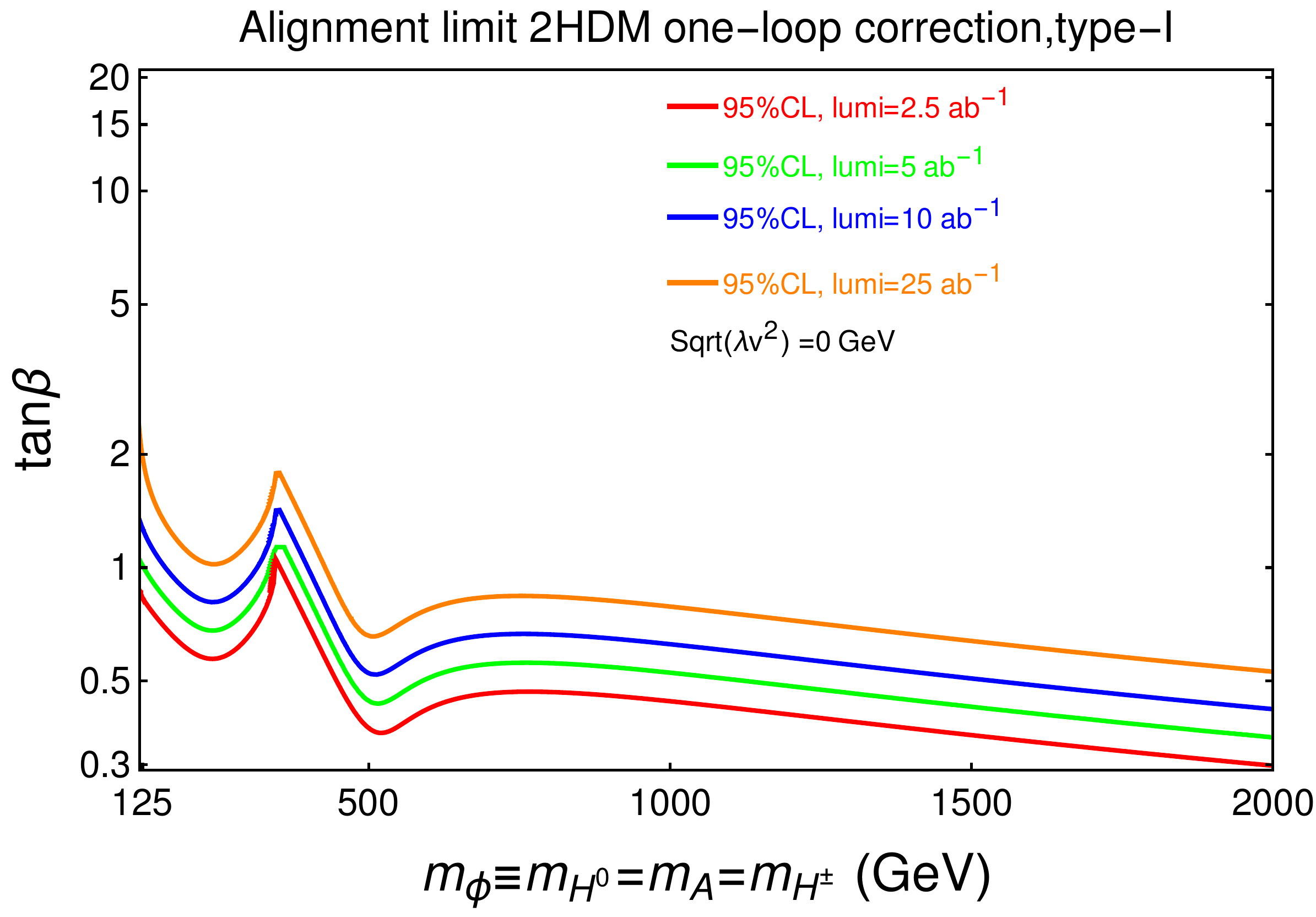}
\includegraphics[width=7.5cm]{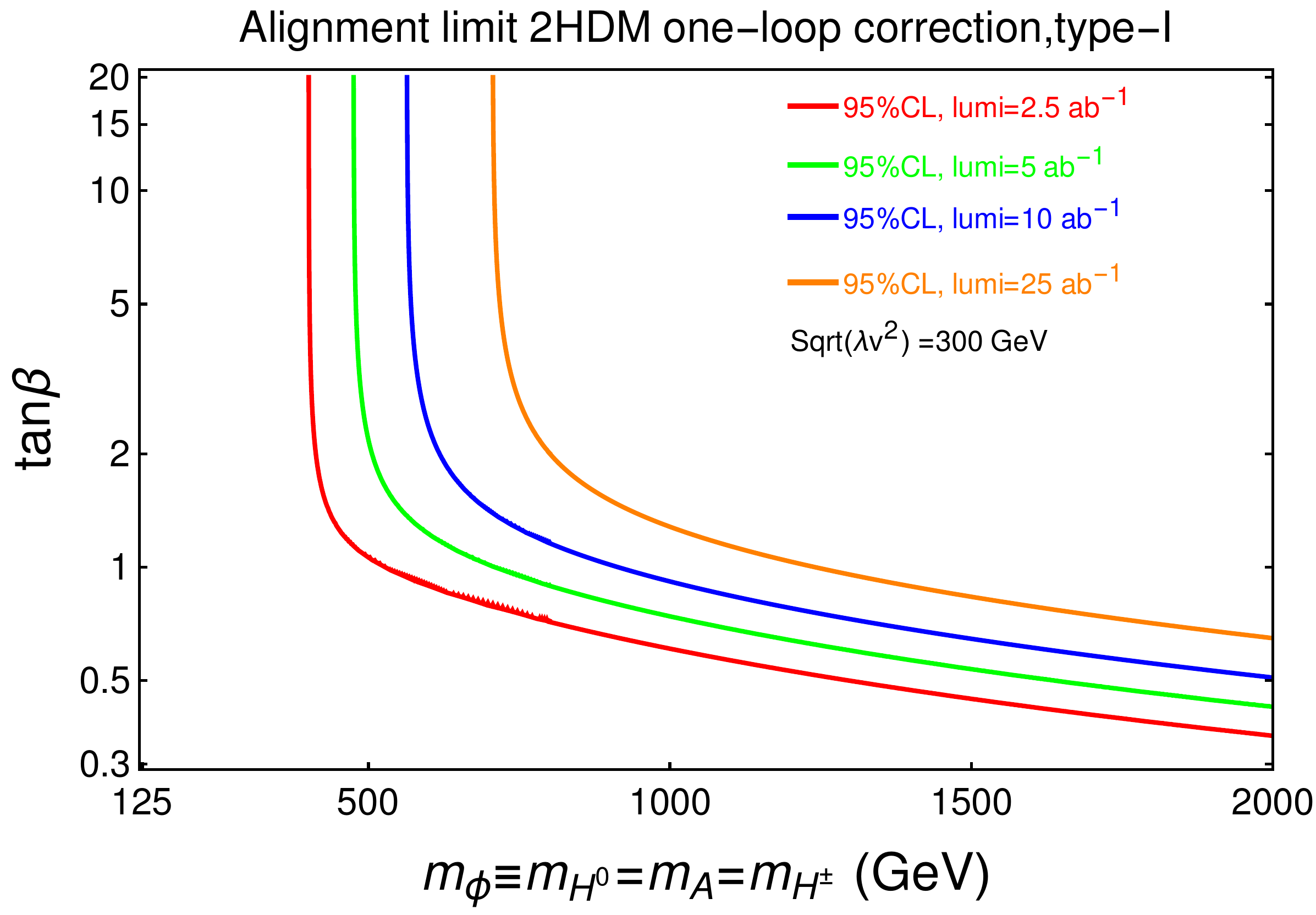}\\\vspace{3mm}
\includegraphics[width=7.5cm]{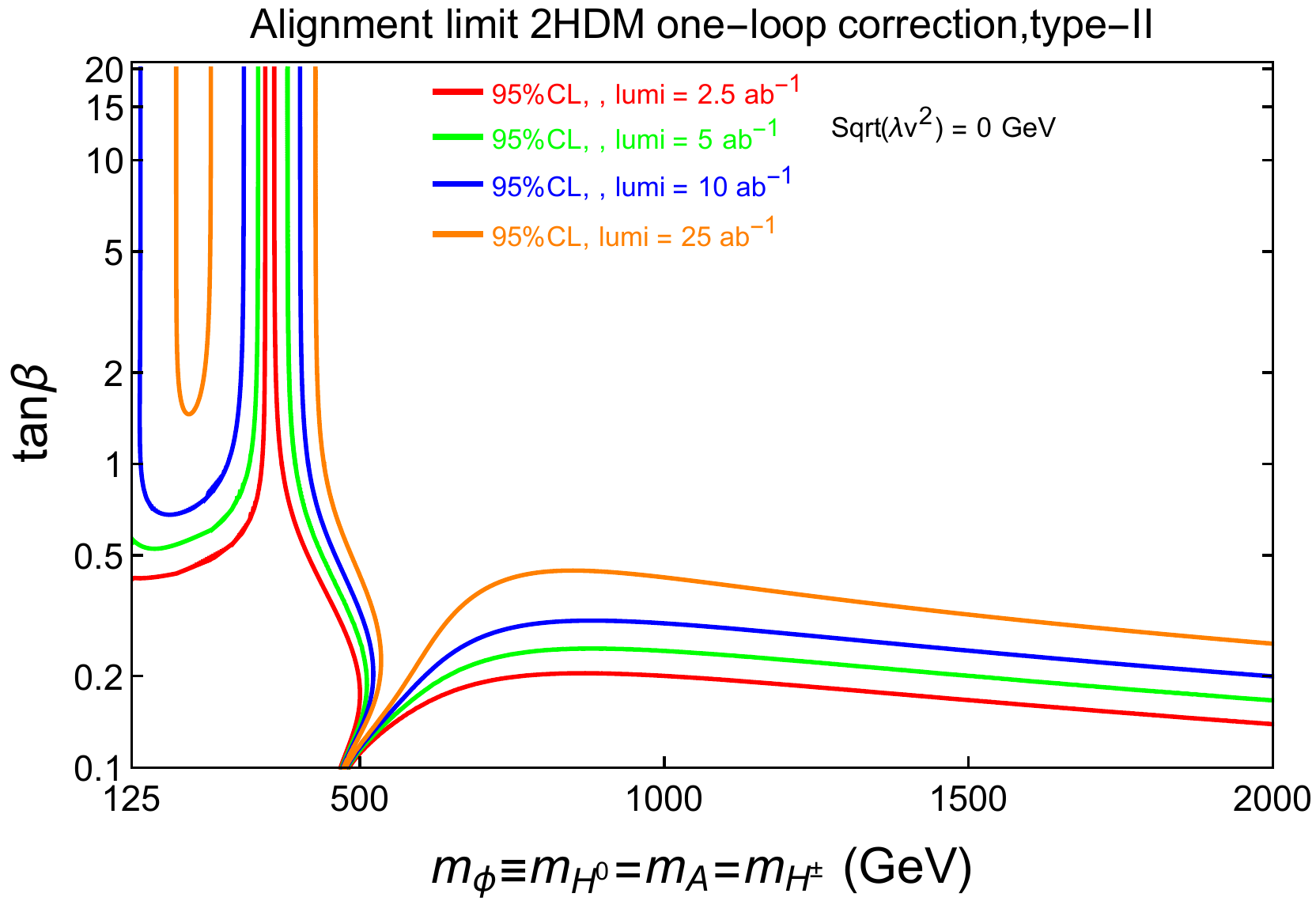}
\includegraphics[width=7.5cm]{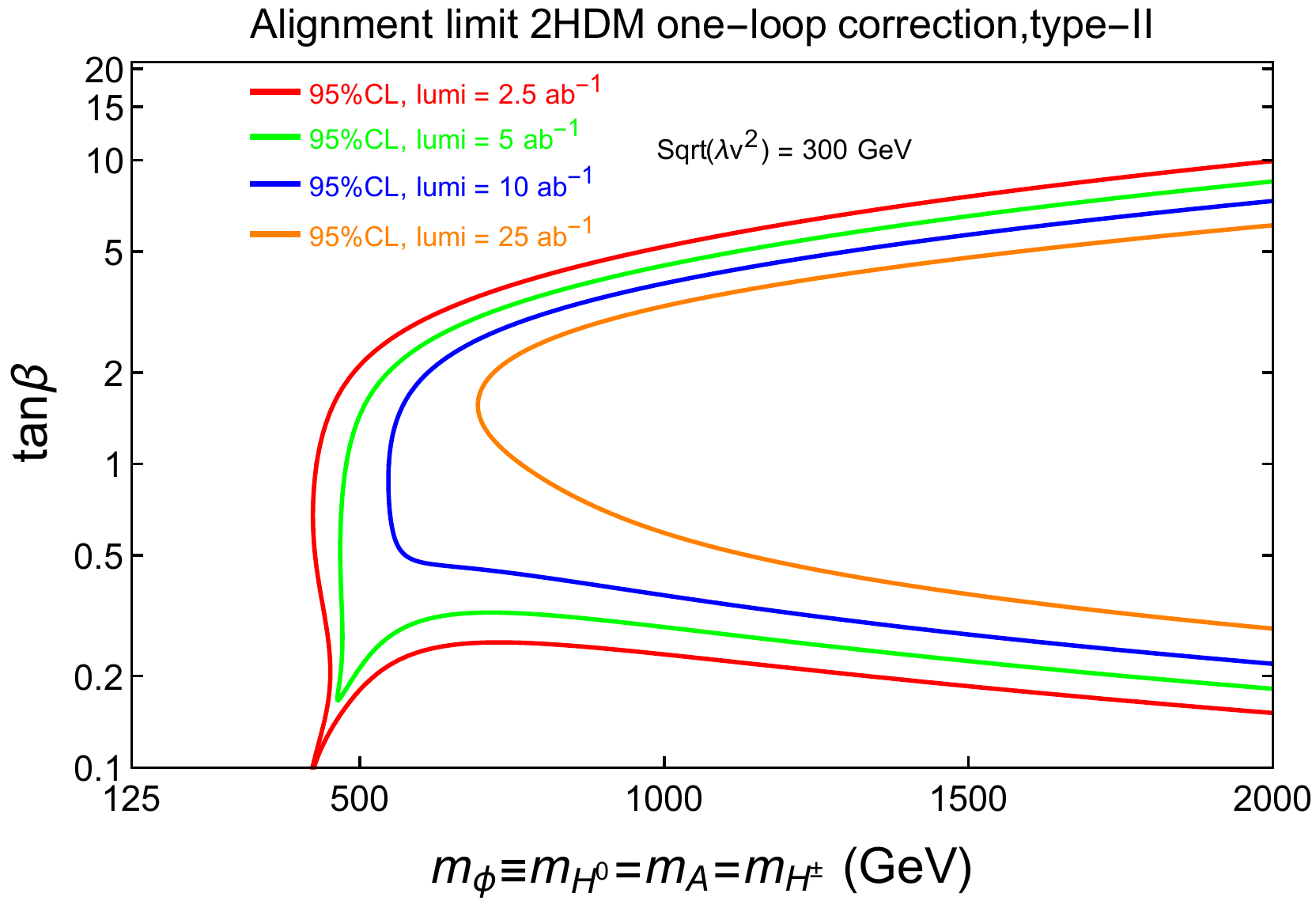}
\caption{95\% C.L. Higgs precision constraints on $\tan\beta$ vs. $m_{\phi}$  plane, assuming   integrated luminosity of 2.5, 5, 10,  and 25 ${\rm ab}^{-1}$.   Results for Type-I (top panels) and Type-II (bottom panels) are presented, for two different benchmarks of $\sqrt{\lambda v^2}=0$ (left panels) and 300 GeV (right panels). }
\label{fig:cepc_lumi}
\end{center}
\end{figure}
To explore the effects of different running scenarios with different integrated luminosity, we presented the 95\% C.L. Higgs precision constraints on $\tan\beta$ vs. $m_{\phi}$ plane in Fig.~\ref{fig:cepc_lumi}, with a rescaled error bar of different integrated luminosity of 2.5, 5, 10,  and 25 ${\rm ab}^{-1}$, based on CEPC 5 ${\rm ab}^{-1}$ results.    Here we have assumed that statistical error dominates\cite{CEPC-SPPCStudyGroup:2015csa,Fujii:2015jha}.  Both the upper and the lower limits on $\tan\beta$, as well the lower limits for $m_\phi$ at large $m_\phi$ region get stronger with increasing luminosity.  For both Type-I and II with $\sqrt{\lambda v^2}=300$ GeV, the lower limit on $m_\phi$ increases from 400 GeV to 700 GeV when the luminosity increases from  2.5 to 25 ${\rm ab}^{-1}$.

\begin{figure}[h]
\begin{center}
\includegraphics[width=7.5cm]{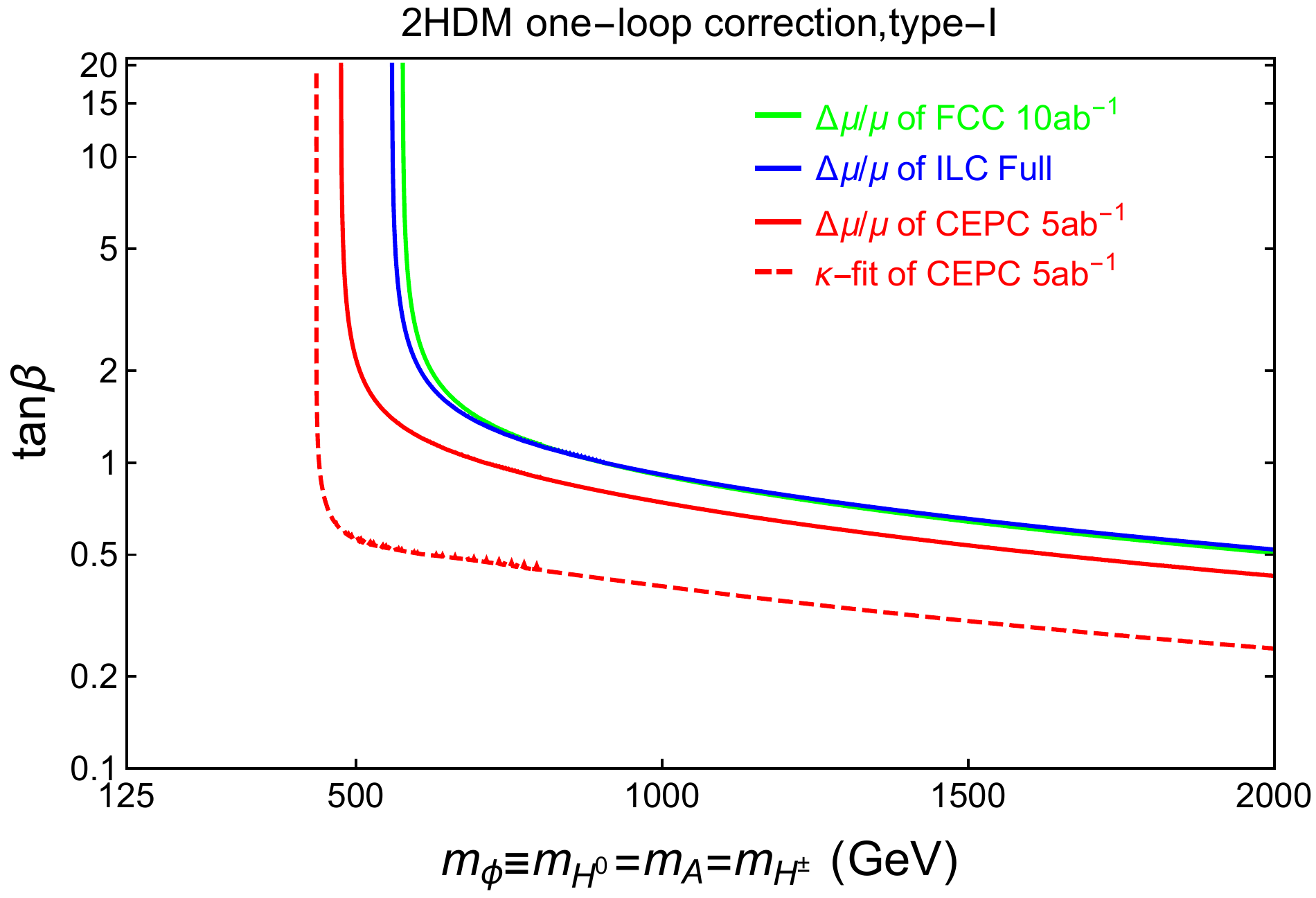}
\includegraphics[width=7.5cm]{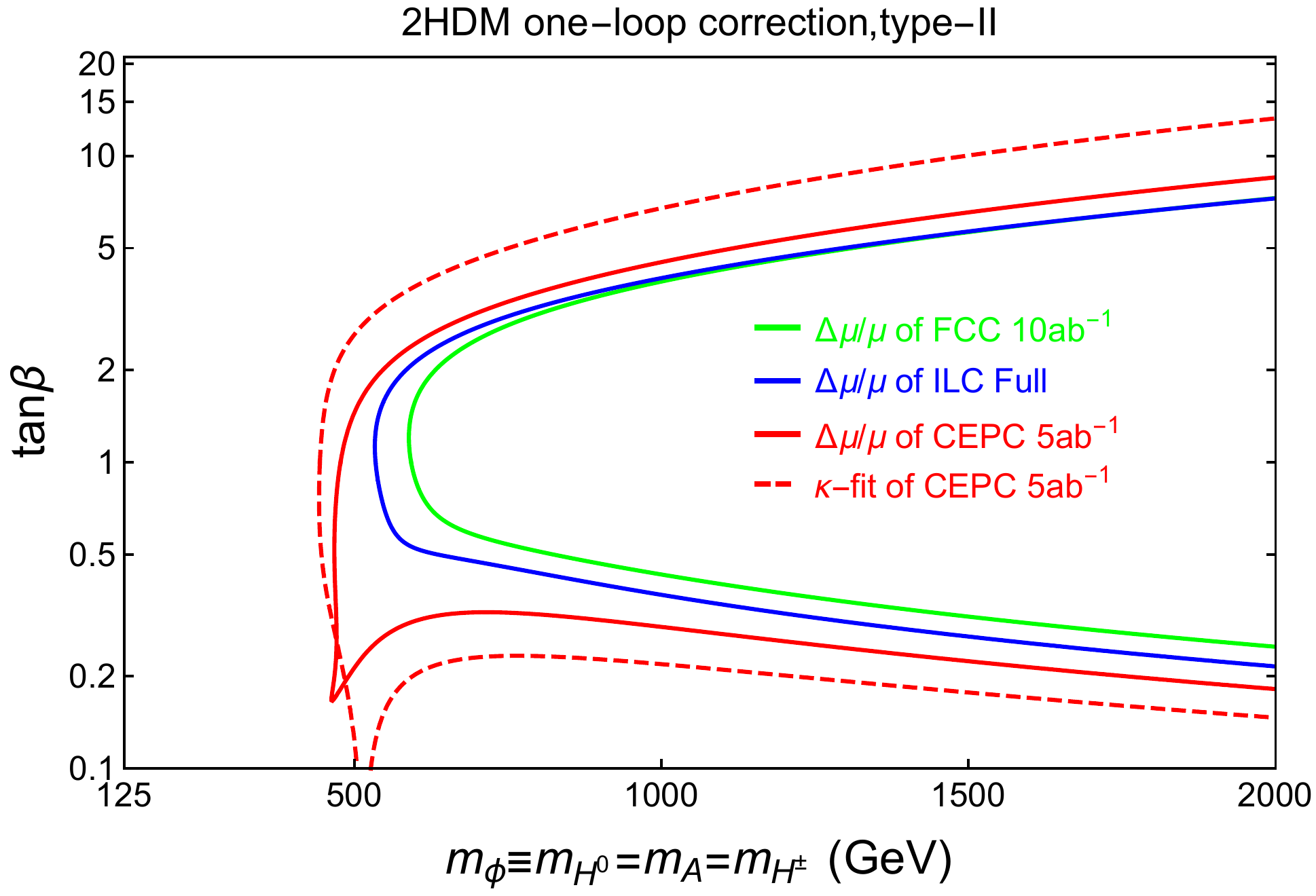}
\caption{95\% C.L. Higgs precision constrained regions in $m_\phi$ vs. $\tan\beta$ for CEPC (red lines), ILC (blue lines), and FCC-ee (green lines). Left panel is for Type-I and right panel is for Type-II with $\sqrt{\lambda v^2}=$ 300 GeV.  The CEPC results using $\kappa$-fit are shown in red dashed lines. }
\label{fig:ilc_loop}
\end{center}
\end{figure}

To compare the sensitivity of different future Higgs factories with the running scenario   listed in Table.~\ref{tab:mu_precision}, as well as the effects of $\Delta\mu/\mu$-fit vs. $\kappa$-fit, we show the 95\% C.L. constrained region in $\tan\beta$ vs. $m_\phi$ plane for CEPC (red lines), ILC (blue lines), and FCC-ee (green lines).  For CEPC precision, the $\Delta\mu/\mu$-fit results are shown in solid lines and $\kappa$-fit results are shown in dashed lines.   ILC and FCC-ee have similar sensitivities, both   better than the CEPC results.   In addition, results with $ {\Delta \mu}/{\mu}$ fit are better than the results with $\kappa$-fit, confirming the tree level fitting results   that including the correlations between different couplings is important.  Overlooking those correlation effects might lead to overly conservative results.

\begin{figure}[h]
\begin{center}
\includegraphics[width=7.5cm]{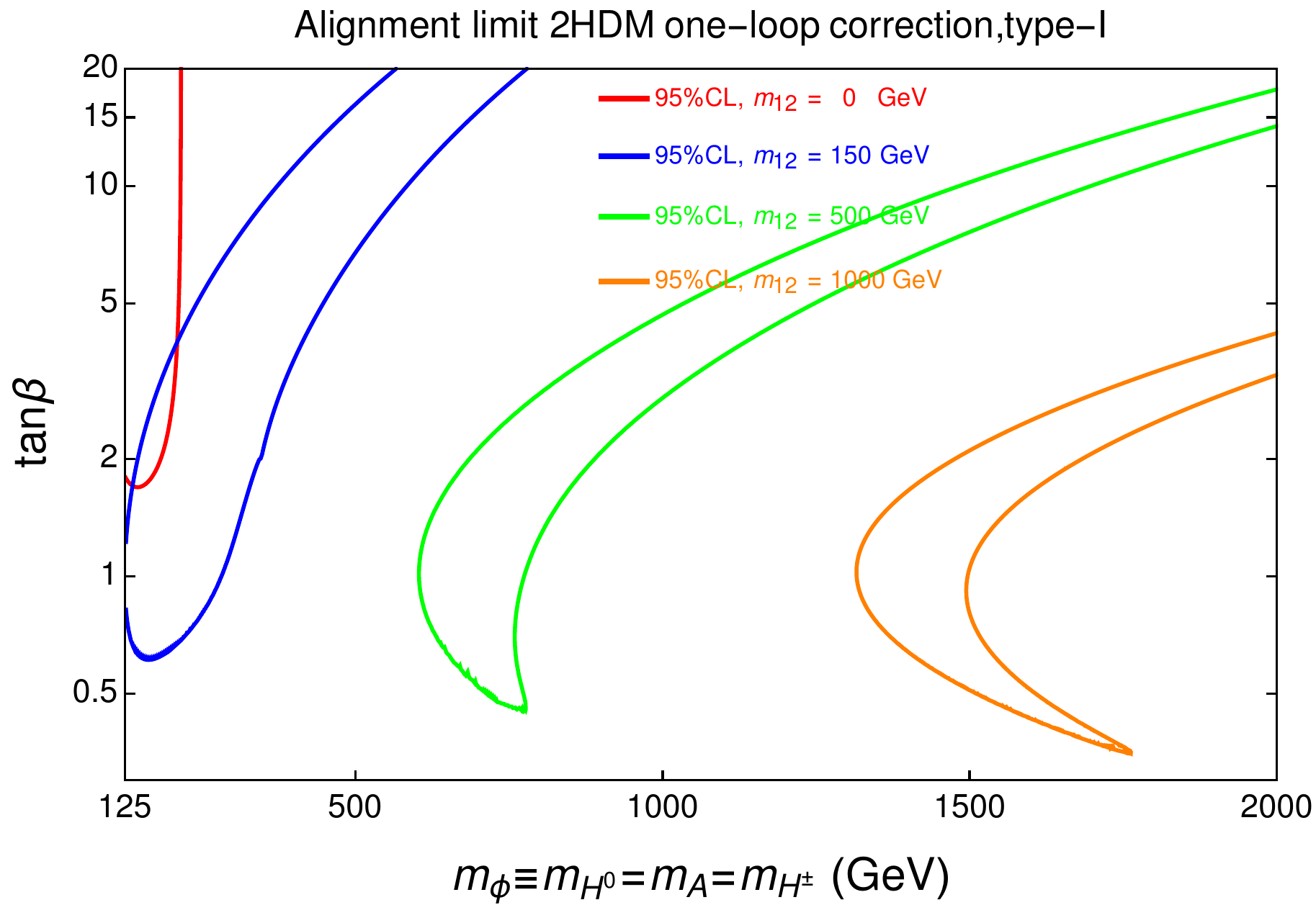}
\includegraphics[width=7.5cm]{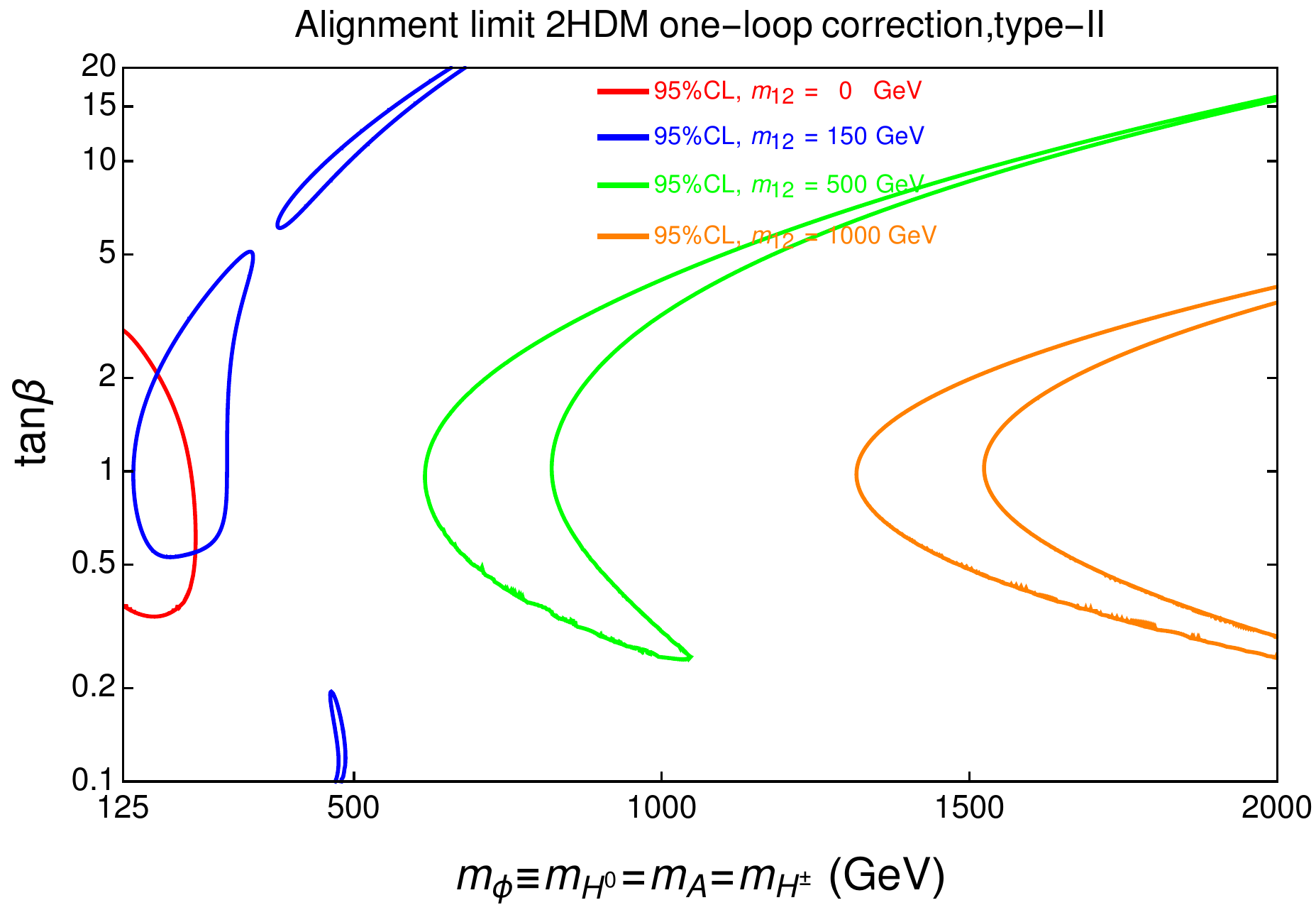}\\\vspace{3mm}
\includegraphics[width=7.5cm]{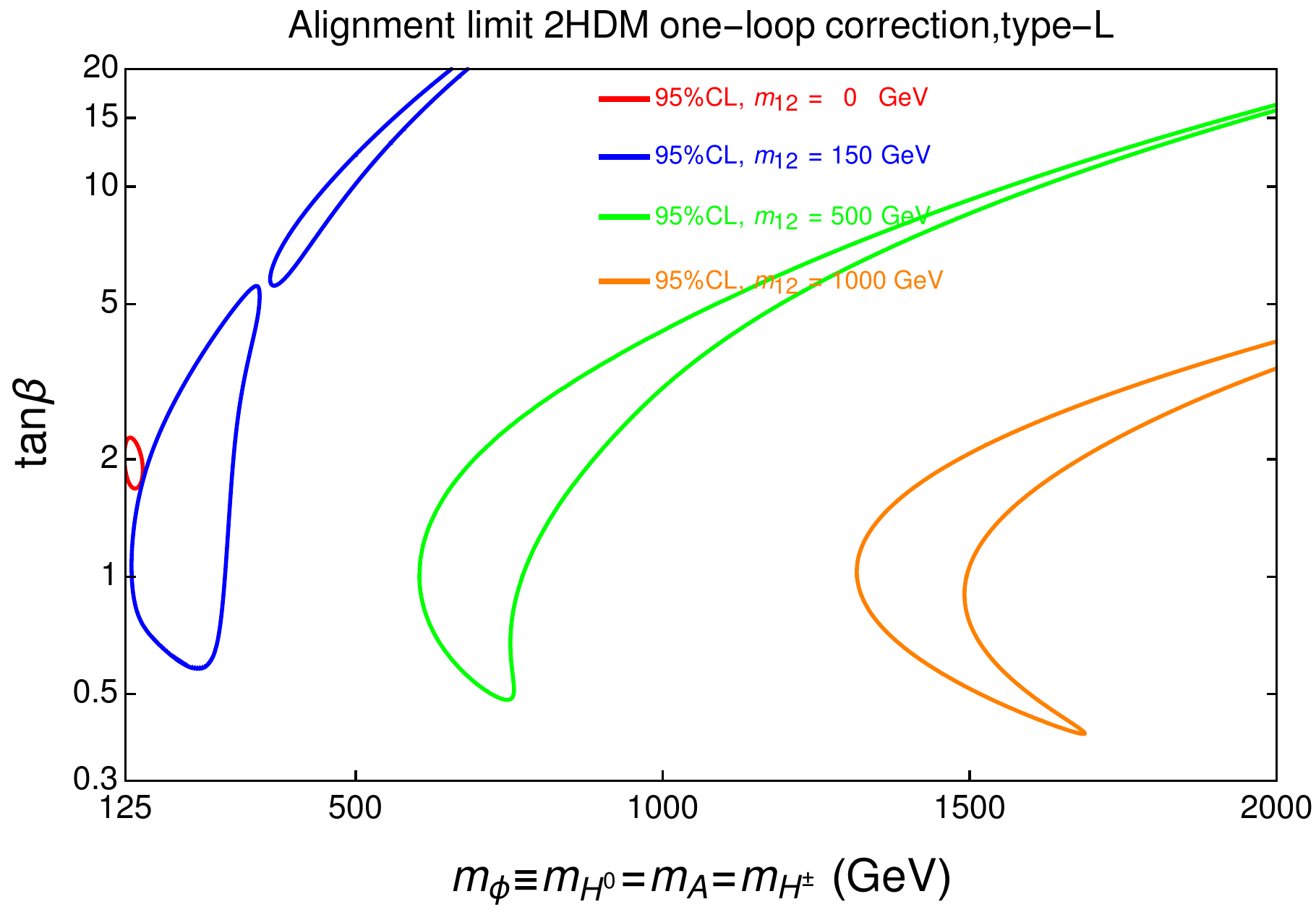}
\includegraphics[width=7.5cm]{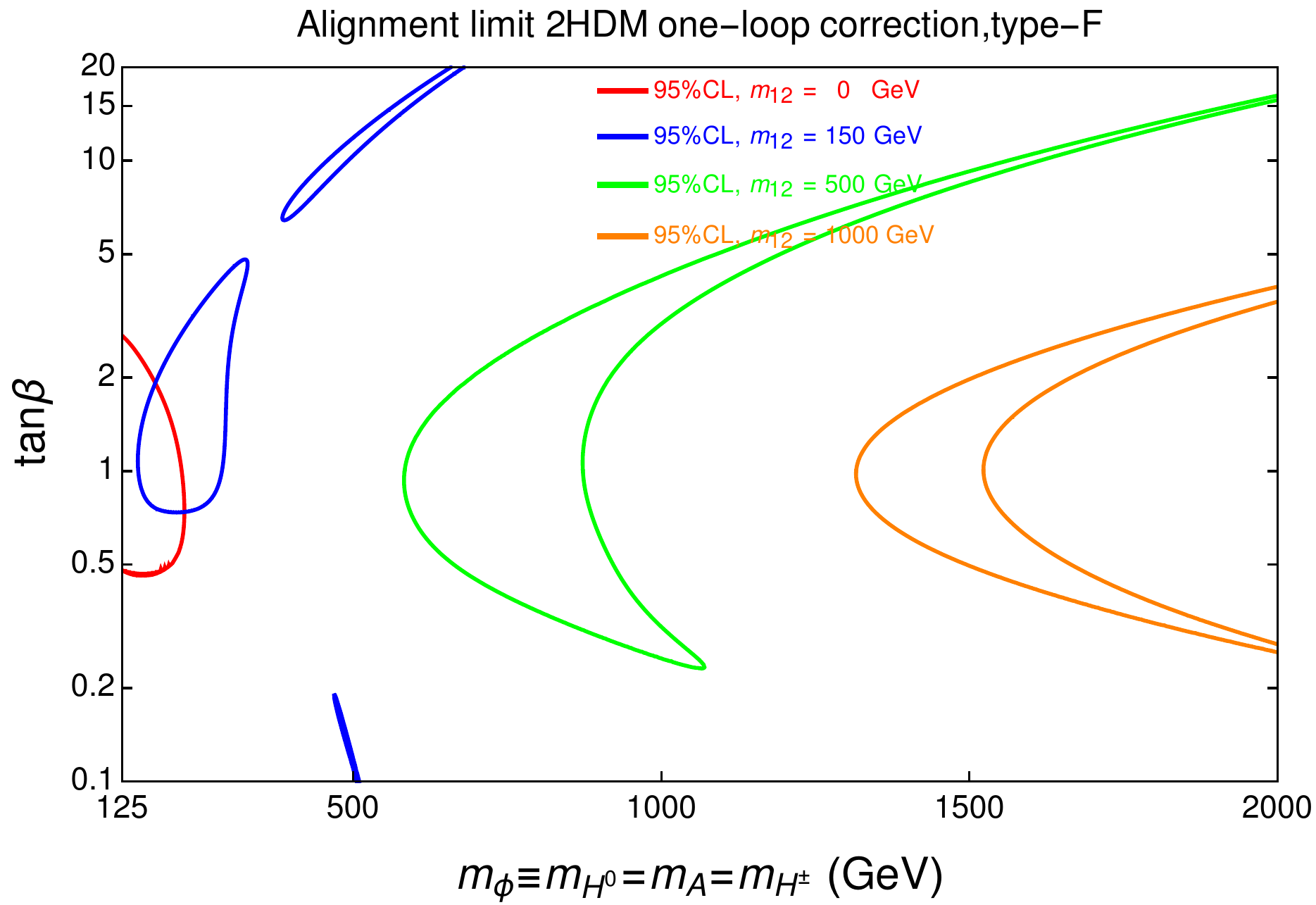}
\caption{The same global fitting results with CEPC as Fig.~\ref{fig:cepc_loop}, with different values of $m_{12}^2$, instead of $\lambda v^2$. The survival regions exhibit (almost) enclosed behavior, with both upper and lower limit on $m_\phi$ for a given value of $\tan\beta$. }
\label{fig:cepc_loopm12}
\end{center}
\end{figure}

Results we presented so far are based on the assumption of  a fixed value of $\lambda v^2= m_{\phi}^2-m_{12}^2/(\sin \beta \cos \beta)$.  It is convenient to do so  since $\lambda v^2$  directly enters the tri-Higgs couplings.    Another common strategy is to fix the value of the soft $Z_2$ breaking parameter $m_{12}^2$.  The global fitting results are shown in Fig.~\ref{fig:cepc_loopm12} for fixed values of $m_{12}^2$ to be 0 (red), $150^2$ (blue), $500^2$ (green) and $1000^2$ (orange) ${\rm GeV}^2$. As before, the top quark threshold effects show up in the blue region. Negative values of $m_{12}^2$ typically lead to large $\lambda v^2$, therefore much worse values of $\chi^2$ and disfavored.  The survival regions now exhibit (almost) enclosed behavior, with both upper and lower limit on $m_\phi$ for a given value of $\tan\beta$.    The location of the allowed $m_\phi$ region also shifts to the larger mass for larger values of $m_{12}$.     For the asymptotic large $m_\phi$ and $\tan\beta$ region, the allowed region centered around $m_\phi^2=m_{12}^2/(\sin\beta\cos\beta)$ line, which minimizes the corresponding $\lambda v^2$.

\begin{figure}[h]
\begin{center}
\includegraphics[width=7.5cm]{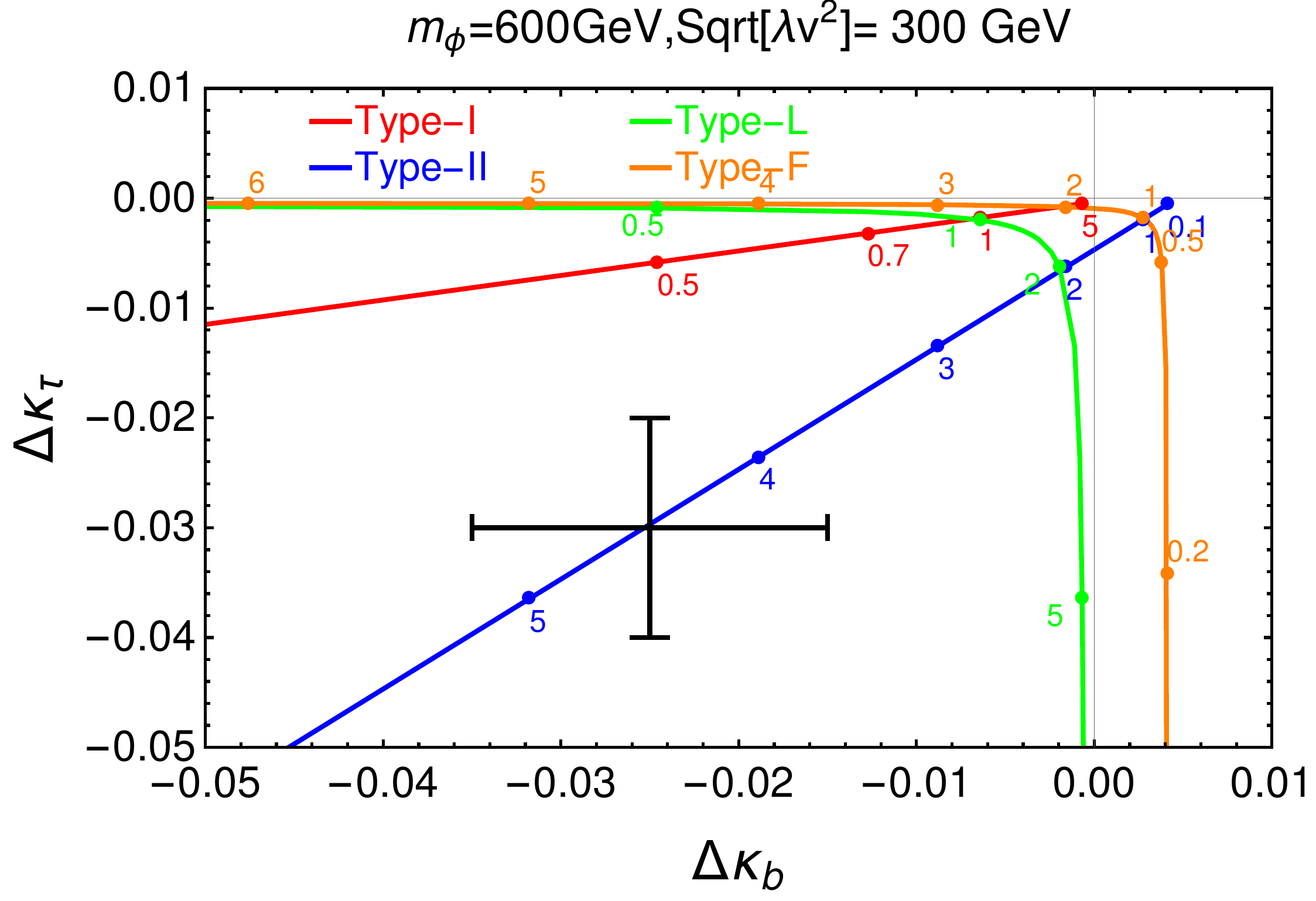}
\includegraphics[width=7.5cm]{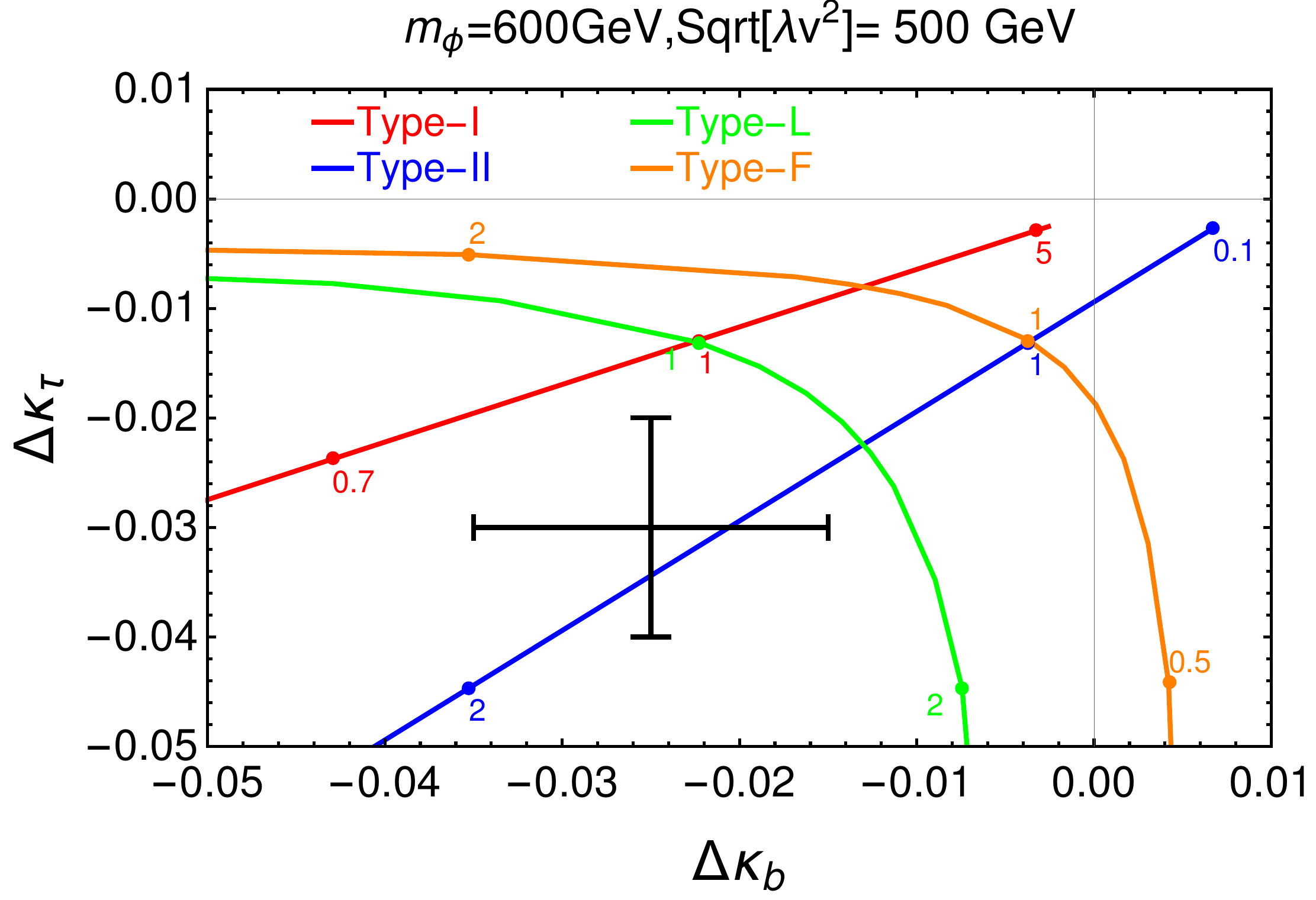}\\
\caption{Loop corrections of $\Delta \kappa_{\tau}$ vs $\Delta \kappa_b$ for four different types of 2HDM, for $m_\phi=600$ GeV, with $\sqrt{\lambda v^2} =$ 300 GeV (left panels), and 500 GeV (right panels).    The values of $\tan\beta$ are indicated in the plots with dots.   The black cross indicate  the estimated experimental errors with a random central point.  }
\label{fig:four_quadrant_loop}
\end{center}
\end{figure}

To study the sensitivity of Higgs precision measurements on the model parameters, as well as to demonstrate how well Higgs precision measurements can be used to distinguish different types of models once certain deviations in the couplings are measured, we show the loop corrections of $\Delta \kappa_{\tau}$ vs $\Delta \kappa_b$ for four different types of 2HDM, for $m_\phi=600$ GeV, with $\sqrt{\lambda v^2} =$ 300 GeV (left panels), and 500 GeV (right panels).  Different values of $\tan\beta$ are also indicated in the curves with dots to show the sensitivity of $\kappa_{b,\tau}$ to the values of $\tan\beta$.

For Type-I, the loop corrections to both $\kappa_b$ and $\kappa_\tau$ are large for small $\tan\beta$, about 4.3\% in $\Delta \kappa_b$ and 2.3\% in $\Delta\kappa_\tau$ for $\tan\beta=$0.7 with $\sqrt{\lambda v^2}=500$ GeV, to be well explored by precision measurements.  The contribution quickly decreases once $\tan\beta$ gets larger.  For Type-II, the dependence on $\tan\beta$ is flipped, with the corrections to $\kappa_\tau$ significantly larger than that of Type-I.  For Type-L, the correction to $\kappa_b$ is large at small $\tan\beta$, while the corrections to $\kappa_\tau$ increases for larger $\tan\beta$.   The opposite behavior occurs for Type-F.   For larger values of $\sqrt{\lambda v^2}$, the loop corrections get even larger.

Since each type of 2HDM swipes out a different region in $\Delta \kappa_b$ vs. $\Delta \kappa_\tau$ space,   simultaneous measurements on $\kappa_b$ and $\kappa_\tau$ with the estimated precision could help   to distinguish four types of 2HDM models.   For CEPC with 5 ab$^{-1}$ 7-parameter fit, the estimated precision is about 1.2\% for $ {\kappa_b}$ and about 1.3\% for $ {\kappa_\tau}$, as indicated by the cross in Fig.~\ref{fig:four_quadrant_loop}, which provide enough precision to separate contributions from different models.    Note that without knowing the value of $\tan\beta$ a priori,  there is degeneracy between Type-L and Type-F, in particular, if only deviations in $\kappa_b$ is observed while $\kappa_\tau$ deviation is small.   However, with the theoretical constraints discussed in Sec.~\ref{sec:2HDMloop} imposed, a constrained range of $\tan\beta$ as  shown in Fig.~\ref{fig:constraint} and Table~\ref{tab:lam_cons} could help to lift the degeneracy.    Since the loop corrections typically get smaller for larger values of $m_\phi$ and smaller values of $\lambda$,  the potential of $\kappa_b$ and $\kappa_\tau$ measurements to determine the values of $\tan\beta$  becomes limited accordingly.

\section{Composite Higgs models}
\label{sec:comp}

The class of BSM physics that features strong dynamics genuinely leads to testable predications through precision measurements.
Composite Higgs Models are very representative class of models  that ties directly to the   hierarchy problem.
 The Higgs boson, instead of being a fundamental particle as in the SM, SUSY model or gauge extensions of the SM, would be a
composite particle as a bound state of additional strong dynamics of the underlying physics model. Higgs now can be viewed as a pseudo-Nambu-Goldstone boson, whose mass is protected by the global symmetry breaking scale parameter $f$ generated by the condensation of strongly interacting particles. The separation between the electroweak scale and the composite scale $f$ is naturally a tuning parameter, representing the fine-tuning of the underlying composite model.

There is a vast range of plausible composite Higgs models. To evaluate the physics reach of the Higgs precision measurements on this broad class of models, we adopted  two approaches in discussing the results. The first approach is to interpret the Higgs precision results  in the Minimal Composite Higgs Model (MCHM)~\cite{Agashe:2004rs} with various embedding of the partners of the SM fermions\footnote{For recent studies focus on the future collider perspective, see, e.g., Refs.~\cite{Thamm:2015zwa, Kanemura:2016tan}.}.
 The second approach is to adopt the language of EFT and followed the discussion of the patterns of strong interactions~\cite{Liu:2016idz}, comparing the implication of the Higgs precisions on the so called Accidentally Light Higgs (ALH), Strongly Interacting Light Higgs (SILH)~\cite{Giudice:2007fh}, and a general SILH (GSILH).  Each of those patterns has different assumptions on the symmetries of the underlying strong dynamics.
We discussed in details about their  underlying assumptions in the following sections.

\subsection{Minimal Composite Higgs Models}
The MCHMs represent a minimal embedding of the Higgs as a pseudo-Nambu-Goldstone boson of the global symmetry $SO(5)/SO(4)$ that respects custodial symmetry.
 We investigated and found a way to best present limits on various tower fermion representations in the MCHMs, following the notation   and calculations detailed in Ref.~\cite{Carena:2014ria}.

In the minimal coset $SO(5)/SO(4)$, gauge invariance fixes the rescaled gauge coupling $\kappa_V$ to be
 \begin{equation}
\kappa_V \equiv \frac{g^{\rm CH}_{hVV}}{g^{\rm SM}_{hVV}} = \sqrt{1-\xi} \,,
\end{equation}
where $\xi \equiv v^2/f^2$ parameterizes the vacuum misalignment.   To simplify the notation, in this section, we referred  to the rescaled couplings of the SM Higgs $\kappa_h^i$ as $\kappa_i$.    The modification to the Yukawa couplings depends on the fermion representations.  In many models, the rescaled Higgs to fermion couplings $\kappa_t$ or $\kappa_b$ can be either
\begin{equation}
F_1\equiv  \frac{1-2\xi}{\sqrt{1-\xi}} \,, \hspace{1cm}  F_2\equiv  \sqrt{1-\xi} \,,  \label{eq:F1F2}
\end{equation}
if summed over all the tower contributions.
In these cases $\xi$ is the only model parameter at leading order for a specific model. For some more complex embedding of the fermions, such as MCHM$_{14-14-10}$, MCHM$_{14-5-10}$ and MCHM$_{5-14-10}$, the  Higgs couplings, are modified in a  more complex way~\cite{Carena:2014ria}.    The corresponding Yukawa couplings are related to functions $F_{3,4,5}$, which  depend on several microscopic Yukawa couplings, with the explicit expressions can be found in Ref.~\cite{Carena:2014ria}.    The resulting  Higgs couplings vary in a certain range, even for a fixed value of $\xi$. In several limiting cases when one of the microscopic Yukawa couplings turns off, these additional coupling functions reduce to simpler functions of $F_1$ and $F_2$.

Note that these simple closed form expressions are obtained by summing over the infinite tower fermions. In reality, the summation is truncated after a few tower fermions, generating a more complicated and scattered relation between model parameters. For simplicity and as the modifications to the Higgs couplings are dominated by the first few tower fermions in most cases, we adopted these simplified formulae to obtain a qualitative physics reach of the Higgs precision program in the MCHMs.

\begin{table}[htbp]\small
\centering
\begin{tabular}{|c|c|c|c|c|c|c|c|} \hline
& 5,~10 & \multirow{4}{*}{10-5-10} & \multirow{4}{*}{5-5-10} &  & \multirow{4}{*}{14-14-10} &  \multirow{4}{*}{14-5-10} &  \multirow{4}{*}{5-14-10}\\
MCHM & 14-1-10 & & & 5-10-10 & & &\\
Reps.& 14-10-10 & & & 5-1-10 & & &\\
& 10-14-10  & & & & & & \\\hline\hline
 $\kappa_t$, $\kappa_g$  &   $F_1$    &   $F_2$   &   $F_1$    &   $F_2$  & $F_3$ & $F_4$ & $F_5$ \\ \hline
 $\kappa_b$  &   $F_1$    &   $F_1$   &   $F_2$    &   $F_2$   & $F_1$ & $F_1$ & $F_1$ \\ \hline \hline
 \multicolumn{8}{|c|}{CEPC} \\ \hline
 $ \xi \times 10^3$  &  2.56 & 2.36 & 4.19 & 3.87   & 2.78 -- 2.56 & 2.71 -- 2.36 & 2.36 -- 2.04 \\ \hline
 $f$\,[TeV] & 4.86 & 5.06 & 3.80 & 3.95  & 4.67 -- 4.86 & 4.72 -- 5.07 & 5.07 -- 5.45 \\ \hline\hline
  \multicolumn{8}{|c|}{ILC} \\ \hline
 $ \xi \times 10^3$  & 2.19 & 2.02 & 3.44 & 3.20   & 2.31 -- 2.19 & 2.06 -- 2.01 & 1.87 -- 1.72 \\ \hline
  $f$\,[TeV] &  5.26 & 5.48 & 4.19 & 4.35  & 5.12 -- 5.26 & 5.42 -- 5.48 & 5.69 -- 5.93 \\\hline\hline
 \multicolumn{8}{|c|}{FCC-ee} \\ \hline
 $ \xi \times 10^3$  & 1.80 & 1.66 & 3.06 & 2.74   & 1.85 -- 1.80 & 1.70 -- 1.66 & 1.66 -- 1.41 \\ \hline
 $f$\,[TeV] &  5.79 & 6.04 & 4.45 & 4.70  & 5.72 -- 5.80 & 5.97 -- 6.05 & 6.05 -- 6.56 \\\hline
\end{tabular}
\caption{95\% C.L. bound on $\xi$ and global symmetry breaking scale parameter $f$ for the MCHMs with various embedding of the fermion content, under CEPC, ILC, and FCC-ee Higgs precisions.     }
\label{tab:xi}
\end{table}

We tabulated in Table~\ref{tab:xi} various MCHM fermion representations and the corresponding leading modifications to $\kappa_t$ (so as to $\kappa_g$) and $\kappa_b$, while the modifications to $\kappa_Z$ always follow function $F_2$.
All the Higgs couplings only depend on one parameter $\xi$ for the first four cases when $\kappa_t$ and $\kappa_b$ is either $F_1$ or $F_2$.  The remaining three embeddings have extra parameter dependence that enters $F_{3,4,5}$.

Adopting the global fitting method as described in Sec.~\ref{sec:fit}, we obtained the 95\% C.L. range  on the $\xi$, which is listed in Table~\ref{tab:xi} for CEPC, ILC, and FCC-ee Higgs measurement precisions.  We further translated the corresponding limits on $\xi$ into the more intuitive limits on composite scale parameter $f$.
 We note that $\kappa_t$ can be well constrained by $h\to gg$ here as we assume the $h\to gg$ process does not receive additional new physics contribution.
We did not make any assumption on the model prediction of $\kappa_c$.  If further assumptions are made ({\it e.g.} $\kappa_c=\kappa_t$), a marginal improvement on the constraint of $\xi$ is expected, as the bounds on $\kappa_V$ and $\kappa_b$ are much more constraining.
At 95\% C.L., $\xi$ can be limited to be less than a few times $10^{-3}$, assuming future measurements follow SM expectations.   The composite scale is   constrained to be bigger than $4~\tev$.      For the embeddings with $\kappa_t$ related to function $F_{3,4,5}$, the 95\% C.L. bounds of $\xi$ and $f$ vary in a certain range,  given the extra parameter dependence.

\begin{figure}[htbp]
\centering
\includegraphics[width=7.5cm]{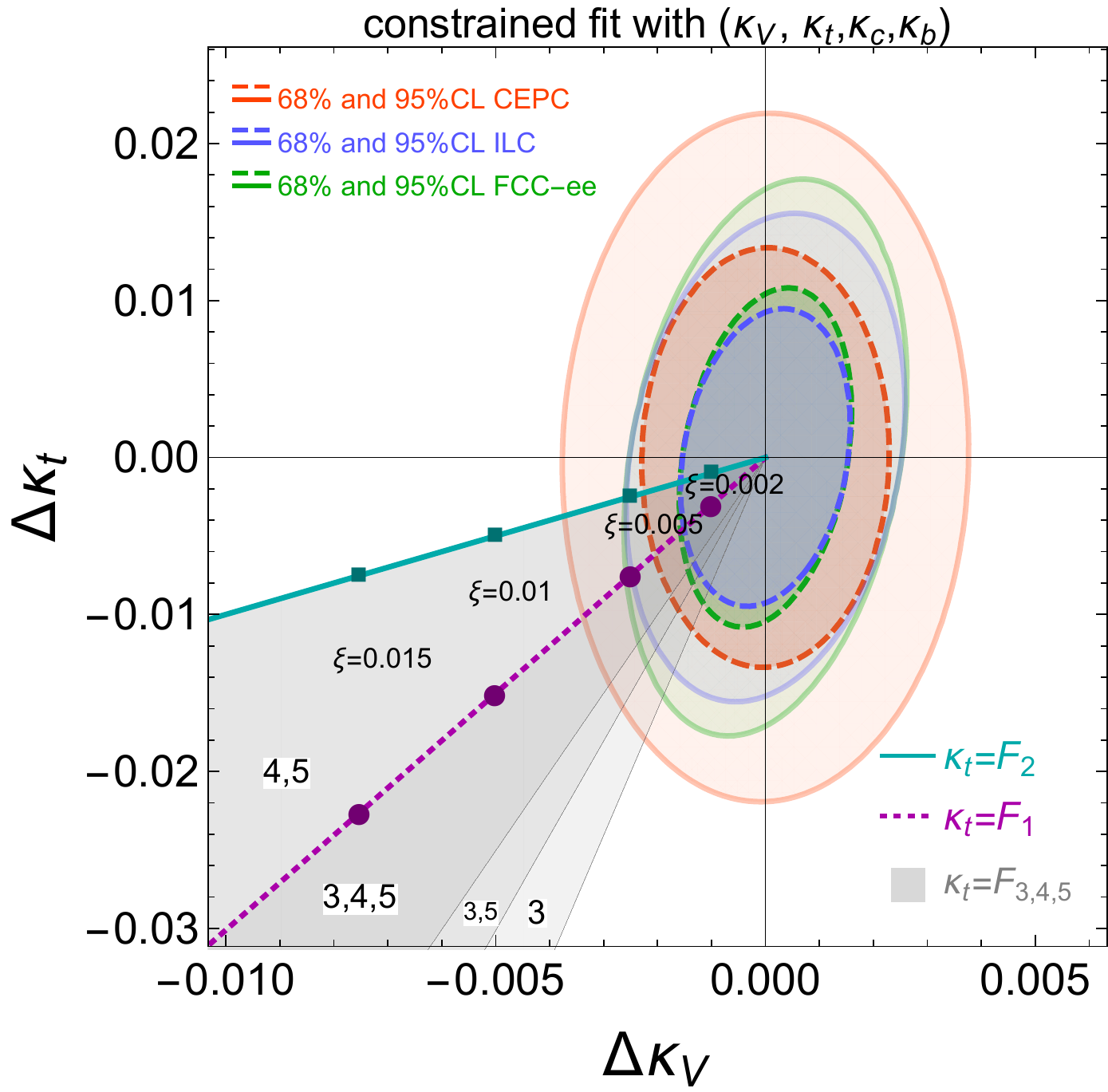}
\includegraphics[width=7.5cm]{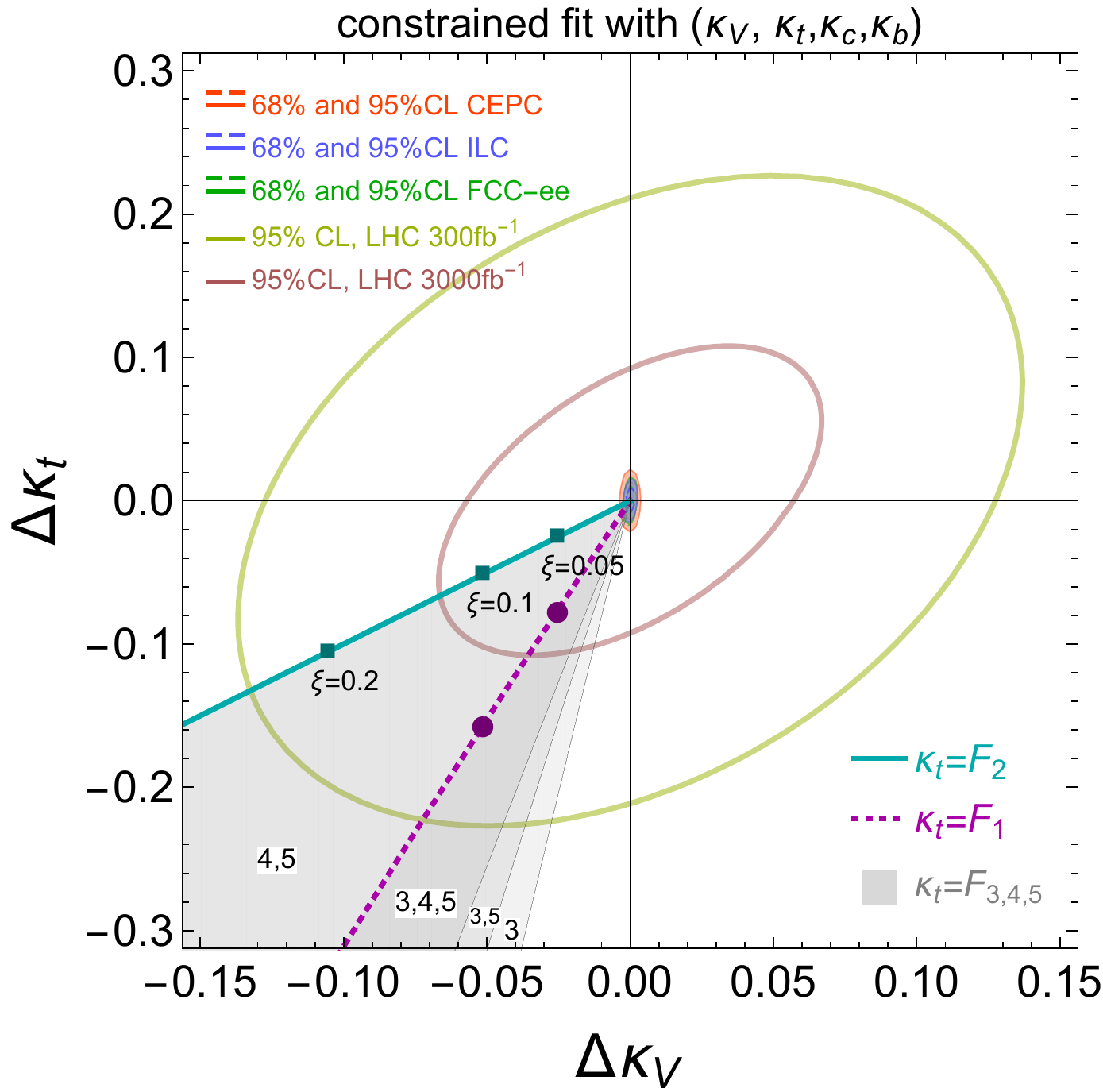}
\caption{The 68\% C.L. (solid lines) and 95\% C.L. (dashed lines) constraints in the $(\Delta \kappa_{V}, \, \Delta \kappa_t)$   plane at various future Higgs factories from a four-parameter fit with $\kappa_V(=\kappa_Z=\kappa_W)$,  $\kappa_t$, $\kappa_c$, and $\kappa_b$.    The left panel shows the constraints for CEPC, FCC-ee and ILC, while the right panel shows the LHC 95\% C.L. constraints as well.   Also shown are the predicted deviation of $\kappa_t$ and $\kappa_V$ for different fermion embeddings, as a function of $\xi$.  The magenta and cyan lines correspond to $\kappa_t = F_1$ and $F_2$, respectively, while $\kappa_t=F_{3,4,5}$ are covered by the gray region.  The labels on each part of the gray region indicate which ones of $\kappa_t=F_{3,4,5}$ cover this part.
 }
\label{fig:vvh1}
\end{figure}

Once deviations of the Higgs couplings are observed at future colliders, different embedding of the fermion tower might be distinguished through  the predicted correlations between Higgs couplings.   In Fig.~\ref{fig:vvh1}, we show in the $\Delta \kappa_V$-$\Delta \kappa_t$ plane the 68\% C.L. and 95\% C.L. exclusion limits of the Higgs precision measurements of CEPC, FCC-ee and ILC, obtained from a global fit in a reduced parameter space of the four parameters $\kappa_V$,  $\kappa_t$, $\kappa_c$, and $\kappa_b$.   The loop contributions of these parameters in the $hgg$, $h\gamma\gamma$, $hZ\gamma$ couplings are also included.  In the right panel, the 95\% C.L. from the LHC $300\infb$ and $3000\infb$ runs are also shown for comparisons with the lepton colliders.

The four-parameter fit is very useful in capturing the main characterization of the MCHMs without making specific assumptions on the fermion representations.  The projection to the $\Delta \kappa_V$-$\Delta \kappa_t$ plane allows us to focus on the two couplings most relevant for the MCHMs.  For specific fermion representations, the parameter space is further constrained, implying certain correlations between $\Delta \kappa_V$ and $\Delta \kappa_t$.  For $\kappa_t = F_1$ or $F_2$ defined in \autoref{eq:F1F2}, both $\Delta \kappa_Z$ and $\Delta \kappa_t$ are fixed by the value of $\xi$, shown by the magenta and cyan lines in Fig.~\ref{fig:vvh1}.  For $\kappa_t=F_{3,4,5}$ with the variation of additional model parameters, the predicted range is covered by the gray region.

 The right panel of Fig.~\ref{fig:vvh1} shows clearly that the Higgs measurements at future lepton colliders can significantly improve the constraints on the MCHMs from the corresponding measurements at the LHC for more than one order of magnitude.   Once deviations of Higgs couplings are measured, different embeddings of Fermion contents could also be tested.  This is, of course, due to the much better determination of the Higgs couplings at lepton colliders.  In particular, the coupling of Higgs to the $Z$ boson is very well constrained by the Higgsstrahlung processes, while at the LHC it is probed by the Higgs decay $h\to 4 \ell$ or the $VH$ production mode and suffers from systematic uncertainties.


\subsection{General Patterns of Strongly Interacting Light Higgs}

Given the richness and depth of strong dynamics with a light Higgs, it would be very informative to understand Higgs factory potentials on strong dynamics models under some generic arguments about the patterns of corresponding EFTs.  In our analyses, we considered three typical patterns of strong dynamics with a light Higgs~\cite{Liu:2016idz}: ALH, SILH and GSILH. These three cases are categorizations of general patterns of strong interactions rather than explicit models.

\begin{table}
\centering
\begin{tabular}{l|l} \hline\hline
$\mathcal{O}_H = \frac{1}{2} (\partial_\mu |H^2| )^2$ &  $\mathcal{O}_{GG} =  g_s^2 |H|^2 G^A_{\mu\nu} G^{A,\mu\nu}$  \\
$\mathcal{O}_{W} =  \frac{ig}{2} (H^\dagger \sigma^a \overleftrightarrow{D}^{\!\mu} H) D^\nu  W^a_{\mu\nu}$  & $\mathcal{O}_{Y_u} = Y_u |H|^2 \bar{Q}_L \tilde{H} u_R$  \\
$\mathcal{O}_{B} = \frac{ig'}{2} (H^\dagger \overleftrightarrow{D}^{\!\mu} H) \partial^\nu  B_{\mu\nu} $ &  $\mathcal{O}_{Y_d} = Y_d |H|^2 \bar{Q}_L H d_R$    \\
$\mathcal{O}_{HW} =  ig(D^\mu H)^\dagger \sigma^a (D^\nu H) W^a_{\mu\nu}$  &  $\mathcal{O}_{Y_e} = Y_e |H|^2 \bar{L}_L H e_R$  \\
$\mathcal{O}_{HB} =  ig'(D^\mu H)^\dagger  (D^\nu H) B_{\mu\nu}$ &      $\mathcal{O}_{3W} = \frac{1}{3!} g \epsilon_{abc} W^{a\,\nu}_\mu W^b_{\nu \rho} W^{c\,\rho\mu} $  \\
$\mathcal{O}_{BB} =  g'^2 |H|^2 B_{\mu\nu} B^{\mu\nu}$  &   \\ \hline\hline
\end{tabular}
\caption{ Operators in the SILH basis considered in our study.
 Assuming no corrections to  electroweak precision observables, only one combination of $O_W$ and $O_B$ is left after imposing the electroweak precision constraints of $c_W+c_B=0$.  While $Y_{u/d/e}$  are $3\times 3$ matrices in general,   we only consider the relevant diagonal ones $y_t$, $y_c$, $y_b$, $y_\tau$ and $y_\mu$.  In this table, $\epsilon_{abc}$ is the totally anti-symmetric tensor for $SU(2)$.
}
\label{tab:opsilh}
\end{table}

To capture the main feature of these cases without the loss of generality, we worked in the EFT framework with dimension-6 (D6) operators, parameterized by
\begin{equation}
\mathcal{L}_6 = \frac{1}{m_*^2} \sum_i c_i \mathcal{O}_i \,,
\end{equation}
where $m_*$ is the new physics scale and the coefficients $c_i$s are generally determined by the coupling of new physics, denoted as $g_*$, as well as the SM gauge couplings
 ($g_s$, $g$ and $g^\prime$, for SU(3), SU(2) and U(1), respectively)
and Yukawa couplings $y_f$.  A collection of the relevant operators $\mathcal{O}_i$ are listed in Table~\ref{tab:opsilh}.
 Amongst these operators, $O_H$ requires renormalization of the Higgs field and shifts the Higgs couplings universally to other SM particles, $O_y$ operators further modifies the Higgs Yukawa couplings. Operators $O_{BB}$, $O_{HB}$, $O_{HW}$, and $O_{GG}$ directly modify Higgs to gauge boson couplings. Operators $O_W$, $O_B$, $O_{HW}$, $O_{HB}$, and $O_{3W}$ also contribute to electro-weak precision observables and anomalous TGCs. Consequently, having additional measurements other than Higgs properties is crucial for developing global constraints on these operators.

\begin{table}
  \centering
  \begin{tabular}{|c||c|c|c|c|c|c|c|c|c|c|c|} \hline
    &   $\mathcal{O}_H$  &  $\mathcal{O}_{W}$  & $\mathcal{O}_{B}$ & $\mathcal{O}_{HW}$ & $\mathcal{O}_{HB}$ & $\mathcal{O}_{BB}$ & $\mathcal{O}_{GG}$ & $\mathcal{O}_{y_u}$ & $\mathcal{O}_{y_d}$  & $\mathcal{O}_{y_e}$  & $\mathcal{O}_{3W}$ \\ \hline
  ALH  &  $g^2_*$  &   1  &   1  &  1  &  1  &  1  &  1  &  $g^2_*$    &  $g^2_*$   &  $g^2_*$   &   $\frac{g^2}{g^2_*}$  \\  \hline
  GSILH  &  $g^2_*$  &   1  &   1  &  1  &  1  &   $\frac{y^2_t}{16\pi^2}$  &   $\frac{y^2_t}{16\pi^2}$  &  $g^2_*$    &  $g^2_*$   &  $g^2_*$   &   $\frac{g^2}{g^2_*}$  \\  \hline
  SILH  &  $g^2_*$  &   1  &   1  &  $\frac{g^2_*}{16\pi^2}$  &  $\frac{g^2_*}{16\pi^2}$  &  $\frac{y^2_t}{16\pi^2}$  &   $\frac{y^2_t}{16\pi^2}$  &  $g^2_*$    &  $g^2_*$   &  $g^2_*$   &   $\frac{g^2}{16\pi^2}$  \\  \hline
  \end{tabular}
  \caption{Estimation of  the magnitudes of $c_i$'s for the operators in \autoref{tab:opsilh} for the three scenarios of ALH, GSILH, and SILH.  Note the difference between this and Table~2 of Ref.~\cite{Liu:2016idz} is due to the different normalization of operators. }
  \label{tab:ci}
  \end{table}

 For three strong dynamics cases that we considered here,   the estimated parametric counting and scaling of the operators are listed in Table~2 of Ref.~\cite{Liu:2016idz}, which are reproduced here in Table~\ref{tab:ci}. For each of these cases, there are only two free parameters: $g_*$ and $m_*$.  However, it should be noted that Table~\ref{tab:ci} provides an estimation on the size of the $c_i$'s rather than their predicted values, which are available only if the model is specified.  We therefore do not assume any relations among the operators in Table~\ref{tab:ci}, but only use the estimated coupling size to derive the scale of new physics $m_*$.

\begin{figure}[h]
\centering
\includegraphics[width=14cm]{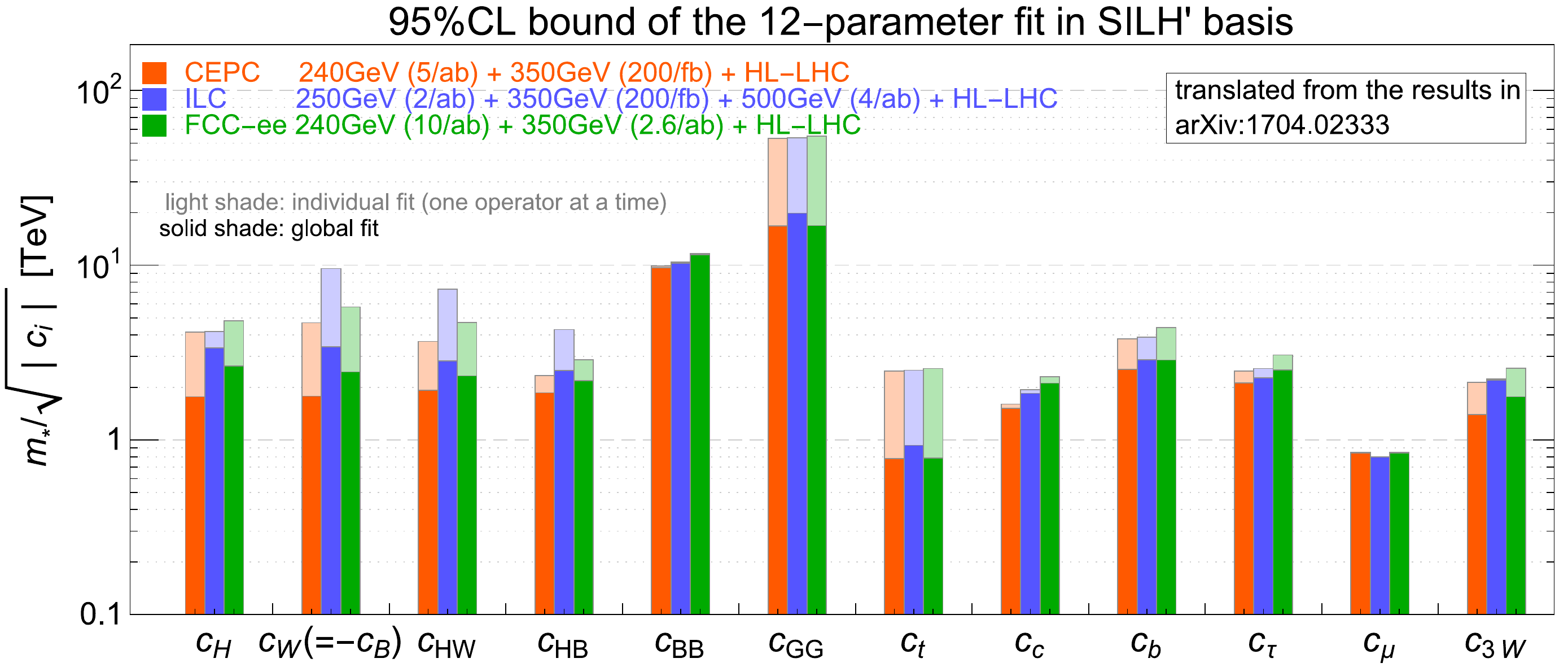}
\caption{The 95\% C.L. constraints on the overall coefficient of  $\mathcal{O}_i$ from Ref.~\cite{Durieux:2017rsg}, which are translated into the SILH basis and presented in the form $m_*/\sqrt{c_i}$.  Estimated Higgs measurement precision from CEPC, ILC, and FCC-ee are used with the inclusion of HL-LHC Higgs precision.}
\label{fig:silh1}
\end{figure}

We derived the reach of the scale of new physics $m_*$ as a function of $g_*$ for three individual scenarios  from the 95\% C.L. constraints on the overall coefficient of EFT operators derived in Ref.~\cite{Durieux:2017rsg}, which was translated to $m_*/\sqrt{c_i}$ as shown in Fig.~\ref{fig:silh1} for  CEPC (red) , ILC (blue), and FCC-ee (green) Higgs precisions.    The Higgs measurements from HL-LHC are included in the total $\chi^2$ to further optimize the reach. The HL-LHC improves the limits through better diphoton statistics
 and $ttH$ processes that compliment  Higgs factory measurements.
 Note that this is different from the global analyses on the 2HDM and MCHM in precious sections, where the inclusion of HL-LHC Higgs measurements does not affect the results much.
 In addition to the Higgs measurement inputs as listed in Section.~\ref{sec:input}, the angular observables in $\eehz$ and the constraints on anomalous TGCs  from measurements of $\eeww$ are also included in Ref.~\cite{Durieux:2017rsg} to help discriminate different EFT parameters and maximize the overall precision reach.   Ref.~\cite{Durieux:2017rsg} also assumed that CEPC is able to collect $200\infb$ data at 350\,GeV.
   Both CEPC and FCC-ee 350\,GeV measurements are obtained by scaling from ILC 350\,GeV, assuming statistical uncertainties dominate.  The difference between unpolarized beams and polarized beams has also been taken into count.

Note that Ref.~\cite{Durieux:2017rsg} only focuses on operators in Table~\ref{tab:opsilh} that can be probed by Higgs processes and diboson production ($\eeww$).   Furthermore, relation $c_W+c_B=0$ is imposed such that there is no additional contributions to the  electroweak precision observables, resulting in only one combination of  $O_W$ and $O_B$ survives.   Ignoring the flavor changing effects and only taken into account the relevant diagonal Yukawa couplings $y_t$, $y_c$, $y_b$, $y_\tau$ and $y_\mu$,  we ended up with 12 coefficients to be constrained by the global fit.

\begin{figure}[h] \vspace{0.8cm}
  \centering
  \includegraphics[width=7cm]{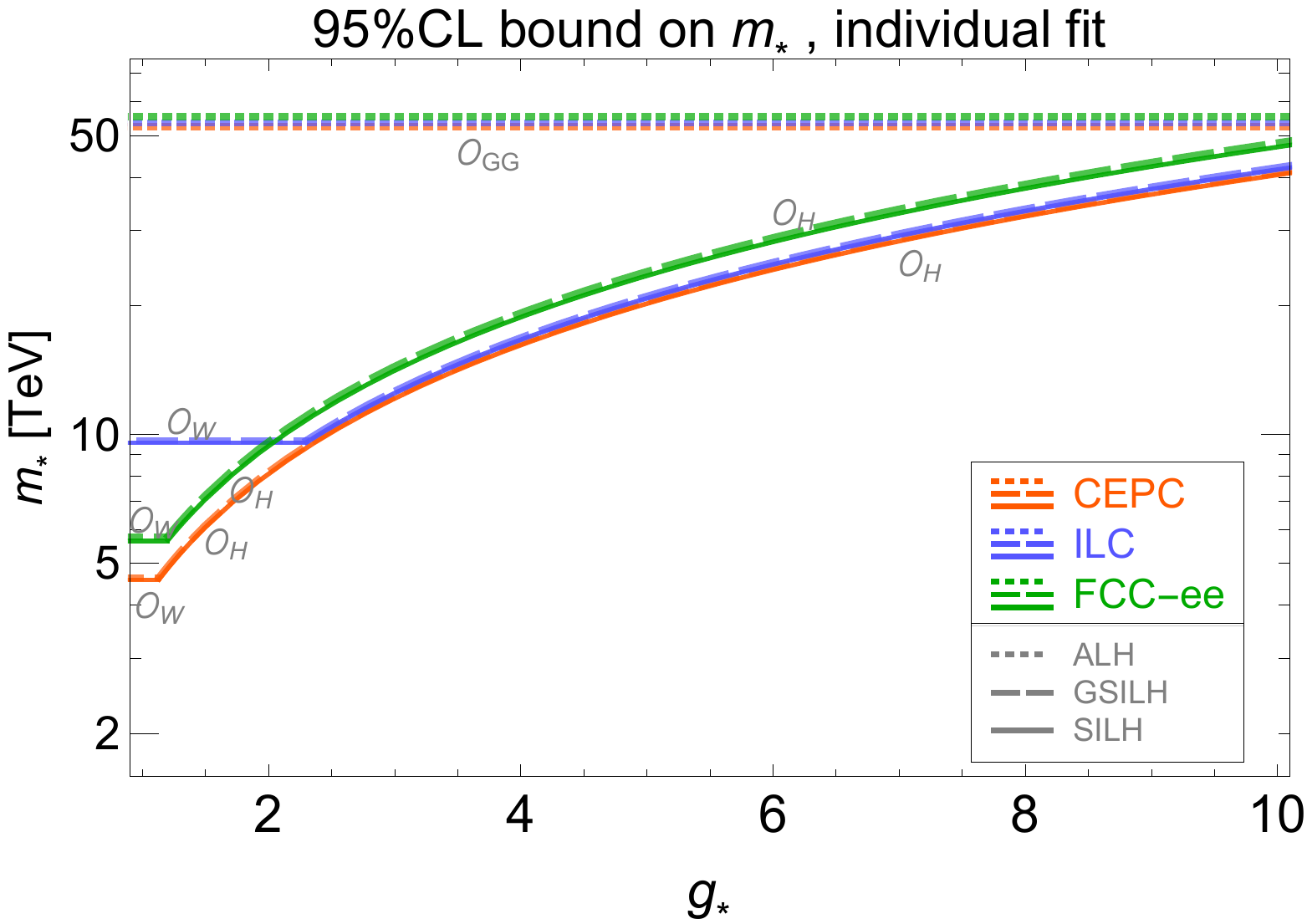}  \hspace{0.5cm}
  \includegraphics[width=7cm]{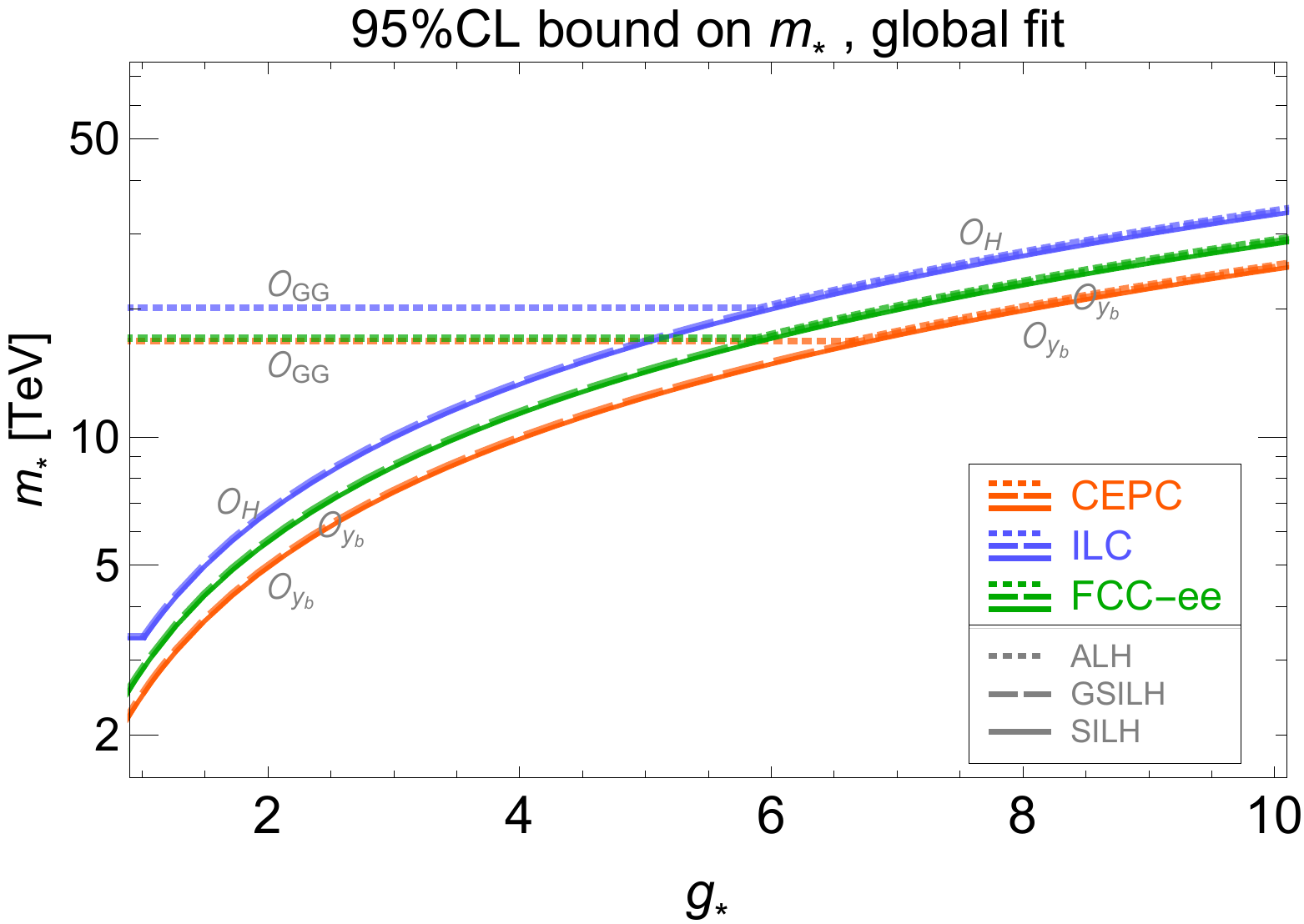}  \hspace{0.5cm}
   \caption{The 95\% C.L. limit on the new physics scale for the three cases of ALH (dotted lines), GSILH (dashed lines) and SILH (solid lines) as a function of  $g_*$ with CEPC (red), ILC (blue) and FCC-ee (green)  precision. The left plot is the results obtained using fit on individual operators, and the right plot is the results obtained using global fit of all 12 operators.     The operator that is most sensitive (therefore determines the best reach) is labelled alongside the curves. }
  \label{fig:silh2}
  \end{figure}

Fig.~\ref{fig:silh1} shows the results from  individual constraints on operators obtained by switching on one of them at a time (light shade) and the other from a 12-parameter global fit (solid shade).  The single operator fit results are typically factors of a few better for operators involving electroweak gauge bosons, due to their correlation in Higgs physics. We also note here the typical scale from a global fit on these operators are around 1$-$3  TeV with $c_i\sim 1$, with the exception of $m_*/c_{BB}$ and $m_*/c_{GG}$, which could reach around 10 TeV.  This is because the leading order contribution from the SM to $h\rightarrow \gamma\gamma, gg$ appears at one-loop level.

Using the bounds on $m_*/\sqrt{c_i}$ in Fig.~\ref{fig:silh1},  the reach of the new physics scale $m_*$ can be derived as a function of the strong coupling $g_*$.  The reaches are presented in \autoref{fig:silh2} at 95\% C.L. for three cases under future  Higgs factories.
 We varied $g_*$ from 1 to 10, to cover the typical range of strong interaction couplings $1<g_*<4\pi$.    Two sets of bounds are shown for each of the three cases of ALH, GSILH and SILH, one from individual fit (left panel) and the other from a  global fit (right panel).
 The   constraints obtained from individual fit provide a optimistic estimation as in realistic models it is unlikely that only one D6 operator is generated.   The global fit, on the other hand, provides a conservative bound, with all coefficients of the relevant D6 operators treated as independent free parameters.
 For each cases, the operator that provides the strongest constraints is labelled alongside the curves.

For $g_*$ varies between 1 to 10, the reach in $m_*$ is about 5 to 50 TeV for individual fit, and about 2 to 30 TeV for global fit.   For individual fit, the CEPC and ILC reach is similar, except for small $g_*$ region, while FCC-ee reach is slightly better.  For global fit, the ILC reach improves over CEPC and FCC-ee, due to the additional measurements of Higgs and diboson processes at higher center of mass energy.   While the reaches in GSILH and SILH are almost identical, the reach in ALH is quite different, in particular, on the $\mathcal{O}_{GG}$ operator where the loop suppression from the shift-symmetry argument is absent.

The dominating Higgs production process,   the Higgsstrahlung process $\eehz$,  is ideal for probing $\mathcal{O}_H$, which provides a universal shift to all the Higgs couplings.  However,   in a global fit its bound suffers from a large degeneracy with the other parameters, in particular with operators that could generate anomalous $hZZ$ couplings with Lorentz structures different from the SM one.  Therefore,  in individual fits $\mathcal{O}_H$ provides the best reach for GSILH and SILH,  while for global fit, the best reach comes from $\mathcal{O}_{y_b}$ instead at CEPC and FCC-ee.  ILC with the run at 500\,GeV is more powerful at discriminating these different operators through the $WW$-fusion production of the Higgs boson and $ZH$ associated production~\cite{Han:2015ofa},
 therefore with a global fit $\mathcal{O}_H$ still has the best reach.  For smaller values of $g_*$,   $\mathcal{O}_W$ provides the best reach since its coefficient is independent of $g_*$.

For ALH, $\mathcal{O}_{GG}$ provides the best sensitivity to $m_*$, around 50 TeV  in the individual operator fit case, and around 20 TeV in the global fit case for $g_* \lesssim 6$.    This is because   $\mathcal{O}_{GG}$ can   be generated at  tree level in ALH, while at one loop level in the SM,  GSILH and SILH.     In the global fit,  the bound on $\mathcal{O}_{GG}$ usually suffers from a flat directions associated with $\mathcal{O}_{y_t}$ which also contributes to the $hgg$ coupling by modifying the top Yukawa coupling in the loop.    Therefore, in ALH, the most sensitive operator in global fitting at large $g_*$ is $\mathcal{O}_{y_b}$ at CEPC/FCC-ee  and $\mathcal{O}_{H}$ operator at ILC, similar to the case of GSILH and SILH.

With these results on the typical patterns of strongly interacting light Higgs, we can see how different components of the Higgs precision programs compliment each other in probing models of strong dynamics.
 In particular, running at different center of mass energies of Higgs factories can be sensitive to a particular set of operators.


\section{Conclusion and outlook}
\label{sec:conclusion}

 In this work, we studied the implication of the Higgs precision measurements from various Higgs factories (CEPC, ILC, FCC-ee) on new physics models.   In particular, we focused on two types of new physics scenarios: perturbative models, and strong dynamics.  For perturbative models, we studied the SM plus a singlet extension as well as the 2HDM as explicit examples, and explored both the tree level effects as well as loop corrections.      For strong dynamics, we studied the MCHM as a concrete example, and also explored the EFT language in three typical patterns of operators that arose from strong interactions.  We took the estimated precision from Higgs factories, as well as LHC results, and performed a global fit in the relevant parameter space to derive the reach in different models.  In the case when no deviation from the SM predictions  is observed, a 95\% C.L. reach in the parameter range was derived.  In the case when deviations from SM predictions are observed, we further explored how different models can be distinguished.

For the SM with a real scalar singlet extension, we obtained the 95\% C.L. limit of the singlet-doublet mixing angle $\sin\theta$ to be   0.062, 0.058 and 0.052 for the CEPC, ILC and FCC-ee, respectively, at 95\% C.L.   We also studied the more general case with the induced $\mathcal{O}_H$ and $\mathcal{O}_6$ operators, and obtained constraints on the Wilson coefficients $c_H/\Lambda^2$ and $c_6/\Lambda^2$.   While  $c_H/\Lambda^2$ is tightly constrained to be around $10^{-2}/v^2$, the limit on $c_6/\Lambda^2$ is about an order of magnitude weaker.

For the 2HDM, we analyzed four different types of 2HDM: Type-I, II, L and F.   For tree level effects, we found that $|\cos(\beta-\alpha)|$ is tightly constrained to be about $10^{-4}$ or smaller at small and/or large $\tan\beta$ for Type-II, L, F, and reached the maximum range of about 0.005 $-$ 0.007 for $\tan\beta \sim 1$.   For Type-I, large $\tan\beta$ region is much less constrained:  $|\cos(\beta-\alpha)| \lesssim 0.08$.  For 2HDM loop effects under the alignment limit,  the lower bounds on the heavy Higgs masses depend on the $\lambda v^2 = m_\phi^2 - m_{12}^2/(s_\beta c_\beta)$: about 350, 450, and 1100 GeV for $\sqrt{\lambda v^2}=100$, 300, 500 GeV, respectively.   For smaller values of $\sqrt{\lambda v^2}\lesssim 100$ GeV, a small mass region of $125\ {\rm GeV} < m_\phi < 350$ GeV is viable as well.   We also explored the constrained parameter space in $\tan\beta$ vs. $m_\phi$ plane when $m_{12}^2$ is varied as an independent parameter.

For the MCHM, the main parameter is $\xi=v^2/f^2$, which is linked to global symmetry breaking scale $f$.  With different embeddings of the fermion content, we obtained the 95\% C.L. limits on $\xi$ to be about a few $\times 10^{-3}$, corresponding to  $f \gtrsim 4$ TeV.    We also studied the ALH, GSILH and SILH as three representative patterns of EFT operators in strong dynamics.  The 95\% C.L. constraints strong dynamics scale $m_*$  as a function of strong coupling $g_*$ were obtained, either with an individual fit to a single operator, or through a global fit of all twelve operators.   For $g_*$ varying between 1 to 10, $m_*$ varies between 5 to 50 TeV for individual fit, and 2 to 30 TeV for global fit.

We also studied the reach for three different Higgs factory machines: the CEPC, ILC and FCC-ee under typical running scenarios as summarized in Table~\ref{tab:mu_precision}.   The reaches of FCC-ee and ILC are slightly better than that of CEPC.  With higher center of mass energy running of ILC, it has advantage of being sensitive to operators that contribute to $hWW$ couplings.

In our analyses, we used the global fit to the signal strength $\mu_i$ for each individual search channel.  This has been shown to provide stronger constraints on parameter spaces  over the usual $\kappa$-scheme when correlations between $\kappa_i$s are not   treated properly.

To summarize, the impressive precisions in the Higgs measurements that can be achieved in future Higgs factories provide us a powerful tool to explore the new physics models through indirect probe.  Those indirect searches to new physics are complementary to direct searches at current and future colliders, and sometime even supersede direct limits given the hadronic environment at the LHC and the limited energy reach of lepton colliders.  Our studies provided concrete examples of how sensitive Higgs precision measurements can be to particular model parameters, new particle masses, or symmetry breaking scales.  While our results are specific towards the SM singlet extension, 2HDM, MCHM, and three particular types of strong dynamics set up, the approaches we took are more general and can be applied to any model set up, as long as Higgs couplings and other relevant interactions are specified.   Our study also provides guidelines to future Higgs factory design, including choices of center of mass energies and the corresponding luminosities.

\begin{acknowledgments}


We would like to thank Christophe Grojean, Tao Han, Da Liu, Manqi Ruan, Jinmin Yang and Lian-Tao Wang for useful discussions,  and Gauthier Durieux for valuable comments on the manuscript.
JG were supported  by the an International Postdoctoral Exchange Fellowship Program between the Office of the National Administrative Committee of Postdoctoral Researchers of China (ONACPR) and DESY.
HL were supported  by the National Natural Science Foundation of China (NNSFC) under grant No. 11635009 and  Natural Science Foundation of Shandong Province under grant No. ZR2017JL006.
SS were supported  by the Department of Energy
under Grant DE-FG02-13ER41976/DE-SC0009913.
WS were supported  by the National Natural Science Foundation of China (NNSFC) under grant No. 11305049 and No. 11675242, by the CAS Center for Excellence in Particle Physics (CCEPP), by the CAS Key Research Program of Frontier Sciences and by the China Scholarship Council.
This manuscript has been authored by Fermi Research Alliance, LLC under Contract No. DE-AC02-07CH11359 with the U.S. Department of Energy, Office of Science, Office of High Energy Physics. The United States Government retains and the publisher, by accepting the article for publication, acknowledges that the United States Government retains a non-exclusive, paid-up, irrevocable, world-wide license to publish or reproduce the published form of this manuscript, or allow others to do so, for United States Government purposes.

\end{acknowledgments}

\appendix


\section{LHC Higgs measurements}
\label{app:lhcinput}

\begin{table}[tb!]
\begin{center}
\renewcommand*{\arraystretch}{1.1}
\begin{tabular}{|c|c|c|c|c|c|}
\hline
Channel   & Production   &   Run-I & Channel   & Production   &   Run-I\\
\hline
$\gamma \gamma$  & $ggh$ & $1.10^{+0.23}_{-0.22} $ &$\tau^+ \tau^-$   & $ggh$ & $1.0^{+0.6}_{-0.6} $
\\ \cline{2-3} \cline{5-6}
         & VBF  &  $1.3^{+0.5}_{-0.5}$ & & VBF  &  $1.3^{+0.4}_{-0.4}$
\\ \cline{2-3}\cline{5-6}
         & $Wh$  &  $0.5^{+1.3}_{-1.2}$ & & $Wh$  &  $-1.4^{+1.4}_{-1.4}$
\\ \cline{2-3}\cline{5-6}
 	& $Zh$  &   $0.5^{+3.0}_{-2.5}$ & & $Zh$  &   $2.2^{+2.2}_{-1.8}$
\\ \cline{2-3}\cline{5-6}
& $t \bar t h$ & $2.2^{+1.6}_{-1.3}$ & & $t \bar t h$ & $-1.9^{+3.7}_{-3.3}$
\\ \hline
$WW^*$   & $ggh$ & $0.84^{+0.17}_{-0.17}$& $b \bar b$ & $Wh$ & $1.0^{+0.5}_{-0.5}$
\\ \cline{2-3}
         & VBF  &  $1.2^{+0.4}_{-0.4}$&   &	$Zh$ & 	$0.4^{+0.4}_{-0.4}$
\\ \cline{2-6}
         & $Wh$  &  $1.6^{+1.2}_{-1.0}$&   &	 $t \bar t h$ & 	$1.15^{+0.99}_{-0.94}$
\\ \cline{2-3}
 	& $Zh$  &   $5.9^{+2.6}_{-2.2}$&   $Z Z^*$   & $ggh$ & $1.13^{+0.34}_{-0.31} $
\\ \cline{2-3}
& $t \bar t h$ & $5.0^{+1.8}_{-1.7}$&  & VBF   & $0.1^{+1.1}_{-0.6}$
\\ \hline
\end{tabular}
\quad
\end{center}
\caption{
\label{tab:HIGGS_datarun12}
The Higgs signal strength in various channels measured at the LHC Run-I~\cite{Khachatryan:2016vau}. Correlations between different Run-I measurements quoted in Fig.~27 of Ref.~\cite{Khachatryan:2016vau} are taken into account.
}
\end{table}

In Table~\ref{tab:HIGGS_datarun12}, we listed the LHC Run-I Higgs measurements for $\gamma\gamma$, $WW$, $\tau\tau$, $bb$ and $ZZ$ for various production channels.    In Table~\ref{tab:HIGGS_datarun3000},  we listed the estimated precision on normalized Higgs signal strength for 300 and 300 ${\rm fb}^{-1}$ integrated luminosity, as well as the corresponding production contamination.   These two tables are used as input for our global fit using LHC data.

\begin{table}[tb!]
\begin{center}
\renewcommand*{\arraystretch}{1.1}
\begin{tabular}{|c|c|c|c|c|c|c|c|c|}
\hline
\multirow{2}{*}{Decay}   & \multirow{2}{*}{Production}   &  \multicolumn{2}{c|}{$\Delta \mu/\mu  $ }&\multicolumn{5}{c|}{True Origin} \\
\cline{3-9}
     & &$300 fb^{-1}$ &$3000 fb^{-1}$ &ggF &VBF &WH &ZH &ttH \\ \hline
\multirow{7}{*}{$\gamma \gamma$ } & comb& 0.09 & 0.04   & $\frac{49.85}{56.92}$  & $\frac{4.18}{56.92}  $ &$\frac{1.5}{56.92}$   & $\frac{0.88}{56.92}$  &$\frac{0.61}{56.92}$
\\ \cline{2-9}
         & 0j  &  0.12  & 0.05 &  $\frac{97}{100}$ & $\frac{3}{100}$  & 0. & 0. & 0.
\\ \cline{2-9}
         & 1j  &  0.14  & 0.23 &  $\frac{86}{100}$  & $\frac{14}{100}$  & 0. & 0. & 0.
\\ \cline{2-9}
         & VBF-like  &  0.43  & 0.15 &  0.3(0.4)  & $0.7(0.6)$  &  0. & 0. & 0.
\\ \cline{2-9}
         & WH-like  &  0.48  & 0.17 &  0. & 0. & $\frac{72}{77}$  & 0. & $\frac{5}{77}$
\\ \cline{2-9}
         & ZH-like  &  0.85  & 0.28 &  0. & 0. & $\frac{10}{102}$  & $\frac{88}{102}$  & $\frac{4}{102}$
\\ \cline{2-9}
         & ttH-like  &  0.36  & 0.12 & 0.  & 0. & $\frac{7}{206}$  & $\frac{12}{206}$   &  $\frac{187}{206}$
\\ \hline
\multirow{5}{*}{ZZ } & comb& 0.07 & 0.04 & $\frac{49.85}{56.92}$  & $\frac{4.18}{56.92}  $ &$\frac{1.5}{56.92}$   & $\frac{0.88}{56.92}$  &$\frac{0.61}{56.92}$
\\ \cline{2-9}
         & VH-like  &  0.34  & 0.12 &  $\frac{22 }{72.5 }$   & $\frac{6.6 }{ 70.5}$  & $\frac{25 }{70.5 }$  & $\frac{8.8 }{ 70.5}$  & $\frac{ 10.1}{70.5 }$
\\ \cline{2-9}
         & ttH-like  &  0.48  & 0.20 & $\frac{ 3.1}{35.4 }$ & $\frac{ 0.6}{35.4 }$ & $\frac{ 0.6}{35.4 }$ & $\frac{ 1.1}{35.4 }$ &   $\frac{ 30}{35.4 }$
\\ \cline{2-9}
         & VBF-like  &  0.33  & 0.16 & ~$\frac{ 41}{97.1}$ ~ & ~$\frac{ 54}{97.1}$~ & ~ $\frac{ 0.7}{97.1}$~& ~$\frac{ 0.4}{97.1}$~ &~$\frac{ 1}{97.1}$~
\\ \cline{2-9}
         & ggF-like  &  0.07  & 0.04 & $\frac{ 3380}{3809}$  &$\frac{ 274}{3809}$  & $\frac{ 77}{3809}$ &$\frac{ 53}{3809}$  &$\frac{ 25}{3809}$
\\ \hline
\multirow{4}{*}{WW } & comb & 0.08 & 0.05   & $\frac{49.85}{56.92}$  & $\frac{4.18}{56.92}  $ &$\frac{1.5}{56.92}$   & $\frac{0.88}{56.92}$  &$\frac{0.61}{56.92}$
\\ \cline{2-9}
         & 0j  &  0.09  & 0.05  & $\frac{ 4085}{4184}$  & $\frac{ 99}{4184}$  & 0. & 0. & 0.
\\ \cline{2-9}
         & 1j  &  0.18  & 0.10 &  $\frac{ 20050}{22375}$  & $\frac{ 2325}{22375}$ &  0. & 0. & 0.
\\ \cline{2-9}
         & VBF-like  &  0.20  & 0.09 &  $\frac{ 9}{59}$  & $\frac{ 50}{59}$ & 0. & 0. & 0.
\\ \hline
Z$\gamma$ & incl. & 0.44 &0.27 &  $\frac{12.79}{28.16}$ & $\frac{15.34}{28.13}$  & 0. & 0. & 0.
\\ \hline
\multirow{3}{*}{bb } & comb & 0.26 & 0.12   & $\frac{49.85}{56.92}$  & $\frac{4.18}{56.92}  $ &$\frac{1.5}{56.92}$   & $\frac{0.88}{56.92}$  &$\frac{0.61}{56.92}$
\\ \cline{2-9}
         & WH-like  &  0.56  & 0.36 &  1 &  0.& 0. & 0. & 0.
\\ \cline{2-9}
         & ZH-like  &  0.29  & 0.13 &  0. & 0. & 0. &$\frac{560}{636}$   & $\frac{76}{636} $
\\ \hline
$\tau \tau$ & VBF-like & 0.18 &0.15 &  $\frac{1641}{2538}$ & $\frac{897}{2538}$  & 0. & 0. & 0.
\\ \hline
\multirow{3}{*}{ $\mu \mu$} & comb & 0.38 & 0.12  & $\frac{49.85}{56.92}$  & $\frac{4.18}{56.92}  $ &$\frac{1.5}{56.92}$   & $\frac{0.88}{56.92}$  &$\frac{0.61}{56.92}$
\\ \cline{2-9}
         & incl. &  0.45  & 0.14   & $\frac{1510}{1725}$ & $\frac{125}{1725}$ & $\frac{45}{1725}$ &$\frac{27}{1725}$&$\frac{18}{1725}$
\\ \cline{2-9}
         & ttH-like  &  0.72  & 0.23 &  0.& 0. & 0. & 0. & 1

\\ \hline
\end{tabular}
\quad
\end{center}
\caption{
\label{tab:HIGGS_datarun3000}
The normalized Higgs signal strength uncertainty~\cite{ATL-PHYS-PUB-2014-016} and their corresponding production contamination information in various channels~\cite{ATL-PHYS-PUB-2013-014,ATL-PHYS-PUB-2014-012,ATL-PHYS-PUB-2014-006,ATL-PHYS-PUB-2014-011,ATL-PHYS-PUB-2014-018,Hartmann:2015aia,ATLAS-collaboration:1484890}
measured at the LHC 300 fb$^{-1}$ and 3000 fb$^{-1}$.  Here the contamination information are   for 3000 fb$^{-1}$, and almost  the same for 300 fb$^{-1}$ except for the VBF-like $H \to \gamma \gamma$ process, which are shown in brackets.
}
\end{table}

\section{2HDM loop corrections}
\label{app:2HDMloop}

The formulas of the one-loop correction from the 2HDM to the Higgs couplings are summarized here, under the alignment limit $\cos (\beta-\alpha)=0$ and with the universal mass assumption $m_{H^{\pm}} = m_{A^0}=m_{H^0}\equiv m_{\phi} $.
 \begin{eqnarray}
 \Delta \kappa^{\rm 2HDM}_{\rm h\gamma \gamma, 1-loop} &=&
 -\frac{2 v \lambda_{h H^+ H^-}}{m_h^2}[1+2 m_{H^{\pm}}^2 C_0(0,0,m_h^2,m_{H^{\pm}}^2,m_{H^{\pm}}^2,m_{H^{\pm}}^2)]\frac{1}{\mathcal{A}^\gamma_{\rm SM}}
 \nonumber\\& \approx & \Big[-\frac{1}{3}(1-\frac{m_{12}^2/(s_{\beta}c_{\beta})}{m_{H^{\pm}}^2})-\frac{1}{90}(23-8\frac{m_{12}^2/(s_{\beta}c_{\beta})}{m_{H^{\pm}}^2})\frac{m_h^2}{m_{H^{\pm}}^2}
\nonumber\\&&
-\frac{1}{1440}(73-9\frac{m_{12}^2/(s_{\beta}c_{\beta})}{m_{H^{\pm}}^2})\frac{m_h^4}{m_{H^{\pm}}^4}\Big]\frac{1}{\mathcal{A}^\gamma_{\rm SM}},
 \end{eqnarray}
with $\mathcal{A}^\gamma_{\rm SM} = 6.53$, representing the one-loop induced $h\gamma\gamma$ coupling in the SM, which includes the W boson, top quark and bottom quark contributions.

 \begin{eqnarray}
&&\Delta \kappa^{\rm 2HDM}_{hff,\rm 1-loop} \simeq - \frac{1}{16\pi^2}\Bigg\{
\frac{2m_{f'}^2}{v^2}\kappa_A^{d}\cot\beta\Big[(m_h^2-2m_f^2)C_{12}(m_f^2,m_f^2,m_h^2;m_{f'},m_\Phi,m_{f'}) \nonumber\\
&&+(2m_{f'}^2-m_f^2)C_0(m_f^2,m_f^2,m_h^2;m_{f'},m_\Phi,m_{f'})+v\lambda_{h \Phi\Phi }C_0(m_f^2,m_f^2,m_h^2;m_{\Phi},m_{f'},m_{\Phi}) \Big]
\nonumber\\
&&+\lambda_{h \Phi\Phi}^2\frac{d}{dp^2}B_0(p^2;m_{\Phi},m_{\Phi})\Big|_{p^2=m_h^2}
-\frac{6m_t^2}{v^2} I_f\kappa_A^f\cot\beta B_0(m_\Phi^2;m_t,m_t)\nonumber\\
&&+\frac{6m_t^4}{v^2(m_\Phi^2-m_h^2)}
I_f\kappa_A^f\cot\beta\left[\left(4-\frac{m_h^2}{m_t^2}\right)B_0(m_h^2;m_t,m_t)-\left(4-\frac{m_\Phi^2}{m_t^2}\right)B_0(m_\Phi^2;m_t,m_t) \right]\nonumber\\
&&+\frac{6\lambda_{h \Phi\Phi}\lambda_{H \Phi\Phi}}{m_\Phi^2-m_h^2}I_f\kappa_A^f
\left[B_0(m_h^2;m_\Phi,m_\Phi)-B_0(m_\Phi^2;m_\Phi,m_\Phi) \right]
\Bigg\},
\end{eqnarray}
where $f'$ is the fermion whose electromagnetic charge is different by one unit from $f$, and
$I_f=+1/2~(-1/2)$ for $f=u$ ($d,e$).
\begin{eqnarray}
 &&\Delta \kappa^{\rm 2HDM}_{hVV,\rm 1-loop} =
- \frac{1}{2 \times 16\pi^2} \frac{d}{d p^2}\left. \Big \{
(\lambda_{hH^+H^-}^{})^2  B_0(p^2; m_{H^\pm}^{},m_{H^\pm}^{})
+(\lambda_{hAA}^{})^2  B_0(p^2; m_{A}^{},m_{A}^{})
\right.\nonumber\\&&\left.
+(\lambda_{hHH}^{})^2  B_0(p^2; m_{H}^{},m_{H}^{})\Big\}
-\frac{1}{16\pi^2} \frac{1}{2v}\Big\{2 \cos^22\theta_W \lambda_{hH^+H^-}^{} B_0 (q; m_{H^\pm}^{},m_{H^\pm}^{})
\right. \nonumber\\&& \left.
+2 \lambda_{hAA}^{} B_0 (q; m_{A}^{},m_{A}^{}) + 2 \lambda_{hHH}^{} B_0 (q; m_{H}^{},m_{H}^{})
\right. \nonumber\\&& \left.
-8 \cos^22\theta_W \lambda_{hH^+H^-}^{} C_{24} (p_1^2,p_2^2,q^2; m_{H^\pm}^{},m_{H^\pm}^{},m_{H^\pm}^{})
\right. \nonumber\\&& \left.
+8\lambda_{hAA}^{}C_{24} (p_1^2,p_2^2,q^2; m_{A}^{},m_{H}^{},m_{A}^{})
+8\lambda_{hHH}^{}C_{24} (p_1^2,p_2^2,q^2; m_{H}^{},m_{A}^{},m_{H}^{})\Big\}
\right. .
\end{eqnarray}
The scalar three-point couplings are given by
\begin{align}
\lambda_{h \Phi\Phi}=\frac{(m_h^2+2m_\Phi^2-2\frac{m_{12}^2}{s_{\beta}c_{\beta}})}{2v},\quad
\lambda_{H \Phi\Phi}=\frac{\frac{m_{12}^2}{s_{\beta}c_{\beta}}-m_\Phi^2}{v}\cot2\beta.
\end{align}

All the $B$ and $C$ function are Passrina-Veltaman Functions \cite{Passarino:1978jh}.

We should emphasize that all the above equations are in the alignment limit $\cos (\beta-\alpha)=0$ and $m_{\phi}\equiv m_{H^{\pm}} = m_{A^0}=m_{H^0}$. Following the fitting techniques  at  tree-level,  $\kappa_W = \kappa_Z $ and $\kappa_{\mu} = \kappa_{\tau}$ still hold at 1-loop.


\bibliographystyle{JHEP}
\bibliography{references2}

\end{document}